\documentclass[twocolumn]{aastex631}

\shorttitle{Maximum attainable energies of supernova remnants}
\shortauthors{Suzuki et al.}
\usepackage[referable]{threeparttablex}
\usepackage{amsmath}
\usepackage{color}
\usepackage{multirow}
\usepackage{amssymb}
\usepackage{url}
\usepackage{breakurl}
\usepackage{here}

\def\approxprop{%
  \def\p{%
    \setbox0=\vbox{\hbox{$\propto$}}%
    \ht0=0.6ex \box0 }%
  \def\s{%
    \vbox{\hbox{$\sim$}}%
  }%
  \mathrel{\raisebox{0.7ex}{%
      \mbox{$\underset{\s}{\p}$}%
    }}%
}

\begin{document}

\title{Observational constraints on the maximum energies of accelerated particles in supernova remnants: low maximum energies and a large variety}

\correspondingauthor{H. Suzuki}
\email{hiromasa050701@gmail.com}

\author[0000-0002-8152-6172]{Hiromasa Suzuki}
\affiliation{Department of Physics, The University of Tokyo, 7-3-1 Hongo, Bunkyo-ku, Tokyo 113-0033, Japan}
\affiliation{Department of Physics, Faculty of Science and Engineering, Konan University, 8-9-1 Okamoto, Higashinada, Kobe, Hyogo 658-8501, Japan}

\author[0000-0003-0890-4920]{Aya Bamba}
\affiliation{Department of Physics, The University of Tokyo, 7-3-1 Hongo, Bunkyo-ku, Tokyo 113-0033, Japan}
\affiliation{Research Center for the Early Universe, The University of Tokyo, 7-3-1 Hongo, Bunkyo-ku, Tokyo 113-0033, Japan}

\author{Ryo Yamazaki}
\affiliation{Department of Physics and Mathematics, Aoyama Gakuin University, 5-10-1 Fuchinobe, Chuo-ku, Sagamihara, Kanagawa 252-5258, Japan}
\affiliation{Institute of Laser Engineering, Osaka University, 2-6 Yamadaoka, Suita, Osaka 565-0871, Japan}

\author[0000-0002-2387-0151]{Yutaka Ohira}
\affiliation{Department of Earth and Planetary Science, The University of Tokyo, 7-3-1 Hongo, Bunkyo-ku, Tokyo 113-0033, Japan}



\begin{abstract}

Supernova remnants (SNRs) are thought to be the most promising sources of Galactic cosmic rays.
One of the principal questions is whether they are accelerating particles up to the maximum energy of Galactic cosmic rays ($\sim$ PeV).
In this work, a systematic study of gamma-ray emitting SNRs is conducted as an advanced study of \cite{suzuki20b}.
Our purpose is to newly measure the evolution of maximum particle energies with increased statistics and better age estimates.
We model their gamma-ray spectra to constrain the particle-acceleration parameters.
Two candidates of the maximum energy of freshly accelerated particles, the gamma-ray cutoff and break energies, are found to be well below PeV.
We also test a spectral model that includes both the freshly accelerated and escaping particles to estimate the maximum energies more reliably, but no tighter constraints are obtained with current statistics.
The average time dependences of the cutoff energy ($\propto t^{-0.81 \pm 0.24}$) and break energy ($\propto t^{-0.77 \pm 0.23}$) cannot be explained with the simplest acceleration condition (Bohm limit), and requires shock-ISM (interstellar medium) interaction.
The average maximum energy during lifetime is found to be $\lesssim 20$~TeV $(t_{\rm M}/1~{\rm kyr})^{-0.8}$ with $t_{\rm M}$ being the age at the maximum, which reachs PeV if $t_{\rm M} \lesssim 10$~yr.
The maximum energies during lifetime are suggested to have a variety of 1.1--1.8 dex from object to object.
Although we cannot isolate the cause of this variety, this work provides an important clue to understand the microphysics of particle acceleration in SNRs.

\end{abstract}



\section{Introduction} \label{sec-intro}

Galactic cosmic rays are high-energy particles which have an energy distribution approximated by a power-law function with the maximum energy of $\approx 10^{15.5}$ eV ($\approx 3$~PeV) and the energy density of $\sim1$~eV~cm$^{-3}$ \citep{gloeckler67}.
Although it has been more than 100 years from their discovery, their acceleration sites are still unclear.
Supernova remnants (SNRs) are thought to be the most promising sources that can provide such a high maximum energy and large energy density.
According to analytical models of diffusive shock acceleration, they are believed to supply particles with energies of $\lesssim$~PeV (e.g., \citealt{bell78, lagage83}).

Gamma-ray observations have revealed that charged particles are accelerated to energies above TeV (10$^{12}$~eV) in young ($\lesssim 2$~kyr) SNRs (e.g., Cassiopeia~A by \citealt{ahnen17}; Tycho by \citealt{giordano12}; RX J1713.7$-$3946 by \citealt{hess18b}).
Most of these SNRs feature, however, spectral turnovers at energies below PeV.
Recent Tibet air-shower array and LHAASO (Large High Altitude Air Shower Observatory) observations have been successfully identifying a number of PeVatron candidates \citep{amenomori21, cao21}, which include only a few SNRs.
These results suggest that very-high-energy particles accelerated in young evolutionary stages have already escaped from the SNRs \citep{ptuskin03, ptuskin05, ptuskin08, caprioli09, ohira10, ohira11a, bell13, nava16, celli19}, or that SNRs simply cannot accelerate particles up to PeV.
Other scenarios have also been proposed in which SNRs in specific conditions are PeVatrons: e.g., SNRs with pulsars by \cite{ohira18}, very young SNRs in dense environments by \cite{schure13, marcowith18, cristofari20, inoue21}.
In any case, evaluation of maximum energies of fleshly accelerated particles by isolating contribution of particles which are no longer accelerated will be essential to examine if SNRs are PeVatrons or not.

Given that evolution of the maximum attainable energies of the particle acceleration processes is unknown, studying individual SNRs is insufficient to determine their maximum energies during lifetime.
And a systematic study with a large number of SNRs is required to extract their global trends.
In such a study, reliability of their ages is very important yet was not taken into account in previous studies \citep{zeng19, suzuki20b}.
\cite{suzuki21a} investigated the reliability of the age estimates and concluded that a systematic uncertainty of a factor of four is associated to the SNR ages except for limited number of systems where more reliable estimations are possible.

In this work, a systematic study of the gamma-ray emitting SNRs is conducted as an advanced study of \cite{suzuki20b} with increased statistics and better age estimates.
We especially aim to evaluate their maximum attainable energies during lifetime and their dependence on environments.

Our sample consists of 38 SNRs. This is described in Section~\ref{sec-sample}.
We analyze the latest data obtained with Fermi-LAT to extract the energy spectra of 15 SNRs out of 38 with the best statistics available. This is described in Appendix~\ref{sec-fermi}.
A systematic spectral modeling of the gamma-ray data is performed for all the 38 SNRs, and then systematic trends and dispersions of the gamma-ray parameters are summarized (Section~\ref{sec-modeling}).
We show that their current maximum energies are indicated to be well below PeV.
Particle-acceleration parameters including the maximum energies and spectral indices are discussed in Section~\ref{sec-discussion}.
Here, we argue that SNRs still can be PeVatrons under certain conditions.
Throughout the paper, errors in text, figures and tables indicate a 1$\sigma$ confidence range unless mentioned otherwise.

\section{Sample and Estimation of Age} \label{sec-sample}
The SNRs considered in this work are selected from the objects either in the 1st Fermi-LAT SNR catalog \citep{acero16b} (30~SNRs) or a preceding systematic gamma-ray study of SNRs \citep{zeng19} (35~SNRs).
Among them, the sample of this work consists of the objects with individually published gamma-ray spectra.
Resultant 38 SNRs are listed in Table~\ref{tab-gamma-sample}.

Two age estimates, plasma age $t_{\rm p}$ and dynamical age $t_{\rm dyn}$, are obtained primarily based on X-ray observations as follows.
Thermal electron number density $n_{\rm e}$ is used if available.\footnote{The number density $n_{\rm e}$ is calculated based on a thermal X-ray emission measure and an assumption of the X-ray emitting volume.}
For those either with ionizing plasmas or recombining plasmas, the ionization or recombination timescales $n_{\rm e}t$, respectively, are also extracted from the literature.
Then, the plasma age $t_{\rm p}$ is calculated by dividing $n_{\rm e}t$ with $n_{\rm e}$.\footnote{The timescale $n_{\rm e}t$ is estimated from an X-ray spectroscopy. Basically, we use $n_{\rm e}t$ for the whole SNR.
For SNRs with $n_{\rm e}t$ measured only for divided regions, we calculate their average values.}
The dynamical age $t_{\rm dyn}$, which is determined based on shock dynamics (from a combination of the parameters, SNR diameter $D$\footnote{The diameter $D$ is calculated based on the Green's SNR catalog \citep{green19a, green19b} with individual publications for distances (see Table~\ref{tab-gevsnrs}).}, shock speed, $n_{\rm e}$, etc.) estimated in previous works are also used.
For objects whose reliable age estimations are available (historical age, light-echo age, kinematic age of associated neutron star, and kinematic age of ejecta knots; summarized in \citealt{suzuki21a}), we adopt these values (referred to as $t_{\rm r}$).

For each object, the most reliable age, that is, the ``best age'', $t_{\rm b}$ is defined, and it is used in our analysis below. If available, we assume $t_{\rm r}$ as the best age estimate ($t_{\rm b} = t_{\rm r}$).
If only $t_{\rm p}$ and $t_{\rm dyn}$ are available and these are consistent within a factor of four, $t_{\rm p}$ is chosen since these two have no significant difference \citep{suzuki21a}.
In the following two cases, these two ages are different by more than a factor of four: RX J1713.7$-$3946 and W 51 C.
For RX J1713.7$-$3946, its plasma age $t_{\rm p}$ is too large considering its large shock velocities observed and bright X-ray and gamma-ray emissions. Thus, $t_{\rm dyn}$ is chosen as the best age.
This is probably because the object has almost completely non-thermal X-ray emission and its thermal X-ray emission has been barely detected only from a small region, and thus the plasma parameters derived may not be reliable \citep{katsuda15}.
For W 51 C, as \cite{sasaki14} mentions, $t_{\rm p}$ is probably too large because of a wrong estimation of the density. Thus $t_{\rm dyn}$ is adopted as $t_{\rm b}$.
In the case where only $t_{\rm dyn}$ is known, it is used as $t_{\rm b}$.
These parameters are summarized in Table~\ref{tab-gevsnrs}.

\begin{figure*}[h!]
\centering
\includegraphics[width=16cm, angle=0]{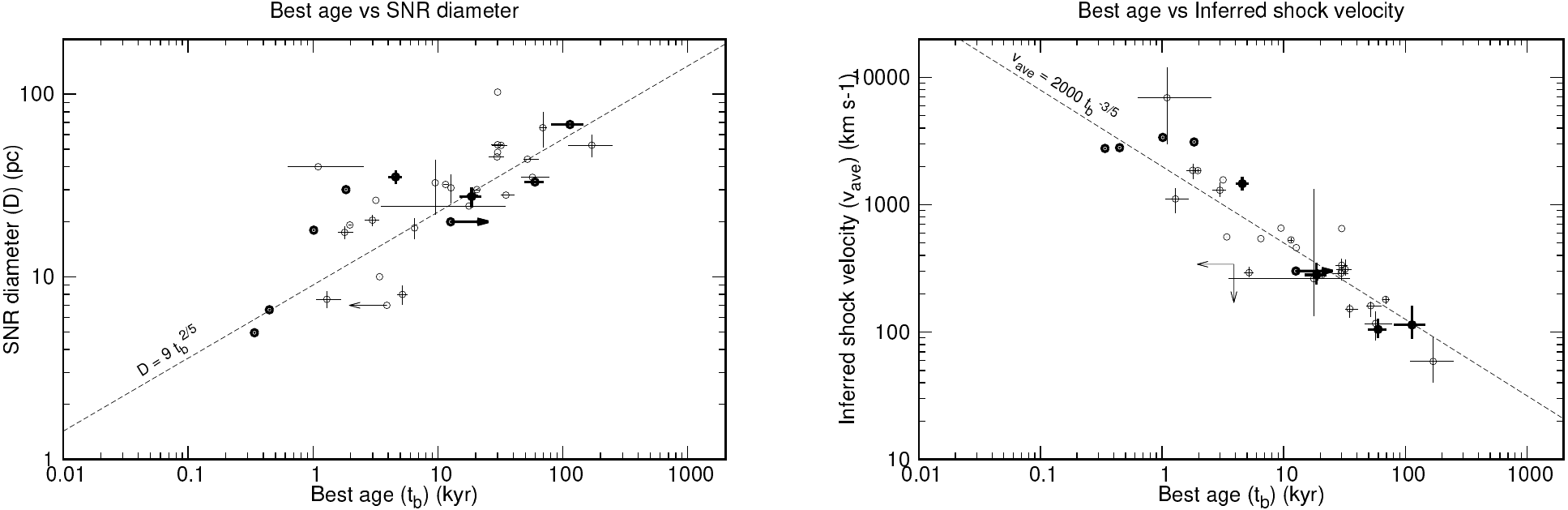}
\caption{
Left and right panels show the SNR diameter $D$ and inferred shock velocity $v_{\rm ave}$ as a function of $t_{\rm b}$, respectively.
See Section~\ref{sec-sample} for the definition of the ``best age'' $t_{\rm b}$.
The velocity $v_{\rm ave}$ is calculated as $v_{\rm ave} = D/5t_{\rm b}$.
Thick and thin crosses represent the SNRs with and without $t_{\rm r}$, respectively.
A representative Sedov model, for which a condition $D = C_D t^{2/5}$ is assumed with a parameter $C_D$ selected by eye to roughly match the observations, is also plotted with black-dashed lines.
\label{fig-diameter-velocity}}
\end{figure*}

Figure~\ref{fig-diameter-velocity} shows the diameter $D$ and inferred shock velocity $v_{\rm ave}$ as a function of $t_{\rm b}$.
We calculate this velocity assuming the Sedov model as $v_{\rm ave} = D/5t_{\rm b}$.
A representative Sedov model, for which a condition $D = C_D t^{2/5}$ is assumed and the parameter $C_D$ is selected by eye to roughly match the observations, is also plotted in Figure~\ref{fig-diameter-velocity}.
Both the $D$--$t_{\rm b}$ and $v_{\rm ave}$--$t_{\rm b}$ plots are roughly consistent with the representative Sedov model within a factor of four, indicating that the estimated age $t_{\rm b}$ is plausible.\footnote{See \cite{suzuki21a} for more details of the reliability of ages.}

\begin{table*}[htb!]
\fontsize{7}{9}\selectfont
\centering
\caption{Gamma-ray emitting SNRs used in this work\tablenotemark{a}} \label{tab-gamma-sample}
\begin{threeparttable}
\begin{tabular}{l l l}\hline\hline
SNR & GeV & TeV \\\hline

Cassiopeia~A~(G111.7$-$2.1) & $\checkmark$ & $\checkmark$ \\
CTB~109~(G109.1$-$1.0) & This work & \nodata \\
CTB~37~B~(G348.7+0.3) & $\checkmark$ & $\checkmark$ \\
Cygnus~loop~(G74.0$-$8.5) & This work & \nodata \\
G349.7+0.2 & $\checkmark$ & $\checkmark$ \\
Gamma-cygni~(G78.2+2.1) & $\checkmark$ & $\checkmark$ \\
Kes~79~(G33.6+0.1) & This work & \nodata \\
MSH~11-62~(G291.0$-$0.1) & This work & \nodata \\
MSH~11-56~(G326.3$-$01.8) & This work & \nodata \\
Puppis~A~(G260.4$-$3.4) & $\checkmark$ & upper limits \\
RCW~103~(G332.4$-$0.4) & This work & \nodata \\
RCW~86~(G315.4$-$2.3) & $\checkmark$ & $\checkmark$ \\
RX~J1713-3946~(G347.3$-$0.5) & $\checkmark$ & $\checkmark$ \\
SN~1006~(G327.6+14.6) & $\checkmark$ & $\checkmark$ \\
Tycho~(G120.1+1.4) & This work & $\checkmark$ \\
W~51~C~(G49.2$-$0.7) & $\checkmark$ & $\checkmark$ \\

3C~391~(G31.9+0.0) & This work & \nodata \\
CTB~37~A~(G348.5+0.1) & $\checkmark$ & $\checkmark$ \\
G166.0+4.3 & This work & \nodata \\
G359.1$-$0.5 & $\checkmark$ & $\checkmark$ \\
HB~21~(G89.0+4.7) & $\checkmark$ & \nodata \\
HB~9~(G160.9+2.6) & This work & \nodata \\
IC~443~(G189.1+3.0) & $\checkmark$ & $\checkmark$ \\
Kes~17~(G304.6+0.1) & This work & \nodata \\
W~28~(G6.4$-$0.1) & $\checkmark$ & $\checkmark$ \\
W~44~(G34.7$-$0.4) & $\checkmark$ & \nodata \\
W~49~B~(G43.3$-$0.2) & $\checkmark$ & $\checkmark$ \\

CTB~33~(G337.0$-$0.1) & $\checkmark$ & \nodata \\
G150.3+4.5 & $\checkmark$ & \nodata \\
G24.7+0.6 & $\checkmark$ & $\checkmark$ \\
G353.6$-$0.7 & $\checkmark$ & $\checkmark$ \\
G73.9+0.9 & This work & \nodata \\
HB~3~(G132.7+1.3) & $\checkmark$ & \nodata \\
Monoceros~nebula~(G205.5+0.5) & This work & \nodata \\
Vela~Jr.~(RX~J0852.0-4622;~G266.2$-$1.2) & $\checkmark$ & $\checkmark$ \\
S~147~(G180.0$-$1.7) & This work & \nodata \\
W~30~(G8.7$-$0.1) & $\checkmark$ & \nodata \\
W~41~(G23.3$-$0.3) & This work & $\checkmark$ \\
\hline
\end{tabular}

\tablenotetext{a}{Checkmarks indicate detections with Fermi-LAT (GeV band) or the ground-based observatories (TeV band), whereas ``This work'' indicates the objects whose Fermi-LAT data are analyzed in Appendix~\ref{sec-fermi}. The references are presented in Table~\ref{tab-gevsnrs}.
}

\end{threeparttable}
\end{table*}
\normalsize

\begin{longrotatetable}
\begin{deluxetable*}{*{7}{l}}
\tablecaption{Properties of the SNRs used in this study} \label{tab-gevsnrs}
\tablewidth{700pt}
\tabletypesize{\scriptsize}
\tablehead{Name & $D$ (pc) & Distance (kpc) & $n_{\rm e}$ (cm$^{-3}$) & $t_{\rm dyn}$ (kyr) & $t_{\rm p}$ (kyr) & $t_{\rm b}$ (kyr)\tablenotemark{a} }
\startdata
Cassiopeia~A~(G111.7$-$2.1) & 4.9 & 3.4 & 6 & 0.48 (0.43--0.52) & 1.1 & 0.340 (historical)  \\ 
CTB~109~(G109.1$-$1.0) & 24 & 2.79 & 1.1 & 14 (12--16) & 18 (3.5--35) & $= t_{\rm p}$   \\ 
CTB~37~B~(G348.7+0.3) & 40 & 13.2 & 2& 4.9 & 1.1 (0.63--2.5) & $= t_{\rm p}$   \\ 
Cygnus~loop~(G74.0$-$8.5) & 31 (25--36) & 0.54 & 2 & 18 (9--36) & 13 & $= t_{\rm p}$   \\ 
G349.7+0.2 & 7.5 (6.7--8.4) & 11.5 & 4.2 & 1.5 & 1.3 (1.1--1.7) & $= t_{\rm p}$   \\ 
Gamma-cygni~(G78.2+2.1) & 26 & 2.5 & 0.24 & 8 (6.8--10) & 3.2 & $= t_{\rm p}$   \\ 
Kes~79~(G33.6+0.1) & 19 & 7 & 1 & 12 (5.4--15) & 2 (1.9--2) & $= t_{\rm p}$   \\ 
MSH~11-62~(G291.0$-$0.1) & 20 (19--22) & 5 & 0.16 & 1.2 & 3 (2.6--3.4) & $= t_{\rm p}$   \\ 
MSH~11-56~(G326.3$-$01.8) & 45 & 4.1 & 0.15 & 16 & 30 (26--34) & $= t_{\rm p}$   \\ 
Puppis~A~(G260.4$-$3.4) & 35 (32--38) & 2.2 & 1 & 3.7 (3.4--4) & 7.4 (7.1--7.9) & 4.58 (4.01--5.15) (NS kinematic)  \\ 
RCW~103~(G332.4$-$0.4) & 10 & 3.3 & 5.7 & 3.2 & 3.4 & $= t_{\rm p}$   \\ 
RCW~86~(G315.4$-$2.3) & 30 & 2.5 & 2 & 1.3 (0.85--5.2) & 1.1 (0.32--1.9) & 1.835 (historical) \\ 
RX~J1713-3946~(G347.3$-$0.5) & 18 (16--19) & 1 & 0.1 & 1.8 (1.6--2.1) & 1.6 (1.3--1.9)$\times 10^2$ & $= t_{\rm dyn}$   \\ 
SN~1006~(G327.6+14.6) & 18 & 2.2 & 0.15 & 1.7 (1.7--1.8) & 5.2 (4.5--6) & 1.014 (historical)  \\ 
Tycho~(G120.1+1.4) & 6.6 & 3 & 0.13 & 1 (0.8--1.3) & 0.073 (0--0.073) & 0.448 (historical)  \\ 
W~51~C~(G49.2$-$0.7) & 48 & 4.3 & 0.07 & 30 & 1.2 (0.61--1.9)$\times 10^2$ & $= t_{\rm dyn}$   \\
\\
3C~391~(G31.9+0.0) & 18 (16--21) & 7.2 & 0.9 & 4.5 (4--5) & 45 (38--49) & $= t_{\rm p}$ \\ 
CTB~37~A~(G348.5+0.1) & 44 & 7.9 & 0.8 & 24 & 52 (48--64) & $= t_{\rm p}$   \\ 
G166.0+4.3 & 66 (51--80) & 4.5 & 0.9 & 24 & 69 (65--75) & $= t_{\rm p}$   \\ 
G359.1$-$0.5 & 28 & 3.29 & 0.7 & 70 & 19 (17--21) & $= t_{\rm p}$   \\ 
HB~21~(G89.0+4.7) & 52 (45--60) & 2.13 & 0.06 & 11 (10--12) & 1.7 (1.1--2.5)$\times 10^2$ & $= t_{\rm p}$   \\ 
HB~9~(G160.9+2.6) & 30 & 0.54 & 0.9 & 10 (0.8--20) & 20 (19--22) & $= t_{\rm p}$   \\ 
IC~443~(G189.1+3.0) & 20 & 1.8 & 1.6 & 4 & 12 (11--13) & $>12.6$ (NS kinematic)  \\ 
Kes~17~(G304.6+0.1) & 35 & 10 & 0.9 & 21 (2--40) & 57 (46--78) & $= t_{\rm p}$   \\ 
W~28~(G6.4$-$0.1) & 28 & 3.45 & 1 & 42 & 35 (32--41) & $= t_{\rm p}$   \\ 
W~44~(G34.7$-$0.4) & 28 (24--31) & 2.66 & 1 & 80 (60--100) & 20 (18--23) & 18.6 (14.9--22.4) (NS kinematic)  \\ 
W~49~B~(G43.3$-$0.2) & 8 (7--9) & 10 & 2.7 & 5.5 (5--6) & 5.2 (4.7--5.7) & $= t_{\rm p}$   \\\
\\
CTB~33~(G337.0$-$0.1) & 5.1 & 11 & \nodata & \nodata & \nodata & \nodata   \\ 
G150.3+4.5 & 19 & 0.4 & \nodata & \nodata & \nodata & \nodata   \\ 
G24.7+0.6 & 33 (22--44) & 2.73 & \nodata & 9.5 & \nodata & $= t_{\rm dyn}$   \\ 
G353.6$-$0.7 & 28 & 3.49 & \nodata & \nodata & \nodata & \nodata   \\ 
G73.9+0.9 & 32 & 4 & \nodata & 12 (11--12) & \nodata & $= t_{\rm dyn}$   \\ 
HB~3~(G132.7+1.3) & 53 & 2.2 & 0.32 & 30 (27--33) & \nodata & $= t_{\rm dyn}$   \\ 
Monoceros~nebula~(G205.5+0.5) & 100 & 1.13 & 0.003 & 30 & \nodata & $= t_{\rm dyn}$   \\ 
Vela~Jr.~(RX~J0852.0-4622;~G266.2$-$1.2) & 7 & 3.8 & 0.03 & 3.9 (0--3.9) & \nodata & $= t_{\rm dyn}$   \\ 
S~147~(G180.0$-$1.7) & 68 & 0.38 & \nodata & \nodata & \nodata & 114 (80.9--147) (NS kinematic)  \\ 
W~30~(G8.7$-$0.1) & 52 & 4.15 & 0.15 & 32 (27--36) & \nodata & $= t_{\rm dyn}$   \\ 
W~41~(G23.3$-$0.3) & 33 & 3.38 & \nodata & 1.3 (0.60--2.0)$\times 10^2$ & \nodata & 59.8 (49.1--70.4) (NS kinematic)  \\ 
\enddata

\tablenotetext{a}{The ``historical'' and ``NS kinematic'' respectively indicate that the age $t_{\rm b}$ is based on a historical document and kinematics of an associated neutron star (see \citealt{suzuki21a} for details).}
\tablecomments{
{Reference for ($D$; $t_{\rm dyn}$; $n_{\rm e}$ and/or $t_{\rm p}$; $t_{\rm b}$; gamma-ray spectrum; distance):}
Cassiopeia~A: (\citealt{green19a}, \citealt{green19b}, \citealt{reed95}; \citealt{patnaude09}; \citealt{murray79}; \citealt{green03}; \citealt{ahnen17}; \citealt{reed95});
CTB~109: (\citealt{green19a}, \citealt{green19b}, \citealt{zhao20}; \citealt{sasaki13}; \citealt{sasaki13}; \citealt{sasaki13}; \citealt{castro12}; \citealt{zhao20});
CTB~37~B: (\citealt{green19a}, \citealt{green19b}, \citealt{caswell75}; \citealt{aharonian08}; \citealt{aharonian08}; \citealt{aharonian08}; \citealt{xin16}; \citealt{caswell75});
Cygnus~loop: (\citealt{green19a}, \citealt{green19b}, \citealt{blair05}; \citealt{rappaport74}; \citealt{miyata94}; \citealt{miyata94}; \citealt{katagiri11}; \citealt{blair05});
G349.7+0.2: (\citealt{green19a}, \citealt{green19b}, \citealt{tian14}; \citealt{slane02}; \citealt{slane02}; \citealt{slane02}; \citealt{ergin15}, \citealt{hess15b}; \citealt{tian14});
Gamma-cygni: (\citealt{green19a}, \citealt{green19b}, \citealt{higgs77}, \citealt{lozinskaya00}; \citealt{leahy13}; \citealt{hui15}; \citealt{hui15}; \citealt{fraija16}, \citealt{aliu13}; \citealt{higgs77}, \citealt{lozinskaya00});
Kes~79: (\citealt{green19a}, \citealt{green19b}, \citealt{case98}; \citealt{giacani09}, \citealt{sun04}; \cite{giacani09}; \cite{giacani09}; \citealt{auchettl14}; \citealt{case98});
MSH~11-62: (\citealt{green19a}, \citealt{green19b}, \citealt{moffett01}, \citealt{moffett02}; \citealt{slane12}; \citealt{slane12}; \citealt{slane12}; \citealt{moffett01}, \citealt{moffett02});
MSH~15-56: (\citealt{green19a}, \citealt{green19b}, \citealt{rosado96}; \citealt{temim13}; \citealt{cesur19}; \citealt{cesur19}; \citealt{temim13}; \citealt{rosado96});
Puppis~A: (\citealt{green19a}, \citealt{green19b}, \citealt{reynoso03}; \citealt{winkler88}; \citealt{petre82}; this work; \citealt{hess15c}; \citealt{reynoso03});
RCW~103: (\citealt{green19a}, \citealt{green19b}, \citealt{carter97}; \citealt{braun19}; \citealt{braun19}; \citealt{braun19}; \citealt{xing14}; \citealt{carter97});
RCW~86: (\citealt{green19a}, \citealt{green19b}, \citealt{rosado96}, \citealt{sollerman03}; \citealt{williams11}; \citealt{lemoine12}; \citealt{green03}; \citealt{yuan14}; \citealt{rosado96}, \citealt{sollerman03});
RX~J1713$-$3946: (\citealt{green19a}, \citealt{green19b}, \citealt{fukui03}; \citealt{tsuji16}; \citealt{katsuda15}; \citealt{tsuji16}; \citealt{hess18b}; \citealt{fukui03});
SN~1006: (\citealt{green19a}, \citealt{green19b}, \citealt{winkler03}; \citealt{winkler14}; \citealt{yamaguchi08}; \citealt{green03}; \citealt{condon17}, \citealt{acero10}; \citealt{winkler03});
Tycho: (\citealt{green19a}, \citealt{green19b}, \citealt{tian11}, \citealt{hayato10}; \citealt{hughes00}; \citealt{hwang02};  \citealt{green03}; \citealt{giordano12}, \citealt{acciari11}; \citealt{tian11}, \citealt{hayato10});
W~51~C: (\citealt{green19a}, \citealt{green19b}, \citealt{koo97a}, \citealt{koo97b}, \citealt{green97}; \citealt{koo95}; \citealt{sasaki14}; \citealt{koo95}; \citealt{jogler16}; \citealt{koo97a}, \citealt{koo97b}, \citealt{green97});
3C~391: (\citealt{green19a}, \citealt{green19b}, \citealt{radhak72}; \citealt{chen01}; \citealt{sato14}; \citealt{sato14}; \citealt{ergin14}; \citealt{radhak72});
CTB~37~A: (\citealt{green19a}, \citealt{green19b}, \citealt{tian12}; \citealt{yamauchi14}; \citealt{yamauchi14}; \citealt{yamauchi14}; \citealt{abdollahi17}; \citealt{tian12});
G166.0+4.3: (\citealt{green19a}, \citealt{green19b}, \citealt{zhao20}; \citealt{burrows94}; \citealt{matsumura17}; \citealt{matsumura17}; \citealt{araya13}; \citealt{zhao20});
G359.1$-$0.5: (\citealt{green19a}, \citealt{green19b}, \citealt{wang20}; \citealt{ohnishi11}; \citealt{suzuki20a}; \citealt{suzuki20a}; \citealt{aharonian08b}, \citealt{hui16}; \citealt{wang20});
HB~21: (\citealt{green19a}, \citealt{green19b}, \citealt{zhao20}; \citealt{lazendic06}; \citealt{suzuki18}; \citealt{suzuki18}; \citealt{ambrogi19}; \citealt{zhao20});
HB~9: (\citealt{green19a}, \citealt{green19b}, \citealt{zhao20}; \citealt{leahy95}; \citealt{sezer19}; \citealt{sezer19}; \citealt{araya14}; \citealt{zhao20});
IC~443: (\citealt{green19a}, \citealt{green19b}, \citealt{zhao20}; \citealt{troja08}; \citealt{matsumura18}; this work; \citealt{ackermann13}; \citealt{zhao20});
Kes~17: (\citealt{green19a}, \citealt{green19b}, \citealt{caswell75}; \citealt{gelfand13}; \citealt{washino16}; \citealt{washino16}; \citealt{gelfand13}; \citealt{caswell75});
W~28: (\citealt{green19a}, \citealt{green19b}, \citealt{wang20}; \citealt{li10}; \citealt{okon18}; \citealt{okon18}; \citealt{cui18}; \citealt{wang20});
W~44: (\citealt{green19a}, \citealt{green19b}, \citealt{wang20}; this work; \citealt{uchida12}; this work; \citealt{ackermann13}; \citealt{wang20});
W~49~B: (\citealt{green19a}, \citealt{green19b}, \citealt{moffett94}; \citealt{zhou18}; \citealt{matsumura18}; \citealt{matsumura18}; \citealt{hess18a}; \citealt{moffett94});
CTB~33: (\citealt{green19a}, \citealt{green19b}, \citealt{sarma97}; \nodata; \nodata; \nodata; \citealt{castro13}; \citealt{sarma97});
G150.3+4.5: (\citealt{green19a}, \citealt{green19b}, \citealt{cohen16}; \nodata; \nodata; \nodata; \citealt{ackermann17}; \citealt{cohen16});
G24.7+0.6: (\citealt{green19a}, \citealt{green19b}, \citealt{wang20}; \citealt{leahy89}; \nodata; \citealt{leahy89}; \citealt{magic19});
G353.6$-$0.7: (\citealt{green19a}, \citealt{green19b}, \citealt{wang20}; \nodata; \citealt{doroshenko17}; \nodata; \citealt{condon17}; \citealt{wang20});
G73.9+0.9: (\citealt{green19a}, \citealt{green19b}, \citealt{wang20}; \citealt{lozinskaya93}; \nodata; \citealt{lozinskaya93}; \citealt{zdziarski16}; \citealt{pavlovic13});
HB~3: (\citealt{green19a}, \citealt{green19b}, \citealt{routledge91}; \citealt{lazendic06}; \citealt{lazendic06}; \citealt{lazendic06}; \citealt{katagiri16a}; \citealt{routledge91});
Monoceros~nebula: (\citealt{green19a}, \citealt{green19b}, \citealt{zhao20}; \citealt{leahy86}; \citealt{leahy86}; \citealt{leahy86}; \citealt{katagiri16b}; \citealt{zhao20});
Vela~Jr.: (\citealt{green19a}, \citealt{green19b}, \citealt{katsuda08}; \citealt{slane01}; \nodata; \citealt{slane01}; \citealt{aharonian07, tanaka11}; \citealt{katsuda08});
S~147: (\citealt{green19a}, \citealt{green19b}, \citealt{zhao20}; \nodata; \nodata; this work; \citealt{katsuta12}; \citealt{zhao20});
W~30: (\citealt{green19a}, \citealt{green19b}, \citealt{wang20}; \citealt{finley94}; \citealt{finley94}; \citealt{finley94}; \citealt{ajello12}; \citealt{wang20});
W~41: (\citealt{misanovic11}; \citealt{tian07}; \citealt{tian07}; this work; \citealt{hess15a}; \citealt{leahy08})
}
\end{deluxetable*}
\end{longrotatetable}
\normalsize

\section{Analysis and results} \label{sec-modeling}

\subsection{Analysis}
Here, a systematic analysis of the gamma-ray spectra obtained in Appendix~\ref{sec-fermi} and those in the literature is conducted.
The analysis is performed as follows.
Here, hadronic emissions originating from protons are assumed to dominate the gamma-ray spectra, which seems to be true at least for several SNRs based on their spectral shapes and energetics (e.g., \citealt{ackermann13}).
The exponential cutoff energies ($E_{\rm cut}$) or break energies ($E_{\rm br}$) of the gamma-ray spectra reflect the maximum energies of freshly accelerated particles \citep{ohira11a, celli19, brose20}.
Assuming that particles which are no longer accelerated do not significantly contribute to gamma-rays\footnote{Note that certain amount of the particles which are no longer accelerated and thus have higher energies than the current maximum energy of particles may still be in downstream, and contribute to the gamma-ray emission. Here, their contribution is ignored.}, we expect exponential-like cutoff features.
On the other hand, if the emission of escaping particles is significant as well, the gamma-ray spectra will be approximated with a broken power-law model.\footnote{Physically, they will have cutoff features above the breaks as well, but these will be hardly visible with current statistics.}
Since we cannot generally distinguish between these two situations without detailed properties of acceleration sites, the gamma-ray spectra are fitted with both an exponential cutoff power-law model,
\begin{equation}\label{eq-cutoffpl}
\frac{dN}{dE} = {\rm A}\, (E/\text{1 GeV})^{-\Gamma_{\rm cut}} {\rm exp}(-E/E_{\rm cut}),
\end{equation}
and a broken power-law model,
\begin{equation}
\frac{dN}{dE} =
\begin{cases}
{\rm A}\, (E/E_{\rm br})^{-\Gamma_{\rm br, low}} \quad (E < E_{\rm br})\\
{\rm A}\, (E/E_{\rm br})^{-\Gamma_{\rm br, high}} \quad (E \ge E_{\rm br}),
\end{cases}
\end{equation}
where A is a normalization factor.

Note that, only the energy spectra are used for modeling instead of using the spatial distributions as well including nearby sources, as in the case of the Fermi-LAT data analysis (Appendix~\ref{sec-fermi}).
This is an approximate way since the populations of spectral data points are all assumed to follow the Gaussian distributions independently.
This approximation is adopted to combine the Fermi-LAT data with those taken with the other gamma-ray observatories\footnote{Most of the data taken with the ground-based gamma-ray observatories are not accessible to the public, so that only the energy spectra made after data reduction and analysis presented in publications are available.}.
In this analysis, upper-limits are also included assuming that they follow a Gaussian probability distribution with the zero mean and standard deviation corresponding to the upper-limit value.

In this work, the Markov Chain Monte-Carlo (MCMC) algorithm is adopted to find the best-fit spectral parameters and their confidence ranges.
MCMCs are particularly useful in Bayesian inference because the posterior distributions are often difficult to work with via analytical examination.
In these cases, MCMCs give approximate aspects of posterior distributions that cannot be directly calculated (e.g., posterior means and standard deviations).

In our spectral analysis below, the ``Random Walk Metropolis-Hasting'' algorithm is used.
Let $\mu^i$ denotes a $d$-dimensional vector of the model parameter values at stage $i$ of the iteration. A candidate (proposal) of the next parameter set $\mu^*$ is defined as
\begin{equation}
\mu^{*}=\mu^{i}+c \Sigma^{1 / 2} W,
\end{equation}
where $c$ and $\Sigma$ are a scaling parameter and covariance matrix, respectively, and $W \sim N (0, I_d)$.
With a parameter $\alpha$ which is defined as
\begin{equation}
\alpha=\min \left\{1,\, p\left(\mu^{*} \mid D\right) / p\left(\mu^{i} \mid D\right)\right\},
\end{equation}
whether the proposal is accepted or rejected is determined as
\begin{equation}
\mu^{i+1}=\left\{\begin{array}{ll}
\mu^{*} & \quad (Z<\alpha) \\
\mu^{i} & \quad \text{(else)},
\end{array}\right.
\end{equation}
where $Z \sim {\rm Unif}(0, 1)$.
Thus, if the likelihood of the proposal state is smaller than the current value, the proposal is rejected with a 100\% probability.
In the opposite case, the proposal is accepted with a certain probability.

The algorithm adopted in this work is constructed utilizing a C++ library MCMCLib\footnote{\url{https://github.com/kthohr/mcmc}}.
In order to sample parameters properly, the scaling parameter of the proposal state $c$ is optimized for each run to get the average acceptance rate of $\approx 0.5$.
The delta-$\chi^2$ method is combined to MCMCs due to difficulties in determining accurate probability distributions of $E_{\rm cut}$, $E_{\rm br}$ and $\Gamma_{\rm br, high}$ with one MCMC run due to broad parameter ranges to be searched and local minima.
In the delta-$\chi^2$ method, searches for confidence regions of parameters are based on the difference between the best and current likelihood, $-2 \log(\Delta L) \approx \Delta\chi^2$.
We run a MCMC for a fixed ($E_{\rm cut}$, $E_{\rm br}$, $\Gamma_{\rm br, high}$) set with the other parameters kept free.
This process is repeated to cover all the sets of ($E_{\rm cut}$, $E_{\rm br}$, $\Gamma_{\rm br, high}$) within their parameter-search ranges.
Considering non-Gaussian probability distributions of $E_{\rm cut}$ and $E_{\rm br}$, we calculate 2$\sigma$ confidence ranges for them.
Iteration number of each MCMC run is $10^5$,  and we use the last $10^4$ iterations to create parameter histograms.

\subsection{Results}\label{sec-results}
The best-fit parameters are presented in Table~\ref{tab-gevresults}.
For $E_{\rm cut}$ and $E_{\rm br}$, 2$\sigma$ errors are presented.
While $E_{\rm cut}$ is constrained well for most of the sample, $E_{\rm br}$ cannot be determined for eight objects.
The objects with bad fits (if the null hypothesis probability is lower than 0.2\%) are excluded because their spectra are regarded as not being explained well.

\begin{figure*}[htb!]
\centering
\includegraphics[width=16cm, angle=0]{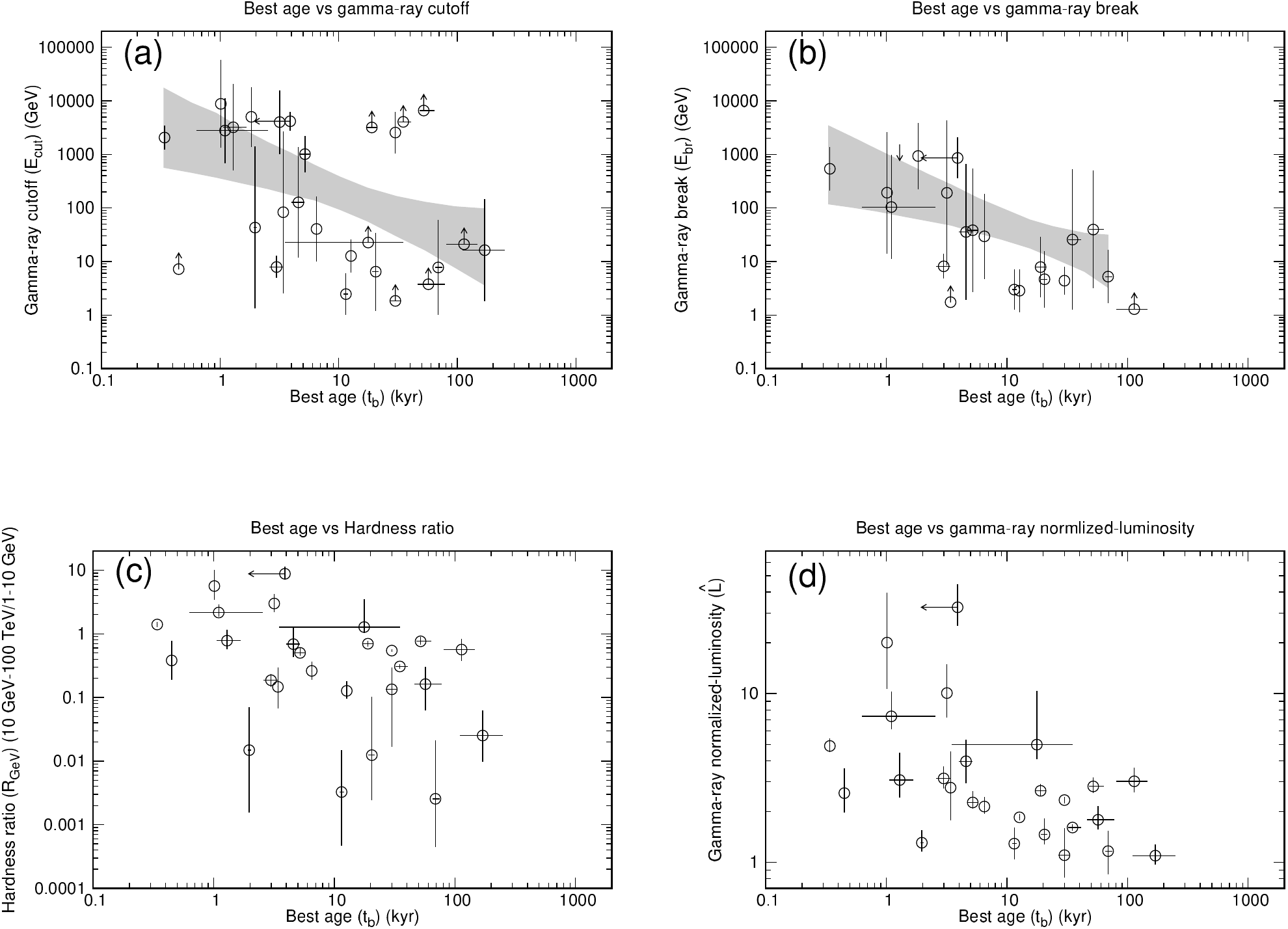}
\caption{Plots of $E_{\rm cut}$, $E_{\rm br}$, $R_{\rm GeV}$ and $\hat{L}$ over $t_{\rm b}$.
See Section~\ref{sec-sample} for the definition of the ``best age'' $t_{\rm b}$.
Grey regions represent the best-fit power-law functions and their $1\sigma$ confidence ranges.
\label{fig-cbhl}}
\end{figure*}

Overall, our results are consistent with \cite{suzuki20b}, and are better constrained with better statistics.
Figures~\ref{fig-cbhl} (a) and (b) show the plots of $E_{\rm cut}$ and $E_{\rm br}$ over the age $t_{\rm b}$, respectively.
Both of them show decreasing trends with age, which are consistent with \cite{zeng19} and \cite{suzuki20b}.
These $E_{\rm cut}$--$t_{\rm b}$ and $E_{\rm br}$--$t_{\rm b}$ plots are fitted with a power-law model, and the best-fit functions are obtained as $E_{\rm cut} = 1.3~(0.67\text{--}2.4)~{\rm TeV}~ (t_{\rm b} / 1~{\rm kyr})^{-0.81 \pm 0.24}$ and $E_{\rm br} = 270~(140\text{--}510)~{\rm GeV}~ (t_{\rm b} / 1~{\rm kyr})^{-0.77 \pm 0.23}$.
These functions and their $1\sigma$ confidence ranges are overplotted with grey regions.
{The cutoff energies might show two distinct populations, one of which is $\sim 0.1$--1 TeV (e.g., RX~J1713.7$-$3946) and the other is $\sim 1$--100 GeV (e.g., Cygnus Loop).}
Note that $E_{\rm cut}$ roughly corresponds to the highest gamma-ray energies whereas $E_{\rm br}$ is sometimes lower especially in the case where another curvature is required below the highest energies.

\begin{figure*}[htb!]
\centering
\includegraphics[width=16cm, angle=0]{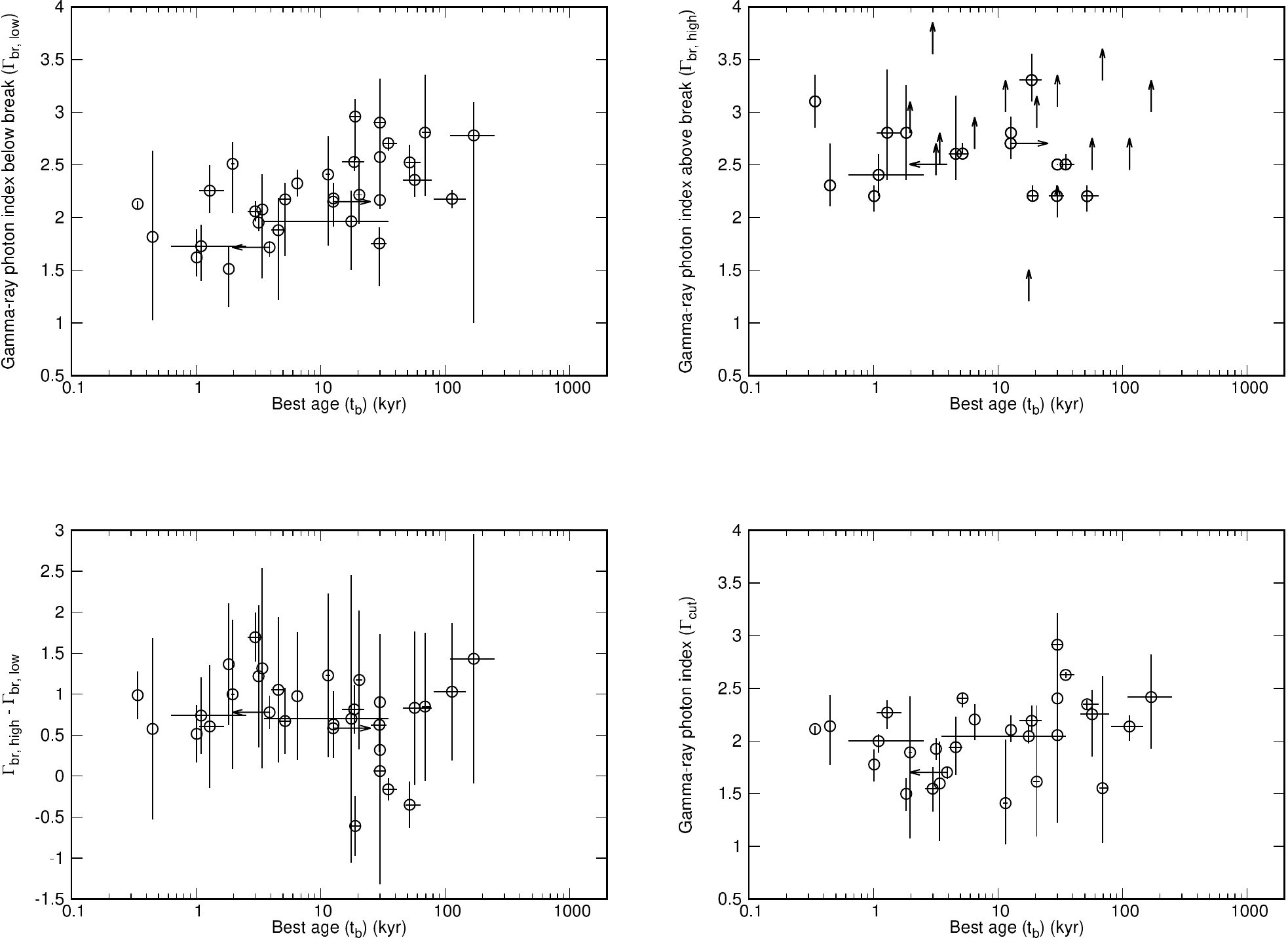}
\caption{Plots of $\Gamma_{\rm br, low}$, $\Gamma_{\rm br, high}$, their difference, and $\Gamma_{\rm cut}$ over $t_{\rm b}$.
See Section~\ref{sec-sample} for the definition of the ``best age'' $t_{\rm b}$.
\label{fig-index}}
\end{figure*}

The spectral indices of the broken power law model, $\Gamma_{\rm br, low}$ and $\Gamma_{\rm br, high}$, are plotted in Figure~\ref{fig-index}.
Distribution of the indices, $\Gamma_{\rm br, low}$ and $\Gamma_{\rm br, high}$, are found to be $2.2 \pm 0.4$ and $2.9 \pm 0.7$, respectively.\footnote{Errors indicate the standard deviations. Same for the other errors in this paragraph.}
The difference between these two, $\Gamma_{\rm br, high} - \Gamma_{\rm br, low}$, is calculated as $0.74 \pm 0.51$.
The spectral index of the cutoff power law model, $\Gamma_{\rm cut}$ have a distribution of $2.0 \pm 0.4$ (Figure~\ref{fig-index}).

In addition to $E_{\rm cut}$ and $E_{\rm br}$, two important parameters are introduced, namely the hardness ratio (ratio of the 10 GeV to 100 TeV and the 1--10 GeV luminosities; hereafter $R_{\rm GeV}$) and the normalized gamma-ray luminosity ($\hat{L}$). The latter is the luminosity (1~GeV--100~TeV) normalized at 1 GeV, i.e.,
\begin{equation}
\hat{L} = \left. \left(\int^{100~{\rm TeV}}_{1~{\rm GeV}} \frac{E\, dN\, (E)}{dE}\, dE\right)\, \middle/ \left. \frac{E^2\, dN (1~{\rm GeV})}{dE} \right. \right. .
\end{equation}
Ideally, the quantity $\hat{L}$ is governed by the maximum energy of the gamma-rays from freshly accelerated particles $E_{\rm max, \gamma}$ with a fixed spectral index of 2.0, and can be approximated as
\begin{equation}
\hat{L} \approx \ln(E_{\rm max, \gamma}/1~\text{GeV}).
\end{equation}
Figure~\ref{fig-cbhl} (c) and (d) show decreasing trends of $R_{\rm GeV}$ and $\hat{L}$ with increasing age $t_{\rm b}$, respectively, corresponding to the decreases of $E_{\rm cut}$ and $E_{\rm br}$.

\begin{figure*}[htb!]
\centering
\includegraphics[width=16cm]{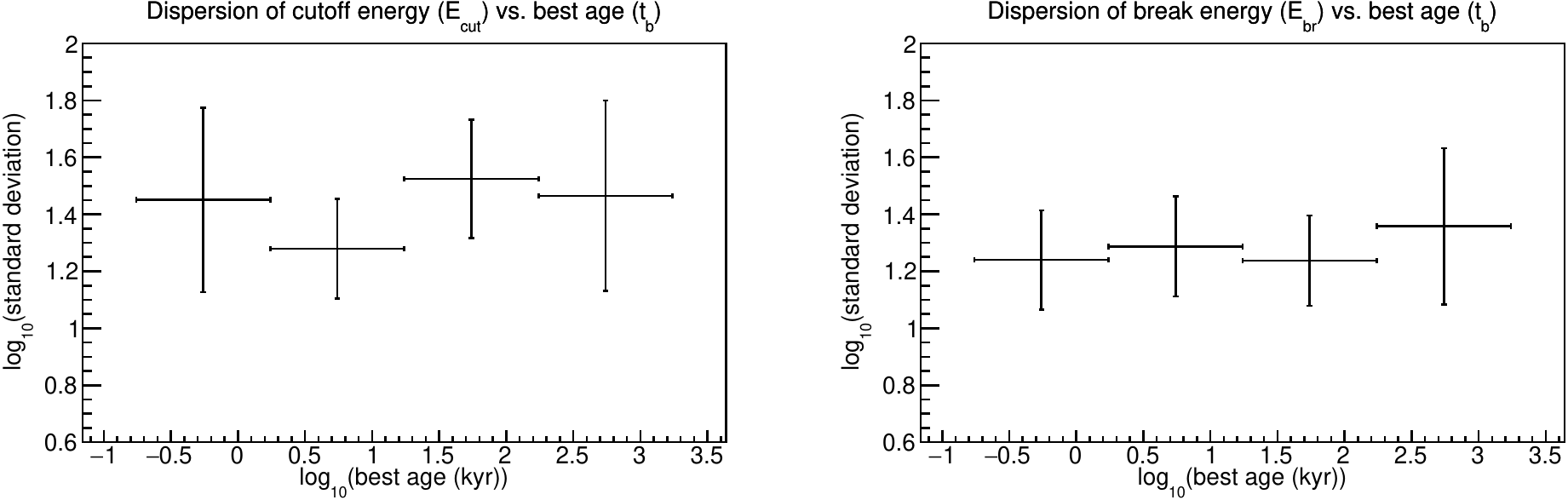}
\caption{Dispersion of $E_{\rm cut}$ and $E_{\rm br}$ as a function of $t_{\rm b}$.
See Section~\ref{sec-sample} for the definition of the ``best age'' $t_{\rm b}$.
The logarithmically divided age bins have a $\sim 1$~dex width each, which is close to the uncertainties associated with the ages for most SNRs.
\label{fig-dispersion}}
\end{figure*}

As seen in Figure~\ref{fig-cbhl}, large dispersions of the gamma-ray parameters at the same ages are indicated.
To evaluate these dispersions quantitatively, distributions of the dispersions as a function of age are plotted in Figure~\ref{fig-dispersion}.
We use the standard deviation of the values (e.g., $E_{\rm cut}$) as a measure of dispersion at individual age bins.
The logarithmically divided age bins have a $\sim 1$~dex width each, which is close to the uncertainties associated with the ages for most SNRs.
These parameter distributions are generated as follows:
randomized data plots (of, e.g., $E_{\rm cut}$--$t_{\rm b}$) considering errors of individual data points are generated.
For each plot, dispersions at individual age bins are calculated. This procedure is repeated for $10^4$ times, and the mean and standard deviation are evaluated for each age bin.
These correspond to the mean and 1$\sigma$ error values plotted in Figure~\ref{fig-dispersion}.
As seen in Figure~\ref{fig-dispersion}, dispersions of $E_{\rm cut}$ and $E_{\rm br}$ are found to be $1.1$--1.8 and $1.1$--1.6 orders of magnitude, respectively.

\begin{longrotatetable}
\begin{deluxetable*}{lllllllllll}
\tablecaption{Gamma-ray spectral parameters of our SNR sample} \label{tab-gevresults}
\tabletypesize{\fontsize{5}{6}\selectfont}
\tablehead{Name & $\Gamma_{\rm cut}$ & $\Gamma_{\rm br, low}$  & $\Gamma_{\rm br, high}$ & $E_{\rm cut}$ (GeV) & $E_{\rm br}$ (GeV) & L$_{\rm1\mathchar`-100 \,GeV}$\tablenotemark{a} & $\hat{L}$  & $R_{\rm GeV}$ & \multicolumn2c{$\chi^2$ (d.o.f.)}\tablenotemark{b} \\
& & & & & & & & & cutoff & broken}
\startdata
CassiopeiaA & 2.11 (2.07--2.15) & 2.13 (2.08--2.16) & 3.10 (2.85--3.36) & 2.3 (1.2--3.5)$\times10^{3}$ & 5.9 (2.1--14)$\times10^{2}$ & 6.86 (6.56--7.17)$\times 10^{34}$ & 4.89 (4.48--5.45) & 1.4 (1.28--1.53) & 22.6 (14) & 21.9 (13) \\ 
CTB109 & 2.04 (1.98--2.15) & 1.96 (1.51--2.26) & 2.10 (1.21--3.96) & $>$23 & --- & 1.19 (1.09--1.29)$\times 10^{34}$ & 4.98 (4.07--10.4) & 1.28 (1.03--3.5) & 9.8 (6) & 8.86 (5) \\ 
CTB37B & 2.00 (1.89--2.06) & 1.73 (1.40--1.93) & 2.41 (2.20--2.61) & 2.5 (0.69--11)$\times10^{3}$ & 60 (11--970) & 3.28 (2.62--3.75)$\times 10^{35}$ & 7.32 (6.13--10.3) & 2.16 (1.85--2.92) & 20.1 (13) & 15.9 (12) \\ 
Cygnus loop & 2.11 (1.99--2.24) & 2.18 (1.92--2.33) & 2.80 (2.55--2.96) & 11 (6.2--26) & 3.1 (1.1--7.2) & 1.55 (1.49--1.64)$\times 10^{33}$ & 1.85 (1.77--1.93) & 0.127 (0.0963--0.179) & 8.14 (5) & 6.86 (4) \\ 
G349.7+0.2 & 2.27 (2.11--2.39) & 2.26 (2.05--2.50) & 2.81 (2.35--3.40) & 1.4 (0.5--20)$\times10^{3}$ & $<$1500 & 4.1 (2.87--5.02)$\times 10^{35}$ & 3.07 (2.42--4.46) & 0.782 (0.569--1.17) & 9.43 (13) & 9.78 (12) \\ 
Gamma-cygni & 1.93 (1.82--2.03) & 1.95 (1.87--2.05) & 2.90 (2.40--3.96) & 3 (1--16)$\times10^{3}$ & 1.1 (8.5--4300) & 1.62 (1.31--1.87)$\times 10^{34}$ & 10.1 (7.24--14.9) & 2.99 (2.21--4.2) & 27.7 (22) & 26.2 (21) \\ 
Kes79 & 1.89 (1.07--2.42) & 2.51 (2.05--2.71) & 3.81 (2.80--3.96) & 3.1 (1.3--1400) & --- & 1.84 (1.61--2.12)$\times 10^{35}$ & 1.31 (1.16--1.55) & 0.0149 (0.00156--0.0699) & 4.97 (5) & 5.04 (4) \\ 
MSH11-62 & 1.55 (1.33--1.75) & 2.06 (1.96--2.16) & 3.91 (3.55--3.96) & 7.4 (4.9--13) & 11 (4.8--14) & 1.22 (1.14--1.28)$\times 10^{35}$ & 3.14 (2.72--3.69) & 0.187 (0.147--0.236) & 7.73 (6) & 11.1 (5) \\ 
MSH11-56 & 2.04 (1.95--2.11) & 1.75 (1.35--1.91) & 2.21 (2.20--2.31) & 80 (27--590) & 1.9 (1.1--3.3) & 7.52 (6.89--8.08)$\times 10^{34}$ & 3.57 (3.24--4.01) & 0.674 (0.518--0.865) & 23.8 (6) & 18.5 (5) \\ 
PuppisA & 1.94 (1.68--2.23) & 1.88 (1.21--2.19) & 2.61 (2.35--3.16) & 55 (12--1400) & 8.1 (1.9--660) & 2.66 (2.2--3.07)$\times 10^{34}$ & 3.97 (2.93--5.33) & 0.689 (0.427--1.25) & 7.06 (10) & 6.76 (9) \\ 
RCW103 & 1.60 (1.05--1.99) & 2.08 (1.42--2.41) & 3.80 (2.50--3.96) & 6.7 (2.5--2700) & $>$1.7 & 3.46 (2.76--3.99)$\times 10^{34}$ & 2.77 (1.77--4.56) & 0.147 (0.0676--0.295) & 7.15 (6) & 6.02 (5) \\ 
RCW86 & 1.50 (1.33--1.65) & 1.51 (1.15--1.73) & 2.80 (2.35--3.25) & 3 (1.4--18)$\times10^{3}$ & 1.3 (0.22--3.9)$\times10^{3}$ & 6.28 (3.71--8.62)$\times 10^{33}$ & 96.5 (43.6--219) & 21.2 (11.3--39.2) & 12.5 (6) & 8.99 (5) \\ 
RXJ1713.7-3946 & 1.69 (1.67--1.71) & 1.46 (1.43--1.52) & 2.30 (2.30--2.31) & 5.8 (4.8--6.8)$\times10^{3}$ & 4.5 (3.4--5.2)$\times10^{2}$ & 1.02 (0.957--1.1)$\times 10^{34}$ & 38.7 (35.1--42.8) & 10.5 (9.66--11.4) & 309 (37) & 274 (36) \\ 
SN1006 & 1.78 (1.61--1.92) & 1.62 (1.44--1.89) & 2.20 (2.05--2.31) & 3 (1.3--59)$\times10^{3}$ & 1.6 (0.14--26)$\times10^{2}$ & 1.66 (0.913--2.18)$\times 10^{33}$ & 20 (10.6--39.8) & 5.64 (3.38--10.1) & 17.8 (13) & 13.2 (12) \\ 
Tycho & 2.14 (1.77--2.44) & 1.82 (1.02--2.64) & 2.31 (2.11--2.71) & $>$7.2 & --- & 6.15 (4.97--7.28)$\times 10^{33}$ & 2.57 (1.97--3.6) & 0.381 (0.191--0.783) & 21.3 (8) & 16.2 (7) \\ 
W51C & 2.40 (2.38--2.43) & 2.17 (2.08--2.29) & 2.50 (2.50--2.51) & 2.3 (1--6.3)$\times10^{3}$ & 3.5 (2.4--7.9) & 3.85 (3.76--3.95)$\times 10^{35}$ & 2.33 (2.23--2.44) & 0.546 (0.522--0.576) & 32.5 (16) & 18.5 (15) \\ 
3C391 & 2.20 (2.01--2.35) & 2.32 (2.20--2.46) & 3.31 (2.65--3.96) & 28 (10--160) & 12 (4.7--180) & 1.92 (1.76--2.08)$\times 10^{35}$ & 2.14 (1.94--2.45) & 0.261 (0.188--0.367) & 9.58 (6) & 8.68 (5) \\ 
CTB37A & 2.35 (2.29--2.38) & 2.52 (2.37--2.69) & 2.20 (2.05--2.31) & $>$6600 & 45 (3.1--490) & 4.24 (3.8--4.55)$\times 10^{35}$ & 2.81 (2.58--3.19) & 0.763 (0.669--0.903) & 23.1 (9) & 13.9 (8) \\ 
G166.0+4.3 & 1.55 (1.03--2.62) & 2.81 (2.20--3.36) & 3.90 (3.30--3.96) & $<$59 & 4.2 (1.7--16) & 5.95 (4.34--7.23)$\times 10^{33}$ & 1.16 (0.848--1.54) & 0.00256 (0.000445--0.021) & 10.4 (4) & 9.36 (3) \\ 
G359.1-0.5 & 2.37 (2.32--2.40) & 2.96 (2.55--3.13) & 2.21 (2.15--2.31) & $>$3200 & 9 (2.1--29) & 5.06 (4.69--5.39)$\times 10^{34}$ & 2.65 (2.48--2.9) & 0.7 (0.627--0.787) & 25.1 (4) & 8.12 (3) \\ 
HB21 & 2.42 (1.93--2.82) & 2.78 (1.00--3.10) & 3.30 (3.00--3.96) & 5.6 (1.8--150) & --- & 1.49 (1.38--1.62)$\times 10^{34}$ & 1.09 (0.97--1.27) & 0.0252 (0.00962--0.0627) & 6.58 (3) & 7.4 (2) \\ 
HB9 & 1.61 (1.09--2.34) & 2.22 (1.94--2.53) & 3.61 (2.85--3.96) & 2.6 (1.2--34) & 3.1 (1.4--16) & 3.52 (3.09--4.38)$\times 10^{32}$ & 1.46 (1.28--1.81) & 0.0124 (0.0024--0.102) & 6.34 (4) & 6.21 (3) \\ 
IC443 & 2.24 (2.21--2.28) & 2.15 (2.06--2.23) & 2.70 (2.65--2.80) & 2.3 (1.2--3.7)$\times10^{2}$ & 8.1 (4.4--24) & 1.28 (1.25--1.3)$\times 10^{35}$ & 2.81 (2.69--2.91) & 0.603 (0.558--0.63) & 73.2 (21) & 38.4 (20) \\ 
Kes17 & 2.25 (1.85--2.49) & 2.36 (2.19--2.56) & 3.50 (2.45--3.96) & $>$3.7 & --- & 1.61 (1.38--1.86)$\times 10^{35}$ & 1.78 (1.56--2.15) & 0.162 (0.0618--0.305) & 10.6 (6) & 7.71 (5) \\ 
W28 & 2.63 (2.60--2.65) & 2.71 (2.63--2.75) & 2.50 (2.46--2.61) & $>$4100 & 23 (1.2--530) & 2.08 (2.01--2.15)$\times 10^{35}$ & 1.61 (1.54--1.68) & 0.306 (0.281--0.333) & 28.1 (11) & 22.9 (10) \\ 
W44 & 2.19 (2.09--2.34) & 2.53 (2.44--2.59) & 3.30 (3.10--3.55) & 7.4 (5.2--13) & 5.1 (2.6--7.5) & 2.1 (2.05--2.18)$\times 10^{35}$ & 1.47 (1.39--1.54) & 0.0632 (0.0523--0.083) & 25 (10) & 23.1 (9) \\ 
W49B & 2.40 (2.32--2.45) & 2.17 (1.63--2.33) & 2.61 (2.60--2.71) & 9.6 (4.7--22)$\times10^{2}$ & 6.1 (2.7--540) & 1.16 (1.08--1.22)$\times 10^{36}$ & 2.26 (2.07--2.63) & 0.502 (0.45--0.604) & 16.3 (13) & 15.4 (12) \\ 
CTB33 & 2.54 (2.30--2.62) & 2.57 (2.40--2.69) & 1.50 (1.00--3.96) & $>$9.1 & --- & 1.33 (1.14--1.45)$\times 10^{36}$ & 1.77 (1.46--2.13) & 0.336 (0.125--0.494) & 6.55 (2) & 5.78 (1) \\ 
G150.3+4.5 & 1.96 (1.82--2.19) & 3.79 (3.27--4.00) & 1.90 (1.85--2.05) & $>$370 & --- & 9.4 (7.32--12.9)$\times 10^{32}$ & 12.3 (2.87--34) & 4.05 (1.11--10.7) & 7.72 (1) & 4.59 (0) \\ 
G24.7+0.6 & 1.99 (1.96--2.03) & 2.01 (1.98--2.04) & 2.90 (2.65--3.51) & 1.9 (1.2--3.7)$\times10^{3}$ & 5.4 (3.1--10)$\times10^{2}$ & 4.63 (4.38--4.8)$\times 10^{34}$ & 7.18 (6.51--7.79) & 2.08 (1.93--2.24) & 79.3 (7) & 63.9 (6) \\ 
G353.6-0.7 & 1.50 (1.39--1.64) & 1.50 (1.34--1.70) & 2.51 (2.35--2.70) & 1.7 (0.91--3.9)$\times10^{3}$ & 5.4 (2.5--15)$\times10^{2}$ & 1.26 (0.882--1.84)$\times 10^{34}$ & 72.4 (35.6--128) & 15.7 (8.94--24.2) & 26.3 (21) & 22.5 (20) \\ 
G73.9+0.9 & 1.41 (1.02--2.01) & 2.41 (1.73--2.77) & 3.90 (3.00--3.96) & $<$6 & 3.5 (1.2--7.2) & 8.64 (7.15--9.89)$\times 10^{33}$ & 1.29 (1.04--1.6) & 0.00326 (0.000472--0.0149) & 7.55 (5) & 6.41 (4) \\ 
HB3 & 2.91 (1.96--3.22) & 2.90 (2.52--3.32) & 2.70 (2.00--3.96) & $>$1.8 & --- & 9.69 (7.35--11.3)$\times 10^{33}$ & 1.1 (0.813--1.59) & 0.134 (0.0167--0.294) & 9.5 (4) & 6.16 (3) \\ 
Monoceros & 2.06 (1.23--2.62) & 2.57 (2.23--2.98) & 3.90 (3.05--3.96) & 3.5 (1.1--310) & $>$2.3 & 8.44 (6.9--9.83)$\times 10^{33}$ & 1.2 (0.972--1.56) & 0.015 (0.00172--0.0466) & 10.9 (2) & 8.19 (1) \\ 
VelaJr. & 1.70 (1.64--1.76) & 1.72 (1.63--1.78) & 2.51 (2.35--2.61) & 4 (2.8--6.2)$\times10^{3}$ & 1.2 (0.36--2.1)$\times10^{3}$ & 1.81 (1.44--2.07)$\times 10^{35}$ & 32.5 (25.3--44.6) & 8.79 (7.19--11.4) & 27.7 (15) & 22.5 (14) \\ 
S147 & 2.14 (2.00--2.24) & 2.18 (2.09--2.26) & 3.91 (2.45--3.96) & $>$21 & $>$1.3 & 6.18 (5.42--6.83)$\times 10^{32}$ & 3.02 (2.59--3.63) & 0.566 (0.379--0.826) & 12.4 (5) & 9.81 (4) \\ 
W30 & 2.56 (2.52--2.59) & 2.76 (2.68--2.83) & 2.31 (2.15--2.36) & $>$390 & 6.7 (4.7--14) & 2.96 (2.86--3.07)$\times 10^{35}$ & 1.8 (1.69--1.92) & 0.376 (0.322--0.419) & 52.1 (2) & 21.2 (1) \\ 
W41 & 2.25 (2.2--2.29) & 1.74 (1.46--1.89) & 2.41 (2.40--2.46) & 1.9 (0.86--4.5)$\times10^{3}$ & 4.6 (2.4--9.6) & 8.27 (7.74--8.89)$\times 10^{34}$ & 3.31 (3.07--3.63) & 0.874 (0.801--0.957) & 59.4 (13) & 33.4 (12) \\ 
\enddata

\tablenotetext{a}{Luminosity in 1--100 GeV energy range calculated from the best-fit parameters of the exponential cutoff power law models.}
\tablenotetext{b}{The ``cutoff'' and ``broken'' indicate the fit statistics with the cutoff power law and broken power law models, respectively.}

\end{deluxetable*}
\end{longrotatetable}
\normalsize

\section{Discussion} \label{sec-discussion}
Here, physical parameters of particle-acceleration environments, in particular the maximum energies of freshly accelerated particles and spectral indices, are discussed.

\subsection{Maximum energies of freshly accelerated particles}\label{sec-emax}

\subsubsection{Simply based on observations}
We have derived the systematic trends approximated with a power-law function, $E_{\rm cut} = 1.3~(0.67\text{--}2.4)~{\rm TeV}~ (t_{\rm b} / 1~{\rm kyr})^{-0.81 \pm 0.24}$ and $E_{\rm br} = 270~ (140\text{--}510)~{\rm GeV}~ (t_{\rm b} / 1~{\rm kyr})^{-0.77 \pm 0.23}$.
With an assumption that either of these corresponds to the maximum energy of the gamma-rays from freshly accelerated particles, we can conclude that the maximum energies of freshly accelerated particles are far below PeV.
It is possible, however, that these results are biased due to the contribution of particles which are no longer accelerated.
We consider this possibility in detail in the next section.

\subsubsection{Based on a combination of observations and theoretical predictions}
\begin{figure*}[htb!]
\centering
\includegraphics[width=16cm, angle=0]{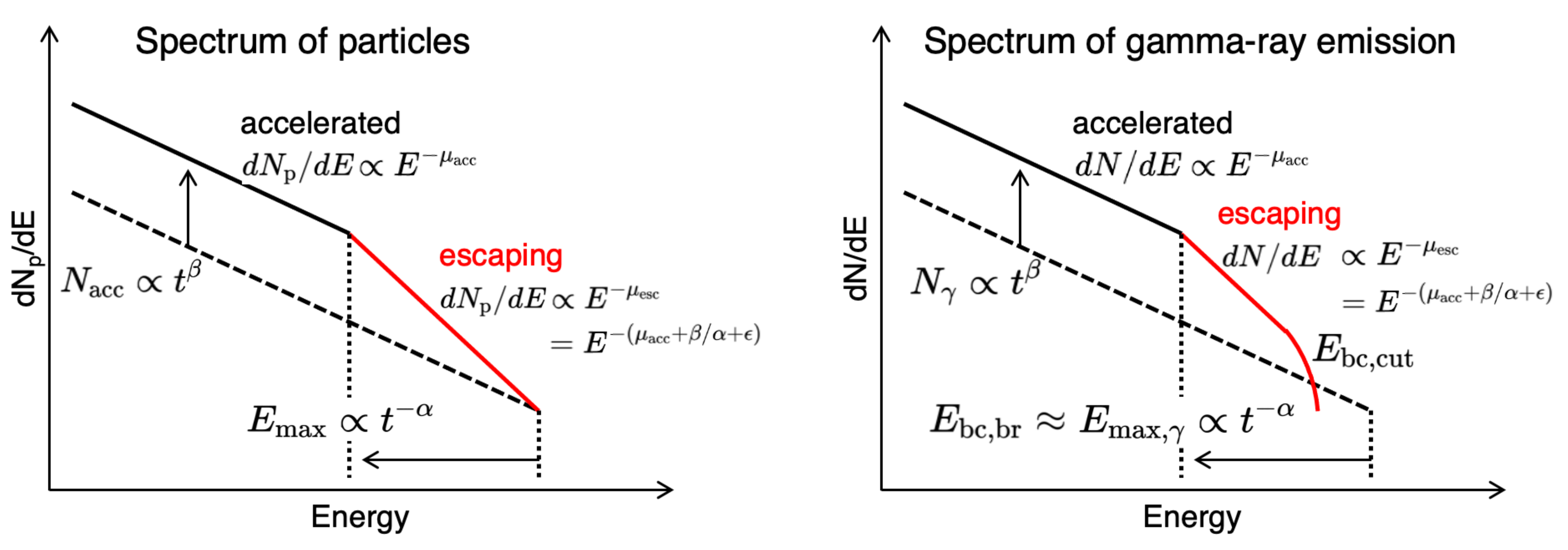}
\caption{Description of our analytical spectral model for escape-limited particle acceleration (based on \citealt{ohira10, ohira11a}).
Left and right panels describe the energy spectra of accelerated particles and their gamma-ray emission, respectively.
Parameters $N_{\rm acc}$, $N_{\gamma}$, $E_{\rm max}$ and $E_{\rm max, \gamma}$ represent the normalizations of freshly accelerated particles' and their gamma-rays' energy spectra, and maximum energies of particles and the gamma-rays from them ($E_{\rm max} \approx 10~ E_{\rm max, \gamma}$), respectively. Spectral indices of accelerated and escaping particles are described with $\mu_{\rm acc}$ and $\mu_{\rm esc}$, respectively.
Physical meanings of $\alpha, \beta$, and $\epsilon$ are explained in Section~\ref{sec-model}.
Break and cutoff energies in the gamma-ray spectrum are indicated with $E_{\rm bc, br}$ and $E_{\rm bc, cut}$.
\label{fig-escape}}
\end{figure*}

Here we use a spectral model which takes into account the contribution of escaping particles to constrain the freshly accelerated particles' maximum energies more reliably.
The model spectrum for total gamma-ray emission is defined as
\begin{equation}\label{eq-brcut}
\frac{dN}{dE} =
\begin{cases}
{\rm A_1}\, E^{-\mu_{\rm acc}}  &\quad (E < E_{\rm bc, br}) \\
{\rm A_2} \, E^{-\mu_{\rm esc}} \exp(-E/E_{\rm bc, cut}) &\quad (E \ge E_{\rm bc, br}),
\end{cases}
\end{equation}
where $\mu_{\rm acc}$, $\mu_{\rm esc}$, $E_{\rm bc, br}$, and $E_{\rm bc, cut}$ are the spectral indices of accelerated particles and escaping particles, and break and cutoff energies, respectively.
A condition ${\rm A_1} = E_{\rm bc, br}^{\mu_{\rm acc} - \mu_{\rm esc}} \exp(-E_{\rm bc, br} /E_{\rm bc, cut}) \, {\rm A_2}$ is kept.
This spectral model is described in Figure~\ref{fig-escape}.
We assume that the break energy $E_{\rm bc, br}$ is the best guess of the maximum energy $E_{\rm max, \gamma}$, but it is still possible that this energy corresponds to that of confined particles which are no longer accelerated \citep{ohira10, ohira11a, celli19, brose20}.
On the other hand, the cutoff energy $E_{\rm bc, cut}$ corresponds to the maximum energy of confined particles, that of escaping particles which contribute to gamma-ray emission, or the highest energy of particles attained during lifetime \citep{ohira10, ohira11a, celli19}.
We cannot distinguish between these possibilities with current low statistics of the gamma-ray spectra and limited spatial resolutions.

\begin{figure*}[htb!]
\centering
\includegraphics[width=10cm, angle=0]{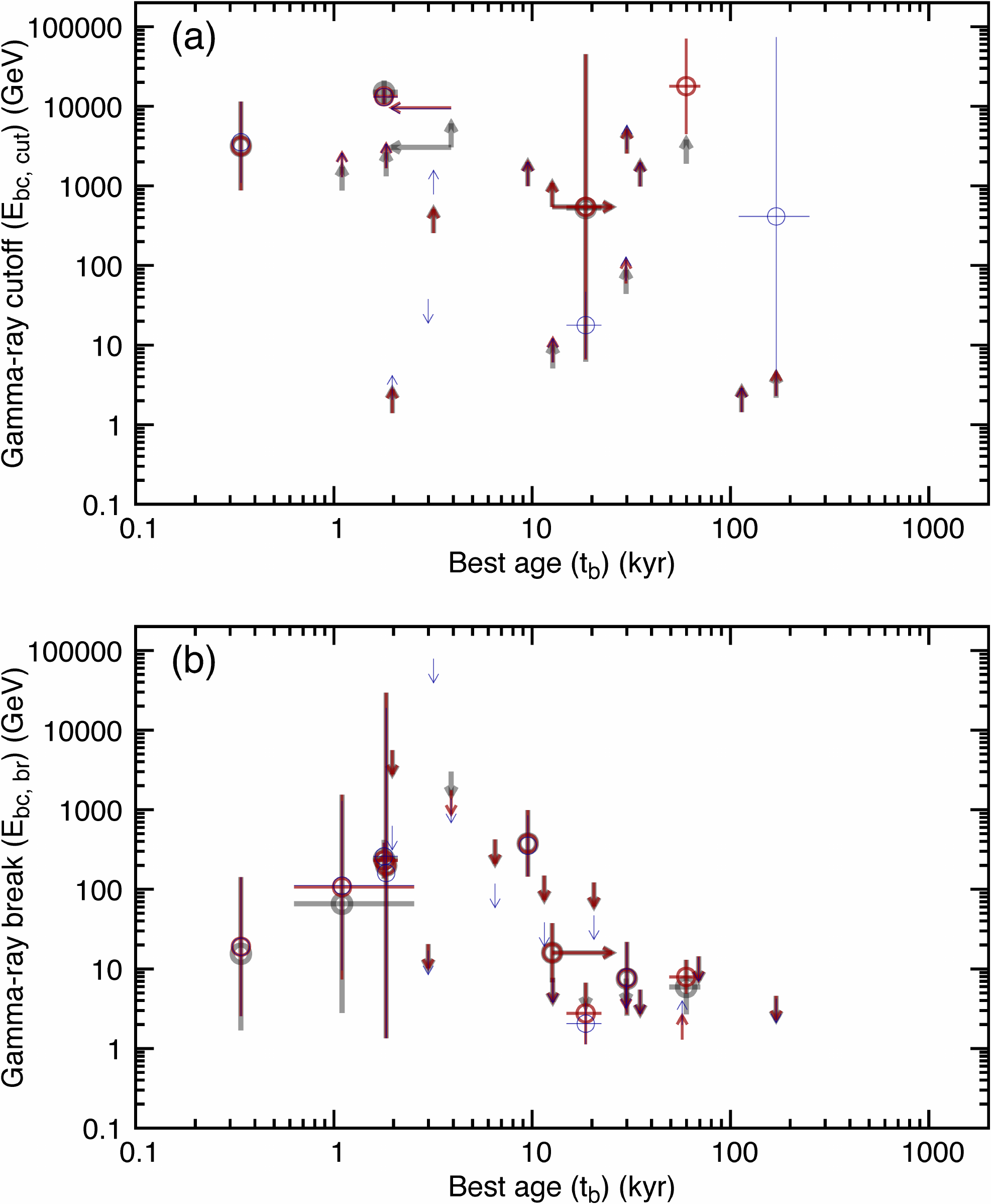}
\caption{
Plots of $E_{\rm bc, cut}$ and $E_{\rm bc, br}$ over $t_{\rm b}$.
See Section~\ref{sec-sample} for the definition of the ``best age'' $t_{\rm b}$.
Black, red, and blue crosses indicate the three cases (A), (B), and (C), respectively.
The conditions for individual cases are explained in Section~\ref{sec-emax}.
\label{fig-brcut}}
\end{figure*}

First, we fit this model to the data and determine the values $E_{\rm bc, br}$ and $E_{\rm bc, cut}$.
In this fit, we set a parameter-search range of $\mu_{\rm acc} = 1.5$--2.5 (hereafter, case (A)).\footnote{This range of $\mu_{\rm acc}$ is selected to explain both the observational gamma-ray index of RX~J1713.7$-$3946 and radio index of Cassiopeia~A. The other parameters' ranges are not bounded.}
The objects with too good fits (if the null hypothesis probability is higher than 75\%) are excluded because their spectra are regarded as being overfitted.
The results are shown in Figure~\ref{fig-brcut} with black crosses.
Here, meaningful constraints on $E_{\rm bc, cut}$ and $E_{\rm bc, br}$ are obtained only for about a half of the sample.
The distributions of $E_{\rm bc, cut}$ and $E_{\rm bc, br}$ are consistent with those of $E_{\rm cut}$ and $E_{\rm br}$, respectively (Figure~\ref{fig-cbhl}), but are less constrained.

Next, we add additional constraints on the parameter-search ranges based on theories, in order to see how the resultant $E_{\rm bc, br}$ and $E_{\rm bc, cut}$ changes.
According to a generalized spectral model composed of both freshly accelerated particles and escaping particles \citep{ohira10, ohira11a}, the spectral index $\mu_{\rm acc}$ is described as 
\begin{equation}
\mu_{\rm esc} = \beta/\alpha + \epsilon, 
\end{equation}
where the parameters $\alpha$, $\beta$, and $\epsilon$ respectively determine the evolutions of $E_{\rm bc, br}$ and particle injection rate, and the energy dependence of the diffusion coefficient around the source.
Hereafter, an additional condition, $\mu_{\rm esc} = 2.0$ if $\mu_{\rm acc} < 2.0$, is required \citep{ptuskin05, ohira10}.
We define the case (B), with parameter-search ranges of $\mu_{\rm acc} = 1.5$--2.5, $\alpha = 0.5$--3.0, $\beta = 0.0$--1.0, and $\epsilon = 0.0$--1.0.
These parameter ranges probably include most of the current theoretical predictions.
The results are shown in Figure~\ref{fig-brcut} with red crosses.

Finally, the case (C) is investigated, in which narrower parameter-search ranges of $\mu_{\rm acc} = 1.5$--2.5, $\alpha = 1.5$--3.0, $\beta = 0.6$ (particle injection by thermal leakage; \citealt{malkov95}), and $\epsilon = 0.0$ (without significant emission from escaping particles outside of the source) are assumed.
The range of $\alpha$ is chosen to satisfy $\mu_{\rm esc} = 2.2$--2.4, which is required to explain the spectral index of Galactic cosmic rays \citep{strong98, strong00, ptuskin06}.
The results under this condition are also presented in Figure~\ref{fig-brcut} with blue crosses.

Overall, the constraints on $E_{\rm bc, br}$ and $E_{\rm bc, cut}$ are similar for all the three cases (A), (B), and (C).
An important finding here is that the distributions of $E_{\rm bc, br}$ and $E_{\rm bc, cut}$ are found to be roughly consistent with those of $E_{\rm cut}$ and $E_{\rm br}$, respectively, and no tighter constraints are obtained.\footnote{Those with relatively well constrained parameters include ``W28-like'' and ``1713-like'' objects defined in Appendix~\ref{sec-classification}.}
Thus, the conclusion is the same as that simply based on $E_{\rm cut}$ and $E_{\rm br}$.
It will be worth trying to quantify the systematic trend of $E_{\rm bc, br}$ as our best guess of the maximum energy $E_{\rm max,\gamma}$ if we have better statistics for quantitative discussions.
The spectral indices $\mu_{\rm acc}$ and $\mu_{\rm esc}$ are poorly constrained for all the cases as well.
With the observatories such as LHAASO \citep{cao10} and CTA (Cherenkov Telescope Array; \citealt{actis11}), the spectral forms and spatial distributions of the gamma-ray emitting SNRs will be more constrained and thus the contribution of freshly accelerated particles will be more clearly extracted from the data.
If we know the absolute energy amount or normalization of the energy distribution of particles emitting gamma-rays, we can get an estimate of the injection parameter $\beta$.
Such an approach will be possible when the target proton densities for hadronic gamma-ray emission can be estimated reliably for a large number of objects.
A recent method on the gas density estimation by \cite{sano21a, sano21b} will be promising in this regard.

\subsection{Comparison of the estimated maximum energies with theoretical calculations} \label{sec-model}
\begin{figure*}[htb!]
\centering
\includegraphics[width=16cm, angle=0]{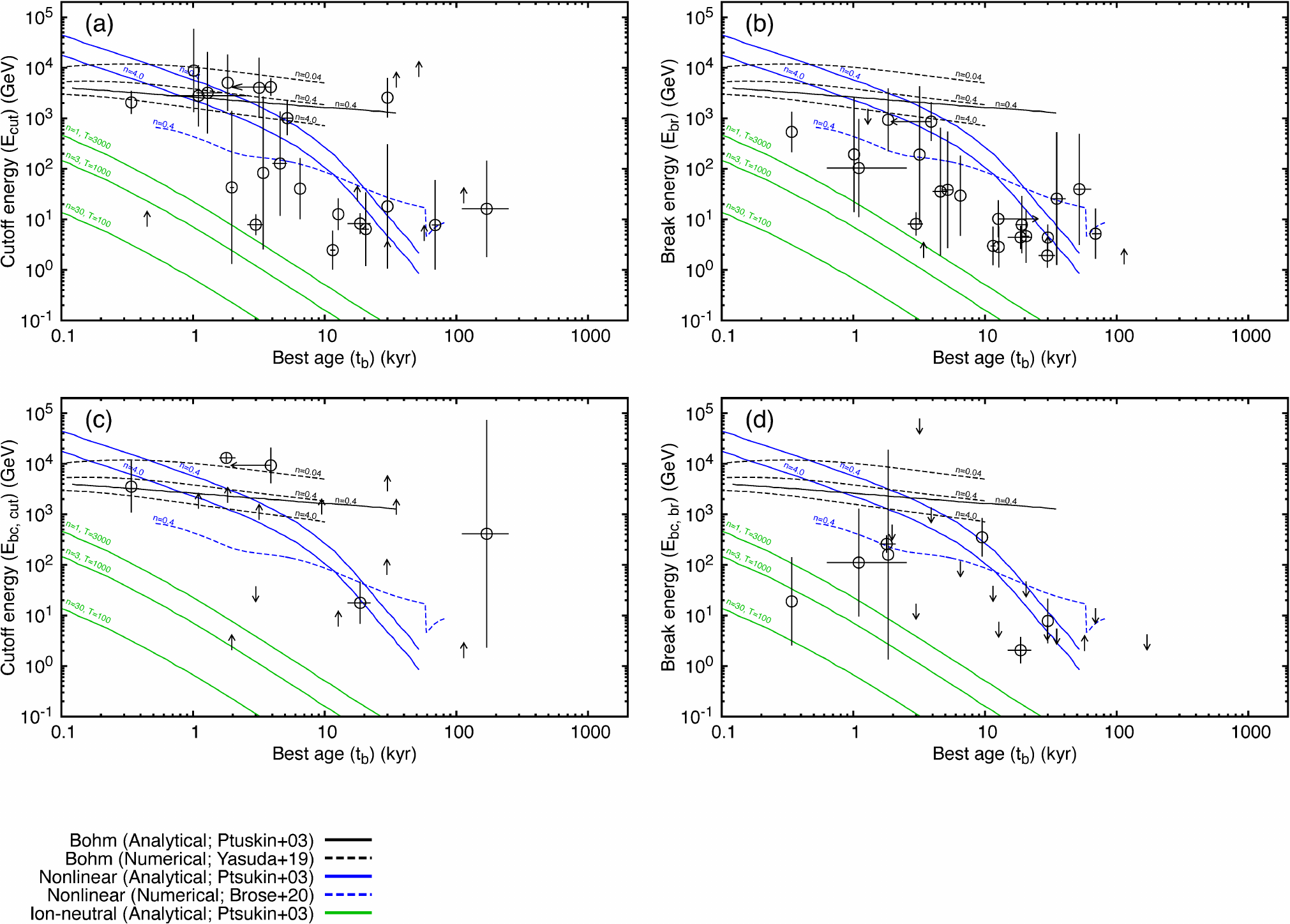}
\caption{
Plots of $E_{\rm cut}$, $E_{\rm br}$, $E_{\rm bc, cut}$, and $E_{\rm bc, br}$ over $t_{\rm b}$ with theoretical calculations.
For the parameters $E_{\rm bc, cut}$ and $E_{\rm bc, br}$, the constraints in the case (C) are plotted.
See Section~\ref{sec-sample} for the definition of the ``best age'' $t_{\rm b}$.
In all the panels, black, blue and green lines correspond to the Bohm limit, nonlinear damping, and ion-neutral damping models, respectively.
The quantities $n$ and $T$ shown beside the model lines stand for the ambient density (cm$^{-3}$) and temperature (K), respectively.
Several cases corresponding to several values of $n$, which are usually thought to be realistic, are plotted.
The ambient density refers to the cloud and ISM density for ``Ion-neutral (Analytical)'' case and for the other cases, respectively.
In the case of the ion-neutral damping, these values are set to satisfy a condition $nT = 3000$ K cm$^{-3}$ \citep{wolfire95}.
\label{fig-plots-with-theories}}
\end{figure*}

Figure~\ref{fig-plots-with-theories} shows comparisons between the observational parameters obtained in Section~\ref{sec-modeling} and several theoretical models.
Figure~\ref{fig-plots-with-theories} includes analytical models for the maximum energies of the gamma-ray emission in the Bohm limit, the acceleration with wave damping by nonlinear wave-wave interactions caused by shock-ISM (interstellar medium) collisions, and the acceleration with the wave damping by shock-cloud collisions \citep{ptuskin03}.
These are the maximum energies of accelerated protons multiplied by 0.1 to approximate those of the emission spectra.
The three cases include wave amplification by accelerated particles.
{Note that the wave damping is negligible and the wave amplification by accelerated particles determines the maximum energies in the fast-shock (young-age) limit of the nonlinear wave-damping case.}
In all the cases, an upstream magnetic field strength ($B_0$) of 5 $\mu$G is assumed.
This value is thought to be reasonable at least for SNRs which are older than 2~kyr (e.g., \citealt{bamba05}).
In the case of ion-neutral damping, the slowing down of a shock in a cloud is considered as
\begin{equation}
v_{\rm c} \approx \frac{v_{0}}{1 + (n_{\rm c}/n_{\rm 0})^{0.5}}, 
\end{equation}
where $v_0$, $v_{\rm c}$, $n_{\rm 0}$ and $n_{\rm c}$ are the shock velocities before and after the collision with clouds, and densities of the intercloud region (assumed to be 1 cm$^{-3}$) and the cloud, respectively \citep{chevalier99}.
In these models, maximum energies represent exact highest energies which freshly accelerated particles at each time can reach.

Figure~\ref{fig-plots-with-theories} includes numerical-calculation results of the maximum energies of accelerated protons with Alfv\'enic diffusion multiplied by 0.1 obtained by \citealt{brose20}.
In this case, the maximum energies are calculated by fitting the time-integrated particle spectra downstream of the shock with an exponential cutoff power-law model, as in the case of the analysis in Section~\ref{sec-modeling}.
Their calculations include nonlinear wave damping, thus the diffusion coefficient is time dependent.
They only consider the resonant amplification of Alfv\'enic turbulence by the accelerated particles.

Figure~\ref{fig-plots-with-theories} also shows numerical calculations of the maximum energies of protons in a uniform ISM case multiplied by 0.1 by \citealt{yasuda19}.
Their simulations assume the Bohm diffusion with $B_0 = 4~\mu$G.
In this case, the maximum energies represent exact highest energies which freshly accelerated particles at each time can reach.

From observations, the average time dependence of $E_{\rm cut}$ and $E_{\rm br}$ are found to be $E_{\rm cut} \propto t^{-0.81 \pm 0.24}$ and $E_{\rm br} \propto t^{-0.77 \pm 0.23}$ (Section~\ref{sec-modeling}).
Individual theoretical curves have time dependences of ``Bohm (Analytical)'': $\propto t^{-0.2}$, ``Bohm (Numerical)'': $\approxprop t^{-0.3}$, ``Nonlinear (Analytical)'': $\propto t^{-0.8}$ (in the fast-shock limit) or $\propto t^{-3.8}$ (in the slow-shock limit), ``Nonlinear (Numerical)'': $\approxprop t^{-0.8}$, and ``Ion-neutral (Analytical)'': $\propto t^{-1.8}$.
Thus, only the nonlinear wave-damping (shock-ISM collision) cases match the observational systematic time-dependence of $E_{\rm cut}$ and $E_{\rm br}$.
Compared to the nonlinear wave-damping models, all the derived parameters, $E_{\rm cut}$, $E_{\rm br}$, $E_{\rm bc, cut}$, and $E_{\rm bc, br}$, exhibit similar or slightly smaller values.
If the observational values are actually smaller than the theoretical predictions, it might suggest that there is a bias that the maximum energies measured from observations are smaller than the highest energies achieved in the systems.\footnote{Such a bias will be expected, for example, if the gamma-ray-dominant regions are interacting with dense gases and have decelerated shock velocities, so that the maximum energies achieved there are lower than those in the other regions.}

\subsection{Are Supernova Remnants PeVatrons ?}
According to our results for $E_{\rm cut}$, $E_{\rm br}$, $E_{\rm bc, cut}$, and $E_{\rm bc, br}$, the maximum energy of the gamma-rays from accelerated particles achieved at 1~kyr is $\lesssim 2$~TeV.
This corresponds to the proton energy of $\lesssim 20$ TeV, which is well below PeV.
These estimates are consistent with the fact that only a few SNRs are potential PeVatrons according to the recent observations at the energies around PeV \citep{cao21, amenomori21}.
If the maximum energy during lifetime is attained at a much smaller age, $t_{\rm M}$ (e.g., \citealt{inoue21}), these estimates can increase to $\lesssim 1$~PeV $(t_{\rm M}/10~{\rm yr})^{-0.8}$ simply based on the measured time dependence of $E_{\rm cut}$ and $E_{\rm br}$.
Also, the actual maximum energies could be larger than our estimates if the possible bias in $E_{\rm cut}$ and $E_{\rm br}$ is taken into account (Section~\ref{sec-model}).

\cite{tsuji21} studied the age dependence of the maximum energies of accelerated electrons.
They found that the maximum energies increased with time and got highest at $\sim 1$~kyr, where an efficient acceleration close to the Bohm limit case seemed to be operating.
If protons are accelerated in the same way, they can reach $\sim 100$~TeV at 1~kyr (see the theoretical calculations for the Bohm limit case in Figure~\ref{fig-plots-with-theories}), which is higher than our estimates ($\lesssim 20$~TeV).
This might be another clue for the potential bias in maximum energies of protons inferred from their gamma-ray emissions.

\subsection{Spectral Indices of Accelerated/Released Particles}
Spectral indices of freshly accelerated particles, which will be close to $\Gamma_{\rm br, low} = 2.2 \pm 0.4$ or $\Gamma_{\rm cut} = 2.0 \pm 0.4$, are consistent with $\approx 2.0$, which is usually predicted by the diffusive shock acceleration theories.
Those of escaping particles, which will be close to $\Gamma_{\rm br, high} = 2.9 \pm 0.7$, are also consistent with those required to explain Galactic cosmic rays ($\approx 2.2$--2.4; \citealt{strong98, strong00, ptuskin06}).\footnote{SNRs with softer indices $\Gamma_{\rm br, high} > 2.4$ can be explained by the softening due to energy-dependent diffusion around the sources (in other words, due to $\epsilon$; see Section~\ref{sec-emax}).}

\subsection{Average Acceleration Environment of Supernova Remnants}
Comparing the time dependence of $E_{\rm cut}$ and $E_{\rm br}$ to the theoretical calculations, we have found that only the shock-ISM collision case with high shock velocities matches the observations (see Section~\ref{sec-model}).
Thus, the acceleration regions where the produced gamma-rays dominate the total emission will be described with the shock-ISM collision models.

It may be natural to assume that $\alpha \approx 0.8 \pm 0.2$ based on the time dependence of $E_{\rm cut}$ and $E_{\rm br}$.
This assumption, however, can be problematic if combined with the natural condition $\beta = 0.6$ (particle injection by thermal leakage), because this combination leads to a condition $\mu_{\rm esc} > \mu_{\rm acc} + 0.6$ and thus the spectral index $\mu_{\rm esc}$ becomes softer than the spectral indices of Galactic cosmic rays ($\approx 2.2$--2.4) if $\mu_{\rm acc} \approx 2.0$.
We hope that this potential problem will also be addressed with future observations.

It should be noted that SNRs with hadronic gamma-ray emissions are often interacting with dense clouds, and thus the acceleration environments might have to be described by the ion-neutral (shock-cloud) collision models rather than those with the shock-ISM collision \citep{ptuskin03, ptuskin05}.
This might contradict our findings, so that some modifications to the inferred acceleration environment might be required (e.g., suppose the presence of clumpy clouds around a shock, they can be ionized (evaporate) due to shock heating; \citealt{white91, slavin17, zhang19}).
Numerical calculations of the acceleration and emission including both the high-velocity and low-velocity regions corresponding to thin and thick environments will be helpful.

\subsection{Variety of Maximum Energies of Accelerated Particles}
The parameters $E_{\rm cut}$ and $E_{\rm br}$ are found to have significant varieties of 1.1--1.8 dex at the same ages (Figure~\ref{fig-dispersion}).
This suggests that the maximum energies during lifetime have certain variety among SNRs.
In principle, this is attributed to fundamental parameters such as kinetic energy of supernova explosion, acceleration efficiency, and ambient density.
However, the exact form of the maximum energy of freshly accelerated particles at an age $t$, $E_{\rm max} (t)$, is complicated and dependent on situations around shocks \citep{ptuskin03, ptuskin05, bell13, recchia21}.
As an example, assuming the nonlinear wave-damping case, parameter dependence of $E_{\rm max} (t)$ at a certain age $t_0$ is
\begin{equation}\label{eq-nonlinear-prop}
E_{\rm max} (t_0) \propto  \xi_{\rm CR} v_{\rm sh}^3 \propto \xi_{\rm CR} E_{\rm SN, kin}^{3/5} n_{\rm e}^{-3/5} t_0^{-4/5}
\end{equation}
in the fast-shock limit and 
\begin{equation}\label{eq-nonlinear-prop2}
E_{\rm max} (t_0) \propto \xi_{\rm CR}^2 v_{\rm sh}^9 B_0^{-3} \propto \xi_{\rm CR}^2 E_{\rm SN, kin}^{8/5} n_{\rm e}^{-8/5} B_0^{-3} t_0^{-19/5}
\end{equation}
in the slow-shock limit, where $\xi_{\rm CR}$, $v_{\rm sh}$, $E_{\rm SN, kin}$, $n_{\rm e}$ and $B_0$ are the acceleration efficiency, shock velocity, kinetic energy of supernova explosion, average ambient density, and the background magnetic field, respectively \citep{ptuskin05}.
The variety of $E_{\rm cut}$ and $E_{\rm br}$ of 1.1--1.8 orders of magnitude is thus easily explained by less than one order-of-magnitude variations of one or more fundamental parameters, and uncertainties associated with the age $t_0$.
As an example, if the parameters $\xi_{\rm CR}$ and $v_{\rm sh}$ are doubled, the maximum energies $E_{\rm max} (t_0)$ change by factors of 16 and 2048 for Eqs.~\ref{eq-nonlinear-prop} and \ref{eq-nonlinear-prop2}, respectively.

\begin{figure*}[htb!]
\centering
\includegraphics[width=16cm, angle=0]{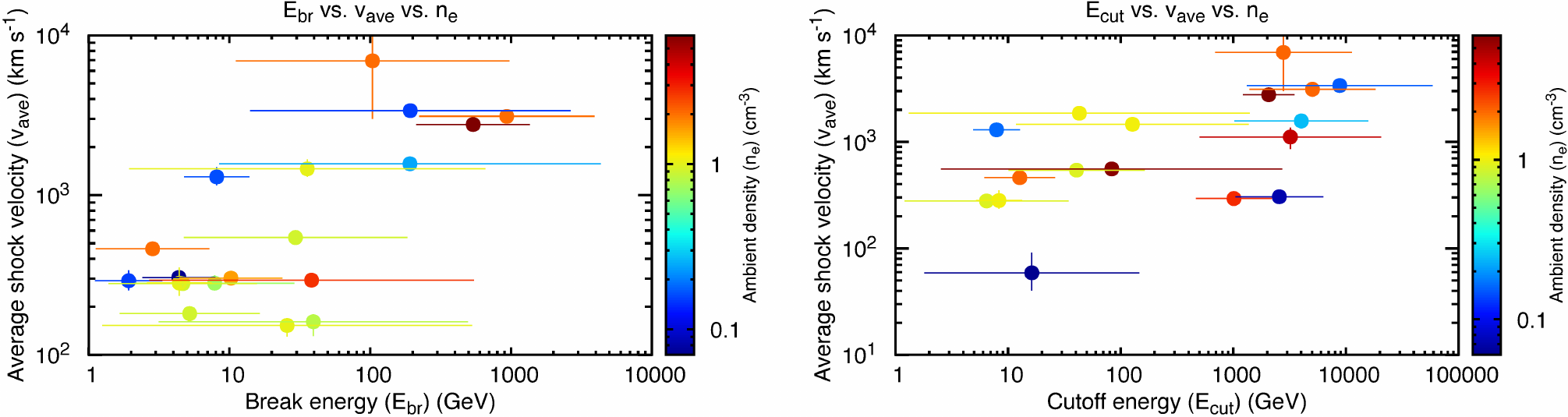}
\caption{
({\it left panel}) Scatter plot for the three parameters, break energy $E_{\rm br}$, average shock velocity $v_{\rm ave}$, and ambient density $n_{\rm e}$.
({\it right panel}) Same for the cutoff energy $E_{\rm cut}$, average shock velocity $v_{\rm ave}$, and ambient density $n_{\rm e}$.
\label{fig-correlation}}
\end{figure*}

We investigate the dependence of the break energy $E_{\rm br}$ and cutoff energy $E_{\rm cut}$ on the shock velocity $v_{\rm sh}$ (substituted by $v_{\rm ave} = D/5 t_{\rm b}$) and ambient density $n_{\rm e}$, to identify the cause of such a variety.
Figure~\ref{fig-correlation} shows the scatter plots of these parameters.
We cannot find the fundamental planes where the break energy $E_{\rm br}$ or cutoff energy $E_{\rm cut}$ can be described as a function of $v_{\rm ave}$ and/or $n_{\rm e}$.\footnote{The positive correlations between $E_{\rm br}$ and $v_{\rm ave}$, and $E_{\rm cut}$ and $v_{\rm ave}$, are due to the correlations between $v_{\rm ave}$ and $t_{\rm b}$ (Figure~\ref{fig-diameter-velocity}), $E_{\rm cut}$ and $t_{\rm b}$ (Figure~\ref{fig-cbhl} (a)), and $E_{\rm br}$ and $t_{\rm b}$ (Figure~\ref{fig-cbhl} (b)).}
This will be at least partly due to inappropriate parameter substitutions (e.g., shock velocities at gamma-ray dominant regions may differ from $v_{\rm ave}$).
Such correlation studies will be meaningful if more appropriate estimates of the parameters such as $v_{\rm sh}$ become available.

The difficulty in constraining physical parameters of particle acceleration is largely due to the uncertainty of the spectral index of freshly accelerated particles, $\mu_{\rm acc}$.
Thus it may be a good idea to use spectral indices of radio continuum emission as $\mu_{\rm acc}$.
However, because the radio spectral indices have spatial variation in a SNR and emission regions for radio and gamma-rays may be different, it may be inappropriate to use the radio spectral indices as $\mu_{\rm acc}$.
In fact, the radio spectral index of Cassiopeia~A ($\approx 2.5$) is different from that in the gamma-ray band ($\approx 2.2$) (e.g.,~\citealt{abeysekara20}).
Thus, in this work, radio spectral indices are not used.
In the future, spatially-resolved gamma-ray spectral shapes will be more tightly constrained with larger photon statistics and better angular resolutions (e.g., with CTA).

\subsection{Contribution of Objects with Possible Inverse Compton Gamma-Ray Emission}
In this work, since the gamma-ray spectral models assume hadronic emissions, contamination of objects with inverse-Compton (IC) gamma-rays can be problematic.
The objects with small $\Gamma_{\rm cut}$ or $\Gamma_{\rm br, low}$ (less than 2.0 at a 1$\sigma$ significance level, i.e., CTB~37~B; MSH 11-56; RCW~103; RCW~86; RX~J1713.7$-$3946; SN~1006; G353.6$-$0.7; Vela~Jr.; W~41) are possibly emitting gamma-rays via IC scattering (e.g., \citealt{ohira12}), and might be unsuitable for this study.
However, average trends and variety of the physical parameters constrained in Sections~\ref{sec-modeling} and \ref{sec-emax} do not change significantly without these objects.
And we note that, even for such objects, the maximum energy of protons should be larger than the maximum energy of their gamma-rays $E_{\rm max, \gamma}$ because the radiative cooling is effective only for electrons. Thus, the lower limit of the maximum energy of protons can still be obtained.

\section{Conclusion}
A systematic analysis of 38 gamma-ray emitting SNRs using their thermal X-ray and gamma-ray properties was performed.
A spectral modeling on their gamma-ray spectra was performed to constrain the particle-acceleration parameters.
Two candidates of the maximum energy of freshly accelerated particles, the gamma-ray cutoff and break energies, were found to be well below PeV for our sample.
We have also tested a simplified spectral model which includes both the freshly accelerated particles and escaping particles to estimate the maximum energies more reliably, but no tighter constraints have been obtained.

The time dependences of the cutoff energy ($E_{\rm cut} \propto t^{-0.81 \pm 0.24}$) and break energy ($E_{\rm br} \propto t^{-0.77 \pm 0.23}$) cannot be explained with the simplest acceleration condition of the Bohm limit, and requires shock-ISM collision based on theoretical calculations \citep{ptuskin03, ptuskin05, yasuda19, brose20}.
The estimated average maximum energies of accelerated particles during lifetime $\lesssim 20$~TeV $(t_{\rm M}/1~{\rm kyr})^{-0.8}$ are well below PeV if the age at the maximum, $t_{\rm M}$, is $\sim 100$--1000~yr.
However, if the maximum energy during lifetime is realized at younger ages such as $t_{\rm M} < 10$~yr, it can become higher to reach PeV.
On the other hand, the maximum energies during lifetime are suggested to have a large variety of 1.1--1.8 orders of magnitude from object to object.
Although we cannot isolate the cause of such a variety, this work provides an important clue to understand the variety in acceleration environments among SNRs.

\begin{acknowledgments}
We appreciate helpful advice by T. Yoshikoshi, K. Asano, M. Hoshino, M. Teshima, and S. Yamamoto, which has improved the paper significantly.
We deeply thank R. Brose and H. Yasuda for providing us with their simulation results.
We are grateful to T. Tanaka for his help in the Fermi-LAT data analysis.
We appreciate the help by H. Odaka and A. Tanimoto about the Markov Chain Monte-Carlo analysis.
This research was partially supported by JSPS KAKENHI grant Nos. 19J11069 and 21J00031 (HS), 19K03908 (AB), 18H01232 (RY), 19H01893 (YO), the Grant-in-Aid for Scientific Research on Innovative Areas ``Toward new frontiers: Encounter and synergy of state-of-the-art astronomical detectors and exotic quantum beams'' (18H05459; AB), Shiseido Female Researcher Science Grant (AB), and Leading Initiative for Excellent Young Researchers, MEXT, Japan (OY).
R.Y. deeply appreciates Aoyama Gakuin University Research Institute for helping our research by the fund.
\end{acknowledgments}

%

\vspace{5mm}
\facilities{Fermi-LAT
}


\software{HEAsoft (v6.27.2; \citealt{heasarc14}), Fermitools (v1.2.23; \url{https://github.com/fermi-lat/Fermitools-conda/}), MCMCLib (\url{https://github.com/kthohr/mcmc})
}



\appendix
\section{Fermi-LAT data analysis} \label{sec-fermi}

An analysis of Fermi-LAT data is performed for the SNRs in our sample which satisfy two conditions: those without recent publications (later than 2017) and those without good statistics such that spectral cutoff energies have been determined.
The resultant 15 objects, which are listed in Table~\ref{tab-fermi}, are the target of this section.

\subsection{Data reduction}
For individual objects, the latest Fermi data are obtained from the Pass 8 database\footnote{\url{https://fermi.gsfc.nasa.gov/cgi-bin/ssc/LAT/LATDataQuery.cgi}}.
Event data are extracted from a circular region with a radius of 20$^\circ$ centered on the target position.
The available time periods for individual sources are shown in Table~\ref{tab-fermi}. The extraction energy range is 0.1--300~GeV.

The tools and databases used in the data analysis below are {\it Fermitools} (v1.2.23)\footnote{\url{https://github.com/fermi-lat/Fermitools-conda/}} installed through {\it anaconda} package, the Instrumental Response File version P8R3\_SOURCE\_V2, the Fermi source list gll\_psc\_v22.fit (4FGL catalog), the Galactic diffuse background model gll\_iem\_v07.fits, and the isotropic background model (instrumental and extragalactic) iso\_P8R3\_SOUCE\_V2.txt.
Following the recommendation by the Fermi team\footnote{\url{https://fermi.gsfc.nasa.gov/ssc/data/analysis/documentation/Pass8_usage.html}}, following event selection is applied: from the SOURCE class, both FRONT and BACK section events are extracted ({\it evclass=128 evtype=3}). Events with zenith angles larger than 90$^\circ$ are rejected from the analysis in order to prevent the contamination from the Earth's bright limb. The energy dispersion correction is enabled for all the model components but the isotropic background model\footnote{\url{https://fermi.gsfc.nasa.gov/ssc/data/analysis/documentation/Pass8_edisp_usage.html}}.

\subsection{Extraction of energy spectra of SNRs with maximum likelihood analysis}
The analysis region is a 20$^\circ$-radius circle centered on the SNR position.
In order to get energy spectrum of a target source, a binned maximum likelihood analysis based on the spatial and energy distributions of the data is conducted.
The spatial bin size is $0\fdg2 \times 0\fdg2$. The data are divided into 35 logarithmic energy bins in the energy range of 100~MeV--300~GeV, which corresponds to $\sim 10$ bins within a decade.
The cataloged sources in the analysis region are considered in our analysis basically with fixed spatial and spectral parameters to the cataloged values.
Spectral parameters of only those within a $8\fdg5$-radius circle centered on the target source are treated as free parameters.

Most of the target SNRs in this work are included in the 4FGL catalog.
For these objects, spectral models for gamma-ray emissions in the catalog are used for the analysis.
Two models are included: a power-law model described as
\begin{equation}
\frac{d N}{d E}=N_{0} \left(\frac{E}{E_{\rm s}}\right)^{-\Gamma},
\end{equation}
and a log parabola model described as
\begin{equation}
\frac{d N}{d E}=N_{0}\left(\frac{E}{E_{\rm s}}\right)^{-\left(a+b \log \left(E / E_{\rm s}\right)\right)},
\end{equation}
where the normalization parameter $N_0$ and the spectral-shape parameters $\Gamma$, $a$, and $b$ are treated as free parameters, and the energy scale $E_{\rm s}$ is fixed to certain values.
In the 4FGL source catalog, MSH~11$-$62 is classified as a pulsar wind nebula (\citealt{slane12}; because it contains a pulsar).
In this study, considering the possibility that the emission is from the SNR, this object is also included, and the possibility of leptonic gamma-ray emission including other SNRs is discussed in Section~\ref{sec-discussion}.
For the objects which are not included in the 4FGL catalog (G166.0+4.3 and RCW~103), a log parabola model is used based on an assumption of the hadronic gamma-ray emission.
The spatial models for G166.0+4.3 and RCW~103 are selected according to the previous studies on individual objects \citep{araya13, xing14}.

\subsubsection{Model fitting for entire energy range}
First, the maximum likelihood analysis is conducted using the model configuration from the 4FGL catalog, and the best-fit source models are obtained.
This procedure is repeated with gradually-smaller fit tolerance values until the difference of the likelihood from that in the last trial becomes less than unity.
Then, in order to check whether the fitting result is acceptable, the likelihood ratio test is applied.
Here, a test-statistic (TS) is calculated based on the best-fit model parameters and data.
Using a likelihood $\mathcal{L}$ corresponding to a hypothesis H, the TS is defined as
\begin{equation}
\mathrm{TS} \equiv 2 \log \left(\mathcal{L}\left(\mathrm{H}_{1}\right) / \mathcal{L}\left(\mathrm{H}_{0}\right)\right),
\end{equation}
with H$_0$ being the null hypothesis in which the best-fit model is assumed and H$_1$ being a hypothesis in which a point source with only one free parameter (normalization) is added.
The TS above is asymptotically distributed as a $\chi^2$ distribution with one degree of freedom \citep{cash79, mattox96}, so that the significance of detection $\sigma \approx \sqrt{\rm TS}$.
A TS map is produced by moving a putative point source through a grid of locations on the sky and maximizing the likelihood at each grid point.
The resulting TS map represents the residuals remaining.
Thus, additional point sources are added at the positions with large residuals, i.e., ${\rm TS} > 25$, which approximately corresponds to $> 5\,\sigma$ significance.
Here, again, the likelihood analysis is done using this updated model configuration.
This process is repeated in some cases, to finally get acceptable TS maps.

\subsubsection{Model fitting for individual energy bins}
After obtaining an acceptable model configuration, the energy bins of the target SNR are determined.
The 100~MeV--300~GeV energy range is divided into 12 logarithmic energy bins, and the likelihood analysis as above is done for each energy bin.
In this process, only the normalizations of the target SNR's spectral model and the Galactic and isotropic background models are treated as free parameters (three free parameters in total).
For each energy bin, if the significance of the target SNR detection is less than $5\sigma$ (${\rm TS} \lesssim 25$), only the upper limit is calculated.
The resultant energy spectra of the SNRs are shown in Figures~\ref{fig-fermi-ss} and \ref{fig-fermi-ss2}.

\subsubsection{Estimation of systematic uncertainties}
\begin{figure*}[htb!]
\centering
\includegraphics[width=14cm, angle=0]{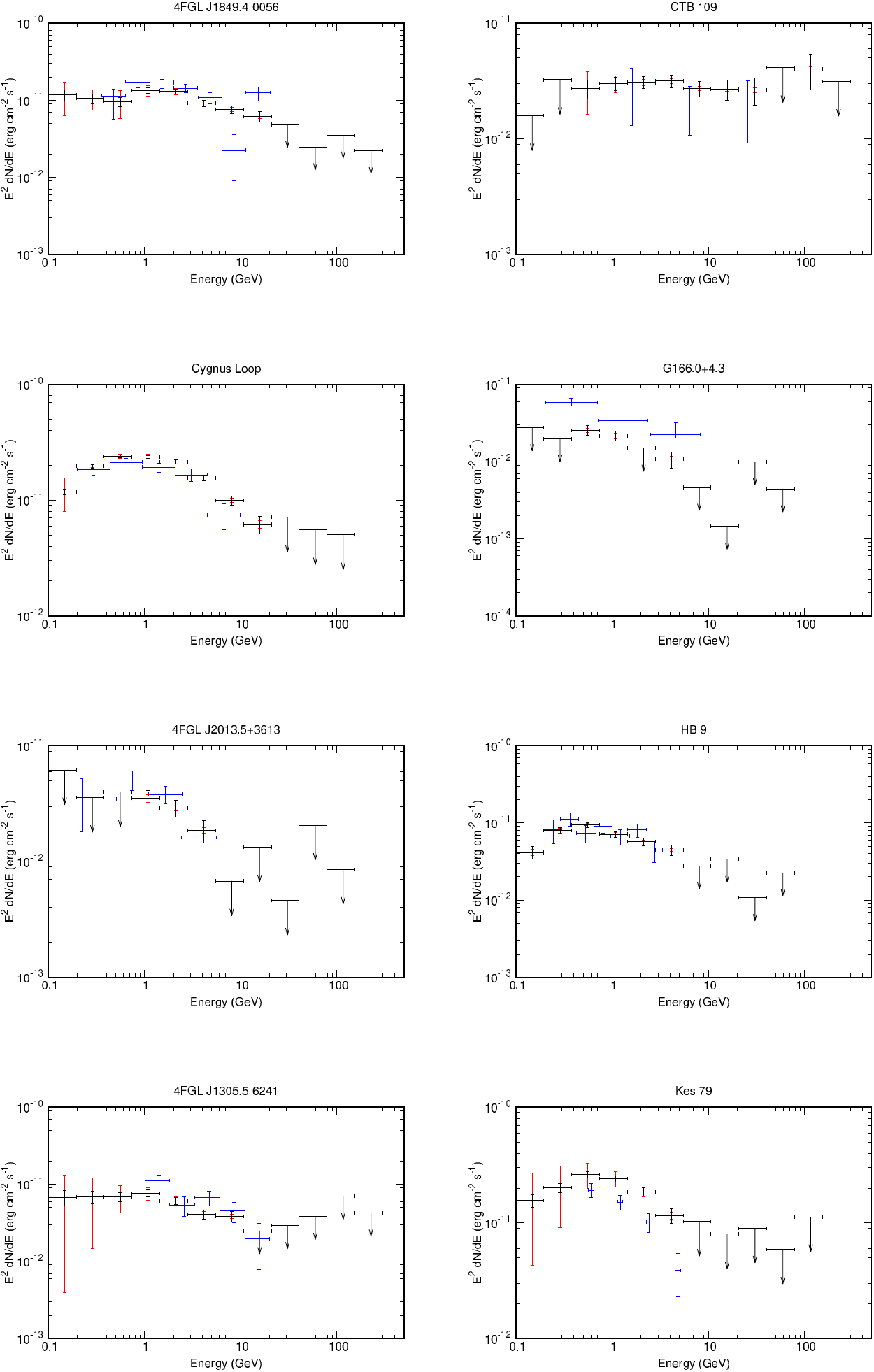}
\vspace{-0.3cm}
\caption
{Fermi-LAT energy spectra of the first half of the SNRs listed in Table~\ref{tab-fermi}. Black and red crosses represent data points with statistical errors and systematic errors, respectively.
Upper limits give 95\% confidence levels for energy bins which are undetected with less than $5 \sigma$ significance.
Blue crosses represent the spectra presented in previous works summarized in Table~\ref{tab-fermi}.
\label{fig-fermi-ss}}
\end{figure*}

\begin{figure*}[htb!]
\centering
\includegraphics[width=14cm, angle=0]{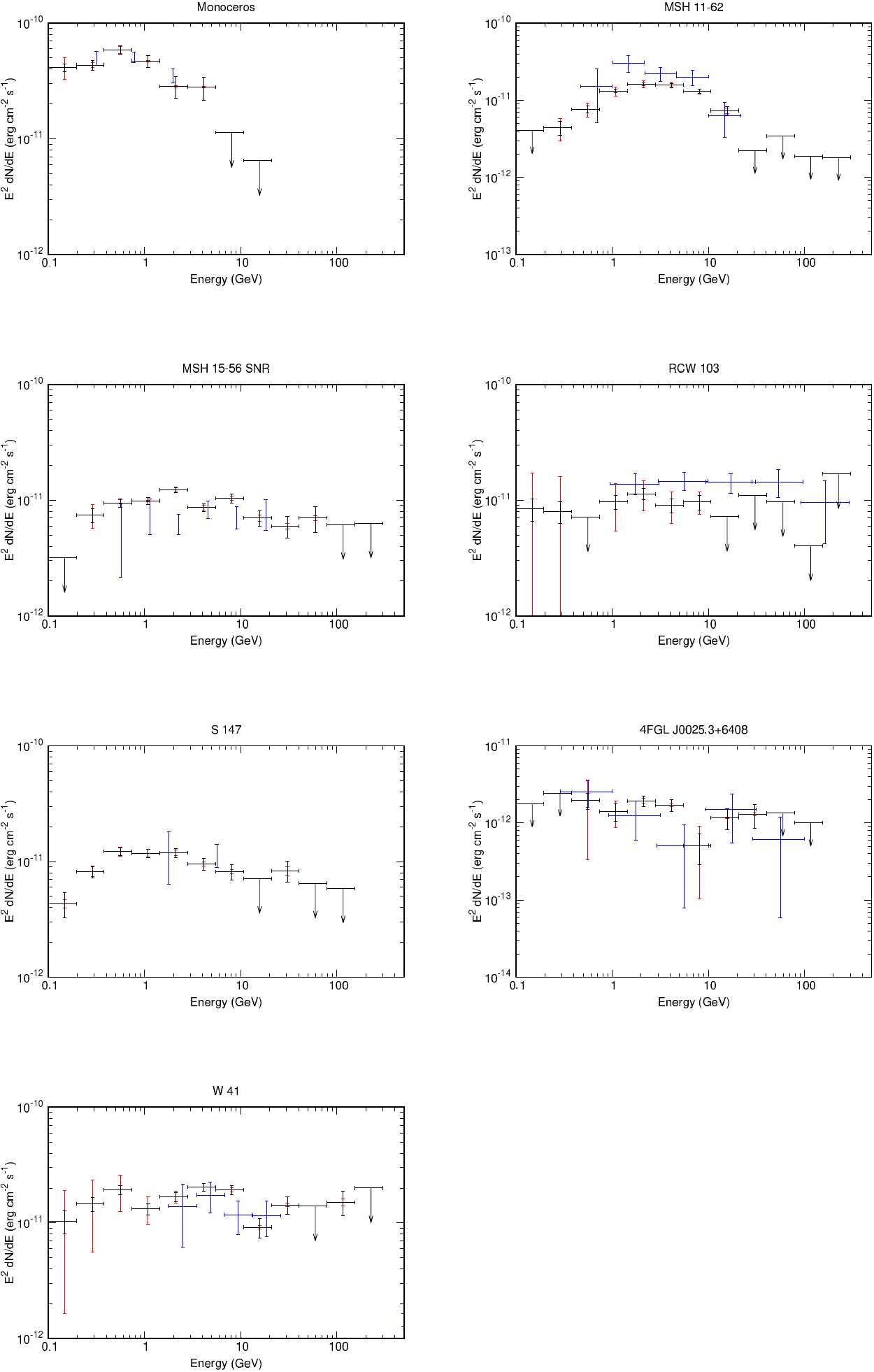}
\vspace{-0.3cm}
\caption
{Same as Figure~\ref{fig-fermi-ss}, but for the second half of the SNRs listed in Table~\ref{tab-fermi}.
\label{fig-fermi-ss2}}
\end{figure*}

Finally, systematic errors on the target SNR spectra are estimated.
Three principal origins of the systematic errors are considered: uncertainties of the LAT effective area and the PSF, and the Galactic diffuse emission model.
The $\pm$ 3\% uncertainties of the effective area and $\pm$ 5\% uncertainties of the PSF are assumed according to the LAT performance for the Pass 8 data\footnote{\url{https://fermi.gsfc.nasa.gov/ssc/data/analysis/LAT_caveats.html}}.
The uncertainties of the Galactic diffuse emission model is treated by changing its normalization by $\pm$ 6\% with respect to the best-fit value (originally performed by \citealt{abdo09} and applied by, e.g., \citealt{castro10, tanaka11}).
It should be noted that this estimation of the uncertainties of the Galactic diffuse emission is different from current standard method, which is comparing several Galactic diffuse emission models (originally performed in \citealt{depalma13} and applied in, e.g., \citealt{hess18a, abdollahi20}). And our estimates of the uncertainties are probably larger than those with this standard method.
The estimated systematic errors on individual energy bins are also shown in Figures~\ref{fig-fermi-ss} and \ref{fig-fermi-ss2} with red solid crosses.

Comparing the individual spectra obtained in this work with the previous studies, we can see that these are consistent with each other in most cases.
The spectra obtained in this work show better statistics (the previous works' energy bins with smaller errors than this work's are present in some cases, probably because they did not include the systematic errors which are considered in this work).
In the case of e.g., G166.0+4.3, Kes~79, and MSH~11$-$62, slight discrepancies can be seen. These are less than a factor of two.
Such discrepancies will be explained by the difference in the Galactic diffuse and/or isotropic background models, because different Galactic diffuse background models sometimes yield differences larger than factor of two in fluxes of diffuse sources \citep{depalma13}.

\startlongtable
\begin{deluxetable}{ l l l l }
\fontsize{8}{9}\selectfont
\tablecaption{Fermi-LAT observation logs of the 15 SNRs.\label{tab-fermi}}
\tablehead{
\colhead{Supernova remnant} & \colhead{4FGL name} & \colhead{Start date} & \colhead{Stop date}
}
\startdata
3C391 & 4FGL J1849.4-0056 &  2008-08-04  &  2012-07-19 \\
&&  2012-07-19  &  2016-07-14  \\
&&  2016-07-14  &  2020-06-02  \\
\hline
CTB109 & CTB 109 &  2008-08-04  &  2009-12-10 \\
&&  2009-12-10  &  2011-04-28  \\
&&  2011-04-28  &  2012-09-13  \\
&&  2012-09-13  &  2014-01-30  \\
&&  2014-01-30  &  2015-06-18  \\
&&  2015-06-18  &  2016-11-03  \\
&&  2016-11-03  &  2018-01-25  \\
&&  2018-01-25  &  2019-03-23  \\
&&  2019-04-18  &  2020-06-02  \\
\hline
Cygnus Loop & Cygnus Loop &  2008-08-04  &  2009-12-10 \\
&&  2009-12-10  &  2011-04-28  \\
&&  2011-04-28  &  2012-09-13  \\
&&  2012-09-13  &  2014-01-30  \\
&&  2014-01-30  &  2015-06-18  \\
&&  2015-06-18  &  2016-11-03  \\
&&  2016-11-03  &  2018-01-25  \\
&&  2018-01-25  &  2019-03-23  \\
&&  2019-04-18  &  2020-06-02  \\
\hline
G166.0+4.3 & \nodata\tablenotemark{a} &  2008-08-04  &  2010-02-04 \\
&&  2010-02-04  &  2011-08-18  \\
&&  2011-08-18  &  2013-02-28  \\
&&  2013-02-28  &  2014-09-11  \\
&&  2014-09-11  &  2016-03-24  \\
&&  2016-03-24  &  2017-10-05  \\
&&  2017-10-05  &  2019-01-15  \\
&&  2019-02-21  &  2020-06-02  \\
\hline
G73.9+0.9 & 4FGL J2013.5+3613 &  2008-08-04  &  2011-01-05 \\
&&  2011-01-06  &  2013-06-20  \\
&&  2013-06-20  &  2015-12-03  \\
&&  2015-12-03  &  2018-03-16  \\
&&  2018-04-03  &  2020-06-02  \\
\hline
HB9 & HB 9 &  2008-08-04  &  2010-02-04 \\
&&  2010-02-04  &  2011-08-18  \\
&&  2011-08-18  &  2013-02-28  \\
&&  2013-02-28  &  2014-09-11  \\
&&  2014-09-11  &  2016-03-24  \\
&&  2016-03-24  &  2017-10-05  \\
&&  2017-10-05  &  2019-01-15  \\
&&  2019-02-21  &  2020-06-02  \\
\hline
Kes17 & 4FGL J1305.5-6241 &  2008-08-04  &  2020-06-02 \\
\hline
Kes79 & Kes 79 &  2008-08-04  &  2012-07-19 \\
&&  2012-07-19  &  2016-07-14  \\
&&  2016-07-14  &  2020-06-02  \\
\hline
Monoceros Nebula & Monoceros &  2008-08-04  &  2014-07-16 \\
&&  2014-07-17  &  2020-06-02  \\
\hline
MSH11-62 & 4FGL J1111.8-6039 &  2008-08-04  &  2010-07-22 \\
&&  2010-07-22  &  2012-07-19  \\
&&  2012-07-19  &  2014-07-16  \\
&&  2014-07-17  &  2016-07-14  \\
&&  2016-07-14  &  2018-07-12  \\
&&  2018-07-12  &  2020-06-02  \\
\hline
MSH15-56 & MSH 15-56 SNR &  2008-08-04  &  2020-06-02 \\
\hline
RCW103 & \nodata\tablenotemark{a} &  2008-08-04  &  2009-12-10 \\
&&  2009-12-10  &  2011-04-28  \\
&&  2011-04-28  &  2012-09-13  \\
&&  2012-09-13  &  2014-01-30  \\
&&  2014-01-30  &  2015-06-18  \\
&&  2015-06-18  &  2016-11-03  \\
&&  2016-11-03  &  2018-03-16  \\
&&  2018-04-03  &  2019-05-30  \\
&&  2019-06-13  &  2020-09-01  \\
\hline
S147 & S 147 &  2008-08-04  &  2011-08-18 \\
&&  2011-08-18  &  2014-09-11  \\
&&  2014-09-11  &  2017-08-10  \\
&&  2017-08-10  &  2020-06-02  \\
\hline
Tycho & 4FGL J0025.3+6408 &  2008-08-04  &  2020-09-01 \\
\hline
W41 & W 41 &  2008-08-04  &  2020-06-02 \\
\enddata

\tablenotetext{a}{Not included in the 4FGL catalog.}
\tablecomments{Latest gamma-ray studies for individual objects:
\citealt{ergin14} (3C391);
\citealt{castro12} (CTB109);
\citealt{katagiri11} (Cygnus Loop);
\citealt{araya13} (G166.0+4.3);
\citealt{zdziarski16} (G73.9+0.9);
\citealt{sezer19} (HB9);
\citealt{gelfand13} (Kes17);
\citealt{auchettl14} (Kes79);
\citealt{katagiri16a} (Monoceros Nebula);
\citealt{slane12} (MSH11-62);
\citealt{temim13} (MSH15-56);
\citealt{xing14} (RCW103);
\citealt{katsuta12} (S147);
\citealt{acciari11} (Tycho);
\citealt{hess15a} (W41);
}

\end{deluxetable}
\normalsize

\section{Classification of gamma-ray emitting SNRs}\label{sec-classification}
\begin{figure*}[htb!]
\centering
\includegraphics[width=16cm, angle=0]{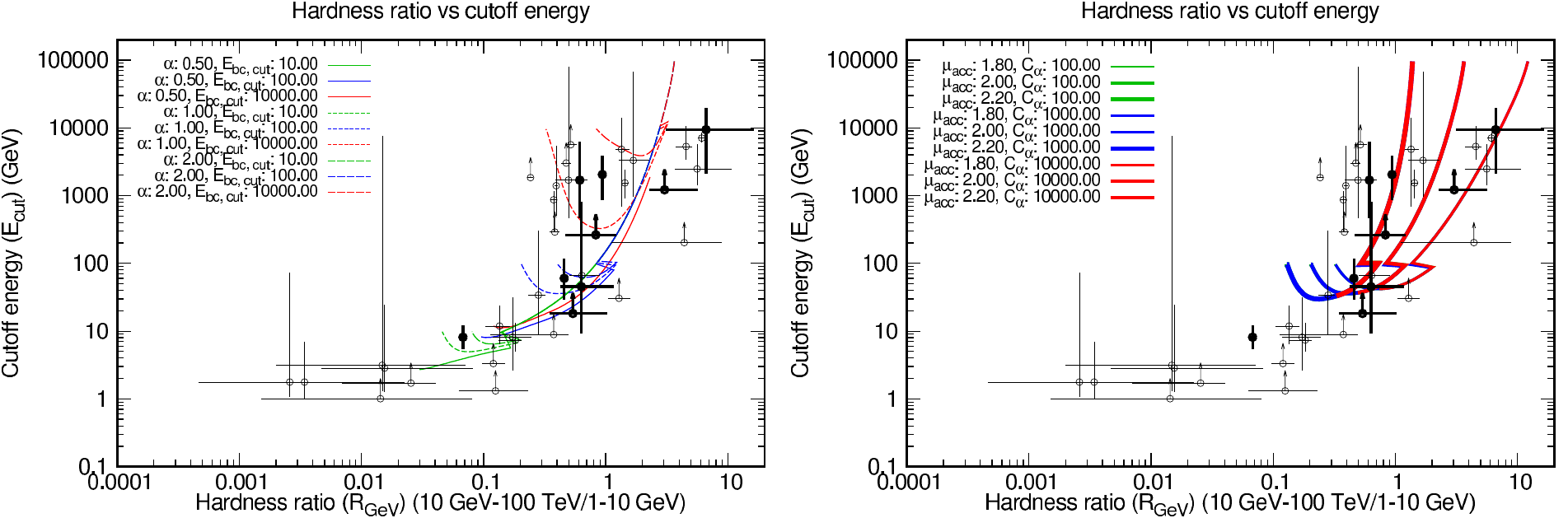}
\caption{
Relation between $E_{\rm cut}$ and $R_{\rm GeV}$ for the data (crosses) and our analytical models (lines).
({\it left panel}) As for the model curves, different colors and line types indicate different model parameters ($\alpha$ and $E_{\rm bc, cut}$) as specified in the figure.
The other parameters are fixed as $\mu_{\rm acc} = 2.0$, $C_{\alpha} = 100$~GeV, $\beta = 0.6$, and $\epsilon = 0.0$.
({\it right panel}) Same as the left panel but with different model parameters.
Different colors and line types indicate different model parameters ($\mu_{\rm acc}$ and $C_{\alpha}$) as specified in the figure.
The other parameters are fixed as $\alpha = 1.0$, $E_{\rm bc, cut} = 100$ GeV, $\beta = 0.6$, and $\epsilon = 0.0$.
For both panels, thick and thin crosses represent the data with and without the reliable ages $t_{\rm r}$.
\label{fig-theoretical-curves}}
\end{figure*}

Here we compare two observational parameters, gamma-ray cutoff energy $E_{\rm cut}$ and hardness ratio $R_{\rm GeV}$, with those of our analytical model for escape-limited particle acceleration (Eq.~\ref{eq-brcut}) to classify the sample based on their constrained parameters.
In this section, we parameterize the break energy as a function of age as $E_{\rm bc, br} = C_{\rm \alpha} t^{-\alpha}$ in our model.
To calculate $R_{\rm GeV}$ and $E_{\rm cut}$ from our analytical model and to compare them with observations, we fit the model spectra with an exponential-cutoff power-law model, as in the case of the observational spectra.
In Figure~\ref{fig-theoretical-curves}, the hardness ratio $R_{\rm GeV}$ is plotted as a function of the cutoff energy $E_{\rm cut}$.
Approximately, ages get larger from top-right to bottom-left in these plots.
Note that our model, in which the cutoff energy $E_{\rm bc, cut}$ is assumed to be time-invariant, may be an extreme case and in reality $E_{\rm bc, cut}$ may change with time significantly.

Regarding the model curves, both the $R_{\rm GeV}$ and $E_{\rm cut}$ basically get smaller with increasing ages, because both of them are governed by $E_{\rm bc, br}$.
However, if $E_{\rm bc, cut}$ is fixed to large values such as 10~TeV, $E_{\rm cut}$ visibly increases again after certain ages whereas $R_{\rm GeV}$ continues to decrease (Figure~\ref{fig-theoretical-curves}).
This is because $E_{\rm cut}$ is close to $E_{\rm bc, br}$ in young phases, but after certain ages where a condition $E_{\rm bc, br} \lesssim E_{\rm bc, cut}$ is satisfied, it in turn becomes similar to $E_{\rm bc, cut}$.
Since the model spectra always get softer with age because escaping particles gradually dominate the emission regardless of whether $E_{\rm cut} \sim E_{\rm bc, br}$ or $E_{\rm cut} \sim E_{\rm bc, cut}$, the hardness ratio $R_{\rm GeV}$ always decreases with age.
All the model curves plotted in Figure~\ref{fig-theoretical-curves} show step-like structures, which are due to such transitions of $E_{\rm cut}$.
Note that such transitions occur in all the cases but are hardly visible in late phases when $E_{\rm bc, cut}$ is small.

\begin{figure}[htb!]
\centering
\includegraphics[width=6cm, angle=90]{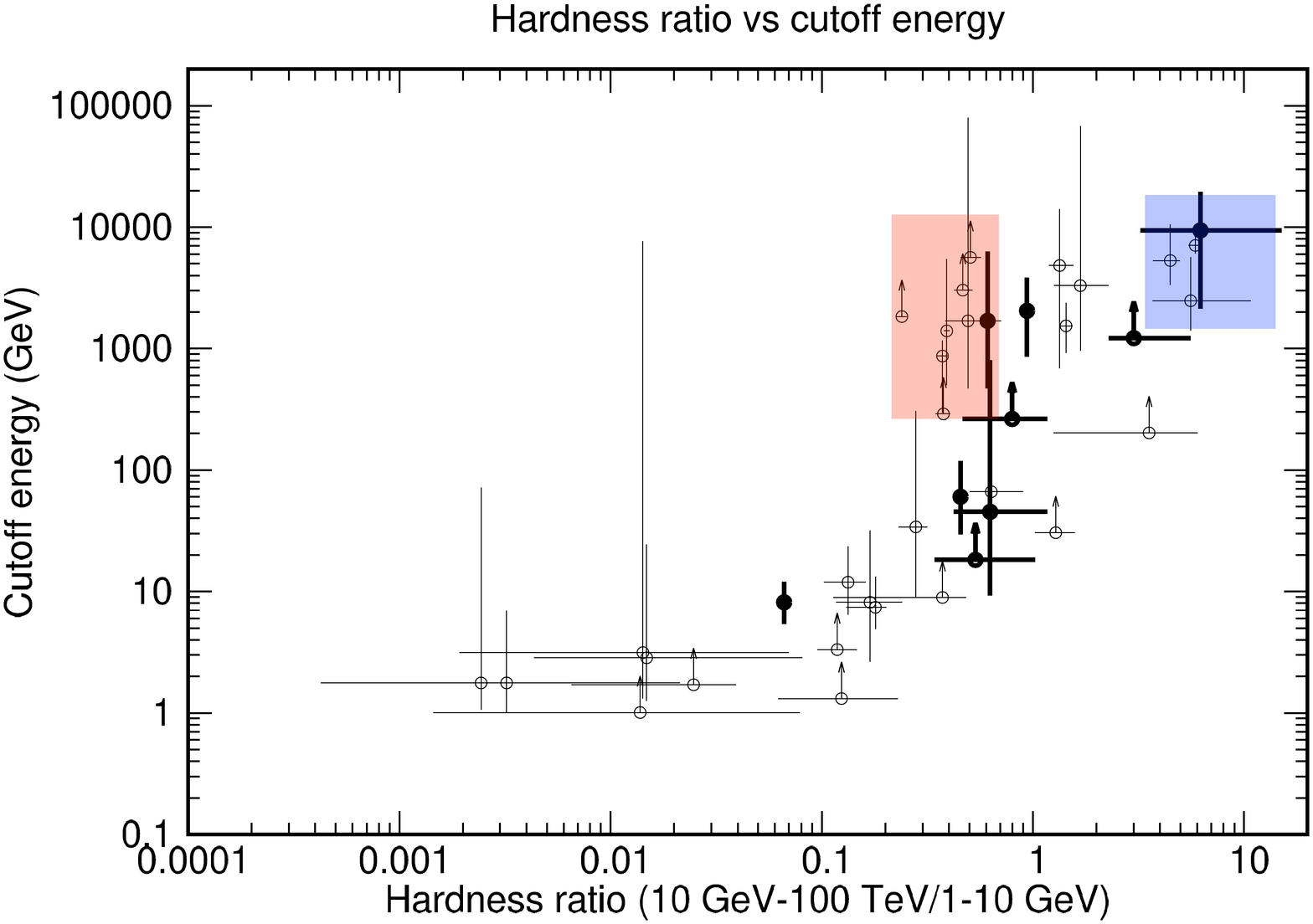}
\caption{
Same as Figure~\ref{fig-theoretical-curves} but without the model curves.
Two SNR groups, W~28-like group and 1713-like group are highlighted with red and blue transparent rectangles, respectively.
\label{fig-hardness-cutoff}}
\end{figure}

Based on Figure~\ref{fig-theoretical-curves}, we can classify the SNRs into three groups as indicated in Figure~\ref{fig-hardness-cutoff}.
The ``W~28-like'' objects, i.e., W~28, CTB~37~A, G359.0$-$0.5, W~49~B, W~30, W~51~C, W~41, and G349.7+0.2, are characterized by moderate $R_{\rm GeV}$ with especially large $E_{\rm cut}$, which will be explained only if the emission is dominated by escaping particles.
For these objects, a condition $E_{\rm cut} \sim E_{\rm bc, cut}$ is satisfied with large values of $E_{\rm bc, cut}$ as can be seen in Figure~\ref{fig-theoretical-curves}.
The ``1713-like'' objects, i.e., RX~J1713.7$-$3946, RCW~86, Vela~Jr., and G353.6$-$0.7, are characterized by especially large $R_{\rm GeV}$ with large $E_{\rm cut}$, which will be explained only if the spectra are very hard and thus contribution of freshly accelerated particles can be separated well from those of escaping particles.
The other objects can be explained by variable parameter sets, so that their parameters are poorly constrained.

\bibliography{/Users/suzuki/Documents/paper_submission/references/references.bib}

\begin{thebibliography}{}
\expandafter\ifx\csname natexlab\endcsname\relax\def\natexlab#1{#1}\fi
\providecommand{\url}[1]{\href{#1}{#1}}
\providecommand{\dodoi}[1]{doi:~\href{http://doi.org/#1}{\nolinkurl{#1}}}
\providecommand{\doeprint}[1]{\href{http://ascl.net/#1}{\nolinkurl{http://ascl.net/#1}}}
\providecommand{\doarXiv}[1]{\href{https://arxiv.org/abs/#1}{\nolinkurl{https://arxiv.org/abs/#1}}}

\bibitem[{Abdo {et~al.}(2009)Abdo, Ackermann, Ajello, Baldini, Ballet,
  Barbiellini, Baring, Bastieri, Baughman, Bechtol, Bellazzini, Berenji,
  Blandford, Bloom, Bonamente, Borgland, Bouvier, Bregeon, Brez, Brigida,
  Bruel, Burnett, Buson, Caliandro, Cameron, Caraveo, Casandjian, Cecchi,
  {\c{C}}elik, Chekhtman, Cheung, Chiang, Ciprini, Claus, Cohen-Tanugi,
  Cominsky, Conrad, Cutini, Dermer, de~Angelis, de~Palma, Digel, Dormody,
  do~Couto~e Silva, Drell, Dubois, Dumora, Farnier, Favuzzi, Fegan, Focke,
  Fortin, Frailis, Fukazawa, Funk, Fusco, Gargano, Gasparrini, Gehrels,
  Germani, Giavitto, Giebels, Giglietto, Giordano, Glanzman, Godfrey, Grenier,
  Grondin, Grove, Guillemot, Guiriec, Hanabata, Harding, Hayashida, Hays,
  Hughes, Jackson, J{\'{o}}hannesson, Johnson, Johnson, Johnson, Kamae,
  Katagiri, Kataoka, Katsuta, Kawai, Kerr, dlseder, Kocian, Kuss, Lande,
  Latronico, Lemoine-Goumard, Longo, Loparco, Lott, Lovellette, Lubrano,
  Makeev, Mazziotta, McEnery, Meurer, Michelson, Mitthumsiri, Mizuno, Moiseev,
  Monte, Monzani, Morselli, Moskalenko, Murgia, Nakamori, Nolan, Norris, Nuss,
  Ohsugi, Okumura, Omodei, Orlando, Ormes, Paneque, Parent, Pelassa, Pepe,
  Pesce-Rollins, Piron, Porter, Rain{\`{o}}, Rando, Razzano, Reimer, Reimer,
  Reposeur, Ritz, Rodriguez, Romani, Roth, Ryde, Sadrozinski, Sanchez, Sander,
  Parkinson, Scargle, Schalk, Sgr{\`{o}}, Siskind, Smith, Smith, Spandre,
  Spinelli, Strickman, Suson, Tajima, Takahashi, Takahashi, Tanaka, Thayer,
  Thayer, Thompson, Tibaldo, Tibolla, Torres, Tosti, Tramacere, Uchiyama,
  Usher, Vasileiou, Venter, Vilchez, Vitale, Waite, Wang, Winer, Wood,
  Yamazaki, Ylinen, \& Ziegler}]{abdo09}
Abdo, A.~A., Ackermann, M., Ajello, M., {et~al.} 2009, The Astrophysical
  Journal, 706, L1, \dodoi{10.1088/0004-637x/706/1/l1}

\bibitem[{{Abdollahi} {et~al.}(2020){Abdollahi}, {Ballet}, {Fukazawa},
  {Katagiri}, \& {Condon}}]{abdollahi20}
{Abdollahi}, S., {Ballet}, J., {Fukazawa}, Y., {Katagiri}, H., \& {Condon}, B.
  2020, \apj, 896, 76, \dodoi{10.3847/1538-4357/ab91b3}

\bibitem[{{Abdollahi} {et~al.}(2017){Abdollahi}, {Fermi-LAT Collaboration},
  {Mizuno}, {Fukazawa}, {Katagiri}, \& {Condon}}]{abdollahi17}
{Abdollahi}, S., {Fermi-LAT Collaboration}, {Mizuno}, T., {et~al.} 2017, in
  International Cosmic Ray Conference, Vol. 301, 35th International Cosmic Ray
  Conference (ICRC2017), 743

\bibitem[{{Abeysekara} {et~al.}(2020){Abeysekara}, {Archer}, {Benbow}, {Bird},
  {Brose}, {Buchovecky}, {Buckley}, {Chromey}, {Cui}, {Daniel}, {Das},
  {Dwarkadas}, {Falcone}, {Feng}, {Finley}, {Fortson}, {Gent}, {Gillanders},
  {Giuri}, {Gueta}, {Hanna}, {Hassan}, {Hervet}, {Holder}, {Hughes},
  {Humensky}, {Kaaret}, {Kar}, {Kelley-Hoskins}, {Kertzman}, {Kieda}, {Krause},
  {Krennrich}, {Kumar}, {Lang}, {Maier}, {Moriarty}, {Mukherjee},
  {Nievas-Rosillo}, {O'Brien}, {Ong}, {Park}, {Petrashyk}, {Pfrang}, {Pohl},
  {Pueschel}, {Quinn}, {Ragan}, {Reynolds}, {Richards}, {Roache}, {Sadeh},
  {Santander}, {Sembroski}, {Shahinyan}, {Sushch}, {Weinstein}, {Wilcox},
  {Wilhelm}, {Williams}, {Williamson}, {Zitzer}, \& {Ghiotto}}]{abeysekara20}
{Abeysekara}, A.~U., {Archer}, A., {Benbow}, W., {et~al.} 2020, \apj, 894, 51,
  \dodoi{10.3847/1538-4357/ab8310}

\bibitem[{{Acciari} {et~al.}(2011){Acciari}, {Aliu}, {Arlen}, {Aune},
  {Beilicke}, {Benbow}, {Bradbury}, {Buckley}, {Bugaev}, {Byrum}, {Cannon},
  {Cesarini}, {Ciupik}, {Collins-Hughes}, {Cui}, {Dickherber}, {Duke},
  {Errando}, {Finley}, {Finnegan}, {Fortson}, {Furniss}, {Galante}, {Gall},
  {Gillanders}, {Godambe}, {Griffin}, {Grube}, {Guenette}, {Gyuk}, {Hanna},
  {Holder}, {Hughes}, {Hui}, {Humensky}, {Kaaret}, {Karlsson}, {Kertzman},
  {Kieda}, {Krawczynski}, {Krennrich}, {Lang}, {LeBohec}, {Madhavan}, {Maier},
  {Majumdar}, {McArthur}, {McCann}, {Moriarty}, {Mukherjee}, {Ong}, {Orr},
  {Otte}, {Pandel}, {Park}, {Perkins}, {Pohl}, {Quinn}, {Ragan}, {Reyes},
  {Reynolds}, {Roache}, {Rose}, {Saxon}, {Schroedter}, {Sembroski}, {Senturk},
  {Slane}, {Smith}, {Te{\v{s}}i{\'c}}, {Theiling}, {Thibadeau}, {Tsurusaki},
  {Varlotta}, {Vassiliev}, {Vincent}, {Vivier}, {Wakely}, {Ward}, {Weekes},
  {Weinstein}, {Weisgarber}, {Williams}, {Wood}, \& {Zitzer}}]{acciari11}
{Acciari}, V.~A., {Aliu}, E., {Arlen}, T., {et~al.} 2011, \apjl, 730, L20,
  \dodoi{10.1088/2041-8205/730/2/L20}

\bibitem[{{Acero} {et~al.}(2010){Acero}, {Aharonian}, {Akhperjanian}, {Anton},
  {Barres de Almeida}, {Bazer-Bachi}, {Becherini}, {Behera}, {Beilicke},
  {Bernl{\"o}hr}, {Bochow}, {Boisson}, {Bolmont}, {Borrel}, {Brucker}, {Brun},
  {Brun}, {B{\"u}hler}, {Bulik}, {B{\"u}sching}, {Boutelier}, {Chadwick},
  {Charbonnier}, {Chaves}, {Cheesebrough}, {Conrad}, {Chounet}, {Clapson},
  {Coignet}, {Dalton}, {Daniel}, {Davids}, {Degrange}, {Deil}, {Dickinson},
  {Djannati-Ata{\"\i}}, {Domainko}, {O'C. Drury}, {Dubois}, {Dubus}, {Dyks},
  {Dyrda}, {Egberts}, {Eger}, {Espigat}, {Fallon}, {Farnier}, {Fegan},
  {Feinstein}, {Fiasson}, {F{\"o}rster}, {Fontaine}, {F{\"u}{\ss}ling},
  {Gabici}, {Gallant}, {G{\'e}rard}, {Gerbig}, {Giebels}, {Glicenstein},
  {Gl{\"u}ck}, {Goret}, {G{\"o}ring}, {Hauser}, {Hauser}, {Heinz},
  {Heinzelmann}, {Henri}, {Hermann}, {Hinton}, {Hoffmann}, {Hofmann},
  {Hofverberg}, {Holleran}, {Hoppe}, {Horns}, {Jacholkowska}, {de Jager},
  {Jahn}, {Jung}, {Katarzy{\'n}ski}, {Katz}, {Kaufmann}, {Kerschhaggl},
  {Khangulyan}, {Kh{\'e}lifi}, {Keogh}, {Klochkov}, {Klu{\'z}niak}, {Kneiske},
  {Komin}, {Kosack}, {Kossakowski}, {Lamanna}, {Lemoine-Goumard}, {Lenain},
  {Lohse}, {Marandon}, {Marcowith}, {Masbou}, {Maurin}, {McComb}, {Medina},
  {M{\'e}hault}, {Moderski}, {Moulin}, {Naumann-Godo}, {de Naurois}, {Nedbal},
  {Nekrassov}, {Nicholas}, {Niemiec}, {Nolan}, {Ohm}, {Olive}, {de O{\~n}a
  Wilhelmi}, {Orford}, {Ostrowski}, {Panter}, {Paz Arribas}, {Pedaletti},
  {Pelletier}, {Petrucci}, {Pita}, {P{\"u}hlhofer}, {Punch}, {Quirrenbach},
  {Raubenheimer}, {Raue}, {Rayner}, {Reimer}, {Renaud}, {de Los Reyes},
  {Rieger}, {Ripken}, {Rob}, {Rosier-Lees}, {Rowell}, {Rudak}, {Rulten},
  {Ruppel}, {Ryde}, {Sahakian}, {Santangelo}, {Schlickeiser}, {Sch{\"o}ck},
  {Sch{\"o}nwald}, {Schwanke}, {Schwarzburg}, {Schwemmer}, {Shalchi}, {Sushch},
  {Sikora}, {Skilton}, {Sol}, {Stawarz}, {Steenkamp}, {Stegmann}, {Stinzing},
  {Superina}, {Szostek}, {Tam}, {Tavernet}, {Terrier}, {Tibolla}, {Tluczykont},
  {van Eldik}, {Vasileiadis}, {Venter}, {Venter}, {Vialle}, {Vincent}, {Vink},
  {Vivier}, {V{\"o}lk}, {Volpe}, {Vorobiov}, {Wagner}, {Ward}, {Zdziarski},
  {Zech}, \& {H.~E.~S.~S. Collaboration}}]{acero10}
{Acero}, F., {Aharonian}, F., {Akhperjanian}, A.~G., {et~al.} 2010, \aap, 516,
  A62, \dodoi{10.1051/0004-6361/200913916}

\bibitem[{{Acero} {et~al.}(2016){Acero}, {Ackermann}, {Ajello}, {Baldini},
  {Ballet}, {Barbiellini}, {Bastieri}, {Bellazzini}, {Bissaldi}, {Blandford},
  {Bloom}, {Bonino}, {Bottacini}, {Brand t}, {Bregeon}, {Bruel}, {Buehler},
  {Buson}, {Caliandro}, {Cameron}, {Caputo}, {Caragiulo}, {Caraveo}, {Casand
  jian}, {Cavazzuti}, {Cecchi}, {Chekhtman}, {Chiang}, {Chiaro}, {Ciprini},
  {Claus}, {Cohen}, {Cohen-Tanugi}, {Cominsky}, {Condon}, {Conrad}, {Cutini},
  {D'Ammando}, {de Angelis}, {de Palma}, {Desiante}, {Digel}, {Di Venere},
  {Drell}, {Drlica-Wagner}, {Favuzzi}, {Ferrara}, {Franckowiak}, {Fukazawa},
  {Funk}, {Fusco}, {Gargano}, {Gasparrini}, {Giglietto}, {Giommi}, {Giordano},
  {Giroletti}, {Glanzman}, {Godfrey}, {Gomez-Vargas}, {Grenier}, {Grondin},
  {Guillemot}, {Guiriec}, {Gustafsson}, {Hadasch}, {Harding}, {Hayashida},
  {Hays}, {Hewitt}, {Hill}, {Horan}, {Hou}, {Iafrate}, {Jogler},
  {J{\'o}hannesson}, {Johnson}, {Kamae}, {Katagiri}, {Kataoka}, {Katsuta},
  {Kerr}, {Kn{\"o}dlseder}, {Kocevski}, {Kuss}, {Laffon}, {Lande}, {Larsson},
  {Latronico}, {Lemoine-Goumard}, {Li}, {Li}, {Longo}, {Loparco}, {Lovellette},
  {Lubrano}, {Magill}, {Maldera}, {Marelli}, {Mayer}, {Mazziotta}, {Michelson},
  {Mitthumsiri}, {Mizuno}, {Moiseev}, {Monzani}, {Moretti}, {Morselli},
  {Moskalenko}, {Murgia}, {Nemmen}, {Nuss}, {Ohsugi}, {Omodei}, {Orienti},
  {Orlando}, {Ormes}, {Paneque}, {Perkins}, {Pesce-Rollins}, {Petrosian},
  {Piron}, {Pivato}, {Porter}, {Rain{\`o}}, {Rando}, {Razzano}, {Razzaque},
  {Reimer}, {Reimer}, {Renaud}, {Reposeur}, {Rousseau}, {Saz Parkinson},
  {Schmid}, {Schulz}, {Sgr{\`o}}, {Siskind}, {Spada}, {Spandre}, {Spinelli},
  {Strong}, {Suson}, {Tajima}, {Takahashi}, {Tanaka}, {Thayer}, {Thompson},
  {Tibaldo}, {Tibolla}, {Torres}, {Tosti}, {Troja}, {Uchiyama}, {Vianello},
  {Wells}, {Wood}, {Wood}, {Yassine}, {den Hartog}, \& {Zimmer}}]{acero16b}
{Acero}, F., {Ackermann}, M., {Ajello}, M., {et~al.} 2016, \apjs, 224, 8,
  \dodoi{10.3847/0067-0049/224/1/8}

\bibitem[{{Ackermann} {et~al.}(2013){Ackermann}, {Ajello}, {Allafort},
  {Baldini}, {Ballet}, {Barbiellini}, {Baring}, {Bastieri}, {Bechtol},
  {Bellazzini}, {Bland ford}, {Bloom}, {Bonamente}, {Borgland }, {Bottacini},
  {Brandt}, {Bregeon}, {Brigida}, {Bruel}, {Buehler}, {Busetto}, {Buson},
  {Caliandro}, {Cameron}, {Caraveo}, {Casandjian}, {Cecchi}, {{\c{C}}elik},
  {Charles}, {Chaty}, {Chaves}, {Chekhtman}, {Cheung}, {Chiang}, {Chiaro},
  {Cillis}, {Ciprini}, {Claus}, {Cohen-Tanugi}, {Cominsky}, {Conrad}, {Corbel},
  {Cutini}, {D'Ammando}, {de Angelis}, {de Palma}, {Dermer}, {do Couto e
  Silva}, {Drell}, {Drlica-Wagner}, {Falletti}, {Favuzzi}, {Ferrara},
  {Franckowiak}, {Fukazawa}, {Funk}, {Fusco}, {Gargano}, {Germani},
  {Giglietto}, {Giommi}, {Giordano}, {Giroletti}, {Glanzman}, {Godfrey},
  {Grenier}, {Grondin}, {Grove}, {Guiriec}, {Hadasch}, {Hanabata}, {Harding},
  {Hayashida}, {Hayashi}, {Hays}, {Hewitt}, {Hill}, {Hughes}, {Jackson},
  {Jogler}, {J{\'o}hannesson}, {Johnson}, {Kamae}, {Kataoka}, {Katsuta},
  {Kn{\"o}dlseder}, {Kuss}, {Lande}, {Larsson}, {Latronico}, {Lemoine-Goumard},
  {Longo}, {Loparco}, {Lovellette}, {Lubrano}, {Madejski}, {Massaro}, {Mayer},
  {Mazziotta}, {McEnery}, {Mehault}, {Michelson}, {Mignani}, {Mitthumsiri},
  {Mizuno}, {Moiseev}, {Monzani}, {Morselli}, {Moskalenko}, {Murgia},
  {Nakamori}, {Nemmen}, {Nuss}, {Ohno}, {Ohsugi}, {Omodei}, {Orienti},
  {Orlando}, {Ormes}, {Paneque}, {Perkins}, {Pesce-Rollins}, {Piron}, {Pivato},
  {Rain{\`o}}, {Rando}, {Razzano}, {Razzaque}, {Reimer}, {Reimer}, {Ritz},
  {Romoli}, {S{\'a}nchez-Conde}, {Schulz}, {Sgr{\`o}}, {Simeon}, {Siskind},
  {Smith}, {Spand re}, {Spinelli}, {Stecker}, {Strong}, {Suson}, {Tajima},
  {Takahashi}, {Takahashi}, {Tanaka}, {Thayer}, {Thayer}, {Thompson},
  {Thorsett}, {Tibaldo}, {Tibolla}, {Tinivella}, {Troja}, {Uchiyama}, {Usher},
  {Vandenbroucke}, {Vasileiou}, {Vianello}, {Vitale}, {Waite}, {Werner},
  {Winer}, {Wood}, {Wood}, {Yamazaki}, {Yang}, \& {Zimmer}}]{ackermann13}
{Ackermann}, M., {Ajello}, M., {Allafort}, A., {et~al.} 2013, Science, 339,
  807, \dodoi{10.1126/science.1231160}

\bibitem[{{Ackermann} {et~al.}(2017){Ackermann}, {Ajello}, {Baldini}, {Ballet},
  {Barbiellini}, {Bastieri}, {Bellazzini}, {Bissaldi}, {Bloom}, {Bonino},
  {Bottacini}, {Brandt}, {Bregeon}, {Bruel}, {Buehler}, {Cameron}, {Caragiulo},
  {Caraveo}, {Castro}, {Cavazzuti}, {Cecchi}, {Charles}, {Chekhtman}, {Cheung},
  {Chiaro}, {Ciprini}, {Cohen}, {Costantin}, {Costanza}, {Cutini}, {D'Ammando},
  {de Palma}, {Desiante}, {Digel}, {Di Lalla}, {Di Mauro}, {Di Venere},
  {Favuzzi}, {Fegan}, {Ferrara}, {Franckowiak}, {Fukazawa}, {Funk}, {Fusco},
  {Gargano}, {Gasparrini}, {Giglietto}, {Giordano}, {Giroletti}, {Green},
  {Grenier}, {Grondin}, {Guillemot}, {Guiriec}, {Harding}, {Hays}, {Hewitt},
  {Horan}, {Hou}, {J{\'o}hannesson}, {Kamae}, {Kuss}, {La Mura}, {Larsson},
  {Lemoine-Goumard}, {Li}, {Longo}, {Loparco}, {Lubrano}, {Magill}, {Maldera},
  {Malyshev}, {Manfreda}, {Mazziotta}, {Michelson}, {Mitthumsiri}, {Mizuno},
  {Monzani}, {Morselli}, {Moskalenko}, {Negro}, {Nuss}, {Ohsugi}, {Omodei},
  {Orienti}, {Orlando}, {Ormes}, {Paliya}, {Paneque}, {Perkins}, {Persic},
  {Pesce-Rollins}, {Petrosian}, {Piron}, {Porter}, {Principe}, {Rain{\`o}},
  {Rando}, {Razzano}, {Razzaque}, {Reimer}, {Reimer}, {Reposeur}, {Sgr{\`o}},
  {Simone}, {Siskind}, {Spada}, {Spandre}, {Spinelli}, {Suson}, {Tak},
  {Thayer}, {Thompson}, {Torres}, {Tosti}, {Troja}, {Vianello}, {Wood}, \&
  {Wood}}]{ackermann17}
{Ackermann}, M., {Ajello}, M., {Baldini}, L., {et~al.} 2017, \apj, 843, 139,
  \dodoi{10.3847/1538-4357/aa775a}

\bibitem[{Actis {et~al.}(2011)Actis, Agnetta, Aharonian, Akhperjanian,
  Aleksi{\'c}, Aliu, Allan, Allekotte, Antico, Antonelli, Antoranz,
  Aravantinos, Arlen, Arnaldi, Artmann, Asano, Asorey, B{\"a}hr, Bais,
  Baixeras, Bajtlik, Balis, Bamba, Barbier, Barcel{\'o}, Barnacka, Barnstedt,
  Barres~de Almeida, Barrio, Basso, Bastieri, Bauer, Becerra, Becherini,
  Bechtol, Becker, Beckmann, Bednarek, Behera, Beilicke, Belluso, Benallou,
  Benbow, Berdugo, Berger, Bernardino, Bernl{\"o}hr, Biland, Billotta, Bird,
  Birsin, Bissaldi, Blake, Blanch, Bobkov, Bogacz, Bogdan, Boisson, Boix,
  Bolmont, Bonanno, Bonardi, Bonev, Borkowski, Botner, Bottani, Bourgeat,
  Boutonnet, Bouvier, Brau-Nogu{\'e}, Braun, Bretz, Briggs, Brun, Brunetti,
  Buckley, Bugaev, B{\"u}hler, Bulik, Busetto, Buson, Byrum, Cailles, Cameron,
  Canestrari, Cantu, Carmona, Carosi, Carr, Carton, Casiraghi, Castarede,
  Catalano, Cavazzani, Cazaux, Cerruti, Cerruti, Chadwick, Chiang, Chikawa,
  Cie{\'s}lar, Ciesielska, Cillis, Clerc, Colin, Colom{\'e}, Compin, Conconi,
  Connaughton, Conrad, Contreras, Coppi, Corlier, Corona, Corpace, Corti,
  Cortina, Costantini, Cotter, Courty, Couturier, Covino, Croston, Cusumano,
  Daniel, Dazzi, Angelis, de~Cea~del Pozo, de~Gouveia Dal~Pino, de~Jager, de~la
  Calle~P{\'e}rez, De~La~Vega, De~Lotto, de~Naurois, de~O{\~n}a~Wilhelmi,
  de~Souza, Decerprit, Deil, Delagnes, Deleglise, Delgado, Dettlaff, Di~Paolo,
  Di~Pierro, D{\'\i}az, Dick, Dickinson, Digel, Dimitrov, Disset,
  Djannati-Ata{\"\i}, Doert, Domainko, Dorner, Doro, Dournaux, Dravins, Drury,
  Dubois, Dubois, Dubus, Dufour, Durand, Dyks, Dyrda, Edy, Egberts,
  Eleftheriadis, Elles, Emmanoulopoulos, Enomoto, Ernenwein, Errando,
  Etchegoyen, Falcone, Farakos, Farnier, Federici, Feinstein, Ferenc,
  Fillin-Martino, Fink, Finley, Finley, Firpo, Florin, F{\"o}hr, Fokitis, Font,
  Fontaine, Fontana, F{\"o}rster, Fortson, Fouque, Fransson, Fraser, Fresnillo,
  Fruck, Fujita, Fukazawa, Funk, G{\"a}bele, Gabici, Gadola, Galante, Gallant,
  Garc{\'\i}a, Garc{\'\i}a~L{\'o}pez, Garrido, Garrido, Gasc{\'o}n, Gasq, Gaug,
  Gaweda, Geffroy, Ghag, Ghedina, Ghigo, Gianakaki, Giarrusso, Giavitto,
  Giebels, Giro, Giubilato, Glanzman, Glicenstein, Gochna, Golev,
  G{\'o}mez~Berisso, Gonz{\'a}lez, Gonz{\'a}lez, Gra{\~n}ena, Graciani, Granot,
  Gredig, Green, Greenshaw, Grimm, Grube, Grudzi{\'n}ska, Grygorczuk, Guarino,
  Guglielmi, Guilloux, Gunji, Gyuk, Hadasch, Haefner, Hagiwara, Hahn, Hallgren,
  Hara, Hardcastle, Hassan, Haubold, Hauser, Hayashida, Heller, Henri, Hermann,
  Herrero, Hinton, Hoffmann, Hofmann, Hofverberg, Horns, Hrupec, Huan, Huber,
  Huet, Hughes, Hultquist, Humensky, Huppert, Ibarra, Illa, Ingjald, Inoue,
  Inoue, Ioka, Jablonski, Jacholkowska, Janiak, Jean, Jensen, Jogler, Jung,
  Kaaret, Kabuki, Kakuwa, Kalkuhl, Kankanyan, Kapala, Karastergiou, Karczewski,
  Karkar, Karlsson, Kasperek, Katagiri, \& Consortium}]{actis11}
Actis, M., Agnetta, G., Aharonian, F., {et~al.} 2011, Experimental Astronomy,
  32, 193, \dodoi{10.1007/s10686-011-9247-0}

\bibitem[{{Aharonian} {et~al.}(2007){Aharonian}, {Akhperjanian}, {Bazer-Bachi},
  {Beilicke}, {Benbow}, {Berge}, {Bernl{\"o}hr}, {Boisson}, {Bolz}, {Borrel},
  {Braun}, {Brown}, {B{\"u}hler}, {B{\"u}sching}, {Carrigan}, {Chadwick},
  {Chounet}, {Coignet}, {Cornils}, {Costamante}, {Degrange}, {Dickinson},
  {Djannati-Ata{\"\i}}, {Drury}, {Dubus}, {Egberts}, {Emmanoulopoulos},
  {Espigat}, {Feinstein}, {Ferrero}, {Fiasson}, {Filipovic}, {Fontaine},
  {Fukui}, {Funk}, {Funk}, {F{\"u}{\ss}ling}, {Gallant}, {Giebels},
  {Glicenstein}, {Goret}, {Hadjichristidis}, {Hauser}, {Hauser}, {Heinzelmann},
  {Henri}, {Hermann}, {Hinton}, {Hiraga}, {Hoffmann}, {Hofmann}, {Holleran},
  {Hoppe}, {Horns}, {Ishisaki}, {Jacholkowska}, {de Jager}, {Kendziorra},
  {Kerschhaggl}, {Kh{\'e}lifi}, {Komin}, {Konopelko}, {Kosack}, {Lamanna},
  {Latham}, {Le Gallou}, {Lemi{\`e}re}, {Lemoine-Goumard}, {Lohse}, {Martin},
  {Martineau-Huynh}, {Marcowith}, {Masterson}, {Maurin}, {McComb}, {Moulin},
  {Moriguchi}, {de Naurois}, {Nedbal}, {Nolan}, {Noutsos}, {Orford}, {Osborne},
  {Ouchrif}, {Panter}, {Pelletier}, {Pita}, {P{\"u}hlhofer}, {Punch},
  {Ranchon}, {Raubenheimer}, {Raue}, {Rayner}, {Reimer}, {Ripken}, {Rob},
  {Rolland}, {Rosier-Lees}, {Rowell}, {Sahakian}, {Santangelo}, {Saug{\'e}},
  {Schlenker}, {Schlickeiser}, {Schr{\"o}der}, {Schwanke}, {Schwarzburg},
  {Schwemmer}, {Shalchi}, {Sol}, {Spangler}, {Spanier}, {Steenkamp},
  {Stegmann}, {Superina}, {Tam}, {Tavernet}, {Terrier}, {Tluczykont}, {van
  Eldik}, {Vasileiadis}, {Venter}, {Vialle}, {Vincent}, {V{\"o}lk}, {Wagner},
  \& {Ward}}]{aharonian07}
{Aharonian}, F., {Akhperjanian}, A.~G., {Bazer-Bachi}, A.~R., {et~al.} 2007,
  \apj, 661, 236, \dodoi{10.1086/512603}

\bibitem[{{Aharonian} {et~al.}(2008{\natexlab{a}}){Aharonian}, {Akhperjanian},
  {Barres de Almeida}, {Bazer-Bachi}, {Behera}, {Beilicke}, {Benbow},
  {Bernl{\"o}hr}, {Boisson}, {Borrel}, {Braun}, {Brion}, {Brucker},
  {B{\"u}hler}, {Bulik}, {B{\"u}sching}, {Boutelier}, {Carrigan}, {Chadwick},
  {Chaves}, {Chounet}, {Clapson}, {Coignet}, {Cornils}, {Costamante}, {Dalton},
  {Degrange}, {Dickinson}, {Djannati-Ata{\"\i}}, {Domainko}, {O'C. Drury},
  {Dubois}, {Dubus}, {Dyks}, {Egberts}, {Emmanoulopoulos}, {Espigat},
  {Farnier}, {Feinstein}, {Fiasson}, {F{\"o}rster}, {Fontaine}, {Funk},
  {F{\"u}{\ss}ling}, {Gabici}, {Gallant}, {Giebels}, {Glicenstein},
  {Gl{\"u}ck}, {Goret}, {Hadjichristidis}, {Hauser}, {Hauser}, {Heinzelmann},
  {Henri}, {Hermann}, {Hinton}, {Hoffmann}, {Hofmann}, {Holleran}, {Hoppe},
  {Horns}, {Jacholkowska}, {de Jager}, {Jung}, {Katarzy{\'n}ski}, {Kaufmann},
  {Kendziorra}, {Kerschhaggl}, {Khangulyan}, {Kh{\'e}lifi}, {Keogh}, {Komin},
  {Kosack}, {Lamanna}, {Latham}, {Lemoine-Goumard}, {Lenain}, {Lohse},
  {Martin}, {Martineau-Huynh}, {Marcowith}, {Masterson}, {Maurin}, {McComb},
  {Moderski}, {Moulin}, {Naumann-Godo}, {de Naurois}, {Nedbal}, {Nekrassov},
  {Nolan}, {Ohm}, {Olive}, {de O{\~n}a Wilhelmi}, {Orford}, {Osborne},
  {Ostrowski}, {Panter}, {Pedaletti}, {Pelletier}, {Petrucci}, {Pita},
  {P{\"u}hlhofer}, {Punch}, {Quirrenbach}, {Raubenheimer}, {Raue}, {Rayner},
  {Renaud}, {Rieger}, {Reimer}, {Ripken}, {Rob}, {Rosier-Lees}, {Rowell},
  {Rudak}, {Ruppel}, {Sahakian}, {Santangelo}, {Schlickeiser}, {Sch{\"o}ck},
  {Schr{\"o}der}, {Schwanke}, {Schwarzburg}, {Schwemmer}, {Shalchi}, {Skilton},
  {Sol}, {Spangler}, {Stawarz}, {Steenkamp}, {Stegmann}, {Superina}, {Tam},
  {Tavernet}, {Terrier}, {van Eldik}, {Vasileiadis}, {Venter}, {Vialle},
  {Vincent}, {Vivier}, {V{\"o}lk}, {Volpe}, {Wagner}, {Ward}, {Zdziarski}, \&
  {Zech}}]{aharonian08}
{Aharonian}, F., {Akhperjanian}, A.~G., {Barres de Almeida}, U., {et~al.}
  2008{\natexlab{a}}, \aap, 486, 829, \dodoi{10.1051/0004-6361:200809655}

\bibitem[{{Aharonian} {et~al.}(2008{\natexlab{b}}){Aharonian}, {Akhperjanian},
  {Barres de Almeida}, {Bazer-Bachi}, {Behera}, {Beilicke}, {Benbow},
  {Bernl{\"o}hr}, {Boisson}, {Bolz}, {Borrel}, {Braun}, {Brion}, {Brown},
  {B{\"u}hler}, {Bulik}, {B{\"u}sching}, {Boutelier}, {Carrigan}, {Chadwick},
  {Chounet}, {Clapson}, {Coignet}, {Cornils}, {Costamante}, {Dalton},
  {Degrange}, {Dickinson}, {Djannati-Ata{\"\i}}, {Domainko}, {O'C. Drury},
  {Dubois}, {Dubus}, {Dyks}, {Egberts}, {Emmanoulopoulos}, {Espigat},
  {Farnier}, {Feinstein}, {Fiasson}, {F{\"o}rster}, {Fontaine}, {Funk},
  {F{\"u}{\ss}ling}, {Gallant}, {Giebels}, {Glicenstein}, {Gl{\"u}ck}, {Goret},
  {Hadjichristidis}, {Hauser}, {Hauser}, {Heinzelmann}, {Henri}, {Hermann},
  {Hinton}, {Hoffmann}, {Hofmann}, {Holleran}, {Hoppe}, {Horns},
  {Jacholkowska}, {de Jager}, {Jung}, {Katarzy{\'n}ski}, {Kendziorra},
  {Kerschhaggl}, {Kh{\'e}lifi}, {Keogh}, {Komin}, {Kosack}, {Lamanna},
  {Latham}, {Lemoine-Goumard}, {Lenain}, {Lohse}, {Martin}, {Martineau-Huynh},
  {Marcowith}, {Masterson}, {Maurin}, {McComb}, {Moderski}, {Moulin},
  {Naumann-Godo}, {de Naurois}, {Nedbal}, {Nekrassov}, {Nolan}, {Ohm}, {Olive},
  {de O{\~n}a Wilhelmi}, {Orford}, {Osborne}, {Ostrowski}, {Panter},
  {Pedaletti}, {Pelletier}, {Petrucci}, {Pita}, {P{\"u}hlhofer}, {Punch},
  {Raubenheimer}, {Raue}, {Rayner}, {Renaud}, {Ripken}, {Rob}, {Rosier-Lees},
  {Rowell}, {Rudak}, {Ruppel}, {Sahakian}, {Santangelo}, {Schlickeiser},
  {Sch{\"o}ck}, {Schr{\"o}der}, {Schwanke}, {Schwarzburg}, {Schwemmer},
  {Shalchi}, {Sol}, {Spangler}, {Stawarz}, {Steenkamp}, {Stegmann}, {Superina},
  {Tam}, {Tavernet}, {Terrier}, {van Eldik}, {Vasileiadis}, {Venter}, {Vialle},
  {Vincent}, {Vivier}, {V{\"o}lk}, {Volpe}, {Wagner}, {Ward}, {Zdziarski}, \&
  {Zech}}]{aharonian08b}
---. 2008{\natexlab{b}}, \aap, 483, 509, \dodoi{10.1051/0004-6361:20079230}

\bibitem[{{Ahnen} {et~al.}(2017){Ahnen}, {Ansoldi}, {Antonelli}, {Arcaro},
  {Babi{\'c}}, {Banerjee}, {Bangale}, {Barres de Almeida}, {Barrio}, {Becerra
  Gonz{\'a}lez}, {Bednarek}, {Bernardini}, {Berti}, {Bhattacharyya},
  {Biasuzzi}, {Biland }, {Blanch}, {Bonnefoy}, {Bonnoli}, {Carosi}, {Carosi},
  {Chatterjee}, {Colak}, {Colin}, {Colombo}, {Contreras}, {Cortina}, {Covino},
  {Cumani}, {Da Vela}, {Dazzi}, {De Angelis}, {De Lotto}, {de O{\~n}a
  Wilhelmi}, {Di Pierro}, {Doert}, {Dom{\'\i}nguez}, {Dominis Prester},
  {Dorner}, {Doro}, {Einecke}, {Eisenacher Glawion}, {Elsaesser},
  {Engelkemeier}, {Fallah Ramazani}, {Fern{\'a}ndez-Barral}, {Fidalgo},
  {Fonseca}, {Font}, {Fruck}, {Galindo}, {Garc{\'\i}a L{\'o}pez},
  {Garczarczyk}, {Gaug}, {Giammaria}, {Godinovi{\'c}}, {Gora}, {Guberman},
  {Hadasch}, {Hahn}, {Hassan}, {Hayashida}, {Herrera}, {Hose}, {Hrupec},
  {Inada}, {Ishio}, {Konno}, {Kubo}, {Kushida}, {Kuve{\v{z}}di{\'c}}, {Lelas},
  {Lindfors}, {Lombardi}, {Longo}, {L{\'o}pez}, {Maggio}, {Majumdar},
  {Makariev}, {Maneva}, {Manganaro}, {Mannheim}, {Maraschi}, {Mariotti},
  {Mart{\'\i}nez}, {Mazin}, {Menzel}, {Minev}, {Mirzoyan}, {Moralejo},
  {Moreno}, {Moretti}, {Neustroev}, {Niedzwiecki}, {Nievas Rosillo}, {Nilsson},
  {Ninci}, {Nishijima}, {Noda}, {Nogu{\'e}s}, {Paiano}, {Palacio}, {Paneque},
  {Paoletti}, {Paredes}, {Pedaletti}, {Peresano}, {Perri}, {Persic}, {Prada
  Moroni}, {Prand ini}, {Puljak}, {Garcia}, {Reichardt}, {Rhode}, {Rib{\'o}},
  {Rico}, {Righi}, {Saito}, {Satalecka}, {Schroeder}, {Schweizer}, {Shore},
  {Sitarek}, {{\v{S}}nidari{\'c}}, {Sobczynska}, {Stamerra}, {Strzys},
  {Suri{\'c}}, {Takalo}, {Tavecchio}, {Temnikov}, {Terzi{\'c}}, {Tescaro},
  {Teshima}, {Torres-Alb{\`a}}, {Treves}, {Vanzo}, {Vazquez Acosta}, {Vovk},
  {Ward}, {Will}, \& {Zari{\'c}}}]{ahnen17}
{Ahnen}, M.~L., {Ansoldi}, S., {Antonelli}, L.~A., {et~al.} 2017, \mnras, 472,
  2956, \dodoi{10.1093/mnras/stx2079}

\bibitem[{{Ajello} {et~al.}(2012){Ajello}, {Allafort}, {Baldini}, {Ballet},
  {Barbiellini}, {Bastieri}, {Bechtol}, {Bellazzini}, {Berenji}, {Blandford},
  {Bloom}, {Bonamente}, {Borgland}, {Bregeon}, {Brigida}, {Bruel}, {Buehler},
  {Buson}, {Caliandro}, {Cameron}, {Caraveo}, {Casandjian}, {Cecchi},
  {Charles}, {Chekhtman}, {Ciprini}, {Claus}, {Cohen-Tanugi}, {Cutini}, {de
  Angelis}, {de Palma}, {Dermer}, {Silva}, {Drell}, {Drlica-Wagner}, {Dubois},
  {Favuzzi}, {Fegan}, {Ferrara}, {Focke}, {Frailis}, {Fukazawa}, {Fukui},
  {Fusco}, {Gargano}, {Gasparrini}, {Germani}, {Giglietto}, {Giommi},
  {Giordano}, {Giroletti}, {Glanzman}, {Godfrey}, {Grove}, {Guiriec},
  {Hadasch}, {Hanabata}, {Harding}, {Hayashi}, {Hays}, {Itoh},
  {J{\'o}hannesson}, {Johnson}, {Kamae}, {Katagiri}, {Kataoka},
  {Kn{\"o}dlseder}, {Kubo}, {Kuss}, {Lande}, {Latronico}, {Lee}, {Lionetto},
  {Longo}, {Loparco}, {Lovellette}, {Lubrano}, {Mazziotta}, {Mehault},
  {Michelson}, {Mizuno}, {Moiseev}, {Monte}, {Monzani}, {Morselli},
  {Moskalenko}, {Murgia}, {Nakamori}, {Naumann-Godo}, {Nishino}, {Nolan},
  {Norris}, {Nuss}, {Ohno}, {Ohsugi}, {Okumura}, {Omodei}, {Orlando}, {Ormes},
  {Paneque}, {Parent}, {Pelassa}, {Pesce-Rollins}, {Pierbattista}, {Piron},
  {Porter}, {Rain{\`o}}, {Rando}, {Reimer}, {Reimer}, {Reposeur}, {Roth},
  {Sadrozinski}, {Sgr{\`o}}, {Siskind}, {Smith}, {Spandre}, {Spinelli},
  {Suson}, {Tajima}, {Takahashi}, {Tanaka}, {Thayer}, {Thayer}, {Tibaldo},
  {Tibolla}, {Torres}, {Tosti}, {Tramacere}, {Troja}, {Uchiyama}, {Uehara},
  {Usher}, {Vand enbroucke}, {Van Etten}, {Vasileiou}, {Vianello}, {Vilchez},
  {Vitale}, {Waite}, {Wang}, {Winer}, {Wood}, {Yamamoto}, {Yamazaki}, {Yang},
  {Yasuda}, {Ziegler}, \& {Zimmer}}]{ajello12}
{Ajello}, M., {Allafort}, A., {Baldini}, L., {et~al.} 2012, \apj, 744, 80,
  \dodoi{10.1088/0004-637X/744/1/80}

\bibitem[{{Aliu} {et~al.}(2013){Aliu}, {Archambault}, {Arlen}, {Aune},
  {Beilicke}, {Benbow}, {Bird}, {Bouvier}, {Bradbury}, {Buckley}, {Bugaev},
  {Byrum}, {Cannon}, {Cesarini}, {Ciupik}, {Collins-Hughes}, {Connolly}, {Cui},
  {Dickherber}, {Duke}, {Dumm}, {Dwarkadas}, {Errando}, {Falcone}, {Federici},
  {Feng}, {Finley}, {Finnegan}, {Fortson}, {Furniss}, {Galante}, {Gall},
  {Gillanders}, {Godambe}, {Gotthelf}, {Griffin}, {Grube}, {Gyuk}, {Hanna},
  {Holder}, {Huan}, {Hughes}, {Humensky}, {Kaaret}, {Karlsson}, {Kertzman},
  {Khassen}, {Kieda}, {Krawczynski}, {Krennrich}, {Lang}, {Lee}, {Madhavan},
  {Maier}, {Majumdar}, {McArthur}, {McCann}, {Millis}, {Moriarty}, {Mukherjee},
  {Nelson}, {O'Faol{\'a}in de Bhr{\'o}ithe}, {Ong}, {Orr}, {Otte}, {Pandel},
  {Park}, {Perkins}, {Pohl}, {Popkow}, {Prokoph}, {Quinn}, {Ragan}, {Reyes},
  {Reynolds}, {Roache}, {Rose}, {Ruppel}, {Saxon}, {Schroedter}, {Sembroski},
  {{\c{S}}ent{\"u}rk}, {Skole}, {Telezhinsky}, {Te{\v{s}}i{\'c}}, {Theiling},
  {Thibadeau}, {Tsurusaki}, {Tyler}, {Varlotta}, {Vassiliev}, {Vincent},
  {Wakely}, {Ward}, {Weekes}, {Weinstein}, {Weisgarber}, {Welsing}, {Williams},
  \& {Zitzer}}]{aliu13}
{Aliu}, E., {Archambault}, S., {Arlen}, T., {et~al.} 2013, \apj, 770, 93,
  \dodoi{10.1088/0004-637X/770/2/93}

\bibitem[{{Ambrogi} {et~al.}(2019){Ambrogi}, {Zanin}, {Casanova}, {De O{\~n}a
  Wilhelmi}, {Peron}, \& {Aharonian}}]{ambrogi19}
{Ambrogi}, L., {Zanin}, R., {Casanova}, S., {et~al.} 2019, \aap, 623, A86,
  \dodoi{10.1051/0004-6361/201833985}

\bibitem[{Amenomori {et~al.}(2021)Amenomori, Bao, Bi, Chen, Chen, Chen, Chen,
  Chen, Cirennima, Cui, Danzengluobu, Ding, Fang, Fang, Feng, Feng, Feng, Gao,
  Gou, Guo, Guo, He, He, Hibino, Hotta, Hu, Hu, Huang, Jia, Jiang, Jin,
  Kasahara, Katayose, Kato, Kato, Kawata, Kihara, Ko, Kozai, Labaciren, Le, Li,
  Li, Li, Lin, Liu, Liu, Liu, Liu, Liu, Lou, Lu, Meng, Munakata, Nakada,
  Nakamura, Nanjo, Nishizawa, Ohnishi, Ohura, Ozawa, Qian, Qu, Saito, Sakata,
  Sako, Shao, Shibata, Shiomi, Sugimoto, Takano, Takita, Tan, Tateyama, Torii,
  Tsuchiya, Udo, Wang, Wu, Xue, Yamamoto, Yang, Yokoe, Yuan, Zhai, Zhang,
  Zhang, Zhang, Zhang, Zhang, Zhang, Zhang, Zhao, Zhaxisangzhu, Zhou, \&
  ASγCollaboration}]{amenomori21}
Amenomori, M., Bao, Y.~W., Bi, X.~J., {et~al.} 2021, Nature Astronomy, 5, 460,
  \dodoi{10.1038/s41550-020-01294-9}

\bibitem[{{Araya}(2013)}]{araya13}
{Araya}, M. 2013, \mnras, 434, 2202, \dodoi{10.1093/mnras/stt1162}

\bibitem[{{Araya}(2014)}]{araya14}
---. 2014, \mnras, 444, 860, \dodoi{10.1093/mnras/stu1484}

\bibitem[{{Auchettl} {et~al.}(2014){Auchettl}, {Slane}, \&
  {Castro}}]{auchettl14}
{Auchettl}, K., {Slane}, P., \& {Castro}, D. 2014, \apj, 783, 32,
  \dodoi{10.1088/0004-637X/783/1/32}

\bibitem[{{Bamba} {et~al.}(2005){Bamba}, {Yamazaki}, {Yoshida}, {Terasawa}, \&
  {Koyama}}]{bamba05}
{Bamba}, A., {Yamazaki}, R., {Yoshida}, T., {Terasawa}, T., \& {Koyama}, K.
  2005, \apj, 621, 793, \dodoi{10.1086/427620}

\bibitem[{{Bell}(1978)}]{bell78}
{Bell}, A.~R. 1978, \mnras, 182, 147, \dodoi{10.1093/mnras/182.2.147}

\bibitem[{{Bell} {et~al.}(2013){Bell}, {Schure}, {Reville}, \&
  {Giacinti}}]{bell13}
{Bell}, A.~R., {Schure}, K.~M., {Reville}, B., \& {Giacinti}, G. 2013, \mnras,
  431, 415, \dodoi{10.1093/mnras/stt179}

\bibitem[{{Blair} {et~al.}(2005){Blair}, {Sankrit}, \& {Raymond}}]{blair05}
{Blair}, W.~P., {Sankrit}, R., \& {Raymond}, J.~C. 2005, \aj, 129, 2268,
  \dodoi{10.1086/429381}

\bibitem[{{Braun} {et~al.}(2019){Braun}, {Safi-Harb}, \& {Fryer}}]{braun19}
{Braun}, C., {Safi-Harb}, S., \& {Fryer}, C.~L. 2019, \mnras, 489, 4444,
  \dodoi{10.1093/mnras/stz2437}

\bibitem[{{Brose} {et~al.}(2020){Brose}, {Pohl}, {Sushch}, {Petruk}, \&
  {Kuzyo}}]{brose20}
{Brose}, R., {Pohl}, M., {Sushch}, I., {Petruk}, O., \& {Kuzyo}, T. 2020, \aap,
  634, A59, \dodoi{10.1051/0004-6361/201936567}

\bibitem[{{Burrows} \& {Guo}(1994)}]{burrows94}
{Burrows}, D.~N., \& {Guo}, Z. 1994, \apjl, 421, L19, \dodoi{10.1086/187177}

\bibitem[{{Cao}(2010)}]{cao10}
{Cao}, Z. 2010, Chinese Physics C, 34, 249, \dodoi{10.1088/1674-1137/34/2/018}

\bibitem[{Cao {et~al.}(2021)Cao, Aharonian, An, Axikegu, Bai, Bai, Bao,
  Bastieri, Bi, Bi, Cai, Cai, Cao, Chang, Chang, Chang, Chen, Chen, Chen, Chen,
  Chen, Chen, Chen, Chen, Chen, Chen, Chen, Chen, Chen, Cheng, Cheng, Cui, Cui,
  Cui, Dai, Dai, Dai, Danzengluobu, della Volpe, D′Ettorre~Piazzoli, Dong,
  Fan, Fan, Fan, Fang, Fang, Feng, Feng, Feng, Feng, Gao, Gao, Gao, Gao, Ge,
  Geng, Gong, Gou, Gu, Guo, Guo, Guo, Guo, Han, He, He, He, He, He, He, Heller,
  Hor, Hou, Hou, Hu, Hu, Hu, Hu, Huang, Huang, Huang, Huang, Huang, Ji, Ji,
  Jia, Jiang, Jiang, Jin, Kuleshov, Levochkin, Li, Li, Li, Li, Li, Li, Li, Li,
  Li, Li, Li, Li, Li, Li, Li, Li, Li, Liang, Liang, Lin, Liu, Liu, Liu, Liu,
  Liu, Liu, Liu, Liu, Liu, Liu, Liu, Liu, Liu, Liu, Liu, Long, Lu, Lv, Ma, Ma,
  Ma, Mao, Masood, Mitthumsiri, Montaruli, Nan, Pang, Pattarakijwanich, Pei,
  Qi, Ruffolo, Rulev, S{\'a}iz, Shao, Shchegolev, Sheng, Shi, Song, Stenkin,
  Stepanov, Sun, Sun, Sun, Tam, Tang, Tian, Wang, Wang, Wang, Wang, Wang, Wang,
  Wang, Wang, Wang, Wang, Wang, Wang, Wang, Wang, Wang, Wang, Wang, Wang, Wang,
  Wang, Wang, Wei, Wei, Wei, Wen, Wu, Wu, Wu, Wu, Wu, Xi, Xia, Xia, Xiang,
  Xiao, Xiao, Xin, Xin, Xing, Xu, Xu, Xue, Yan, Yang, Yang, Yang, Yang, Yang,
  Yang, Yang, Yao, Yao, Ye, Yin, Yin, You, You, Yu, Yuan, Zeng, Zeng, Zeng,
  Zeng, Zha, Zhai, Zhang, Zhang, Zhang, Zhang, Zhang, Zhang, Zhang, Zhang,
  Zhang, Zhang, Zhang, Zhang, Zhang, Zhang, Zhang, Zhang, Zhang, Zhang, Zhang,
  Zhao, Zhao, Zhao, Zhao, Zhao, Zheng, Zheng, Zhou, Zhou, Zhou, Zhou, Zhou,
  Zhou, Zhu, Zhu, Zhu, Zhu, \& Zuo}]{cao21}
Cao, Z., Aharonian, F.~A., An, Q., {et~al.} 2021, Nature, 594, 33,
  \dodoi{10.1038/s41586-021-03498-z}

\bibitem[{{Caprioli} {et~al.}(2009){Caprioli}, {Blasi}, \&
  {Amato}}]{caprioli09}
{Caprioli}, D., {Blasi}, P., \& {Amato}, E. 2009, \mnras, 396, 2065,
  \dodoi{10.1111/j.1365-2966.2008.14298.x}

\bibitem[{{Carter} {et~al.}(1997){Carter}, {Dickel}, \& {Bomans}}]{carter97}
{Carter}, L.~M., {Dickel}, J.~R., \& {Bomans}, D.~J. 1997, \pasp, 109, 990,
  \dodoi{10.1086/133971}

\bibitem[{{Case} \& {Bhattacharya}(1998)}]{case98}
{Case}, G.~L., \& {Bhattacharya}, D. 1998, \apj, 504, 761,
  \dodoi{10.1086/306089}

\bibitem[{{Cash}(1979)}]{cash79}
{Cash}, W. 1979, \apj, 228, 939, \dodoi{10.1086/156922}

\bibitem[{Castro \& Slane(2010)}]{castro10}
Castro, D., \& Slane, P. 2010, The Astrophysical Journal, 717, 372,
  \dodoi{10.1088/0004-637x/717/1/372}

\bibitem[{{Castro} {et~al.}(2013){Castro}, {Slane}, {Carlton}, \&
  {Figueroa-Feliciano}}]{castro13}
{Castro}, D., {Slane}, P., {Carlton}, A., \& {Figueroa-Feliciano}, E. 2013,
  \apj, 774, 36, \dodoi{10.1088/0004-637X/774/1/36}

\bibitem[{{Castro} {et~al.}(2012){Castro}, {Slane}, {Ellison}, \&
  {Patnaude}}]{castro12}
{Castro}, D., {Slane}, P., {Ellison}, D.~C., \& {Patnaude}, D.~J. 2012, \apj,
  756, 88, \dodoi{10.1088/0004-637X/756/1/88}

\bibitem[{{Caswell} {et~al.}(1975){Caswell}, {Murray}, {Roger}, {Cole}, \&
  {Cooke}}]{caswell75}
{Caswell}, J.~L., {Murray}, J.~D., {Roger}, R.~S., {Cole}, D.~J., \& {Cooke},
  D.~J. 1975, \aap, 45, 239

\bibitem[{{Celli} {et~al.}(2019){Celli}, {Morlino}, {Gabici}, \&
  {Aharonian}}]{celli19}
{Celli}, S., {Morlino}, G., {Gabici}, S., \& {Aharonian}, F.~A. 2019, \mnras,
  490, 4317, \dodoi{10.1093/mnras/stz2897}

\bibitem[{{Cesur} {et~al.}(2019){Cesur}, {Sezer}, {de Plaa}, \&
  {Vink}}]{cesur19}
{Cesur}, N., {Sezer}, A., {de Plaa}, J., \& {Vink}, J. 2019, Advances in Space
  Research, 64, 759, \dodoi{10.1016/j.asr.2019.05.010}

\bibitem[{{Chen} \& {Slane}(2001)}]{chen01}
{Chen}, Y., \& {Slane}, P.~O. 2001, \apj, 563, 202, \dodoi{10.1086/323886}

\bibitem[{{Chevalier}(1999)}]{chevalier99}
{Chevalier}, R.~A. 1999, \apj, 511, 798, \dodoi{10.1086/306710}

\bibitem[{{Cohen}(2016)}]{cohen16}
{Cohen}, J.~M. 2016, PhD thesis, University of Maryland, College Park

\bibitem[{{Condon} {et~al.}(2017){Condon}, {Lemoine-Goumard}, {Acero}, \&
  {Katagiri}}]{condon17}
{Condon}, B., {Lemoine-Goumard}, M., {Acero}, F., \& {Katagiri}, H. 2017, \apj,
  851, 100, \dodoi{10.3847/1538-4357/aa9be8}

\bibitem[{{Cristofari} {et~al.}(2020){Cristofari}, {Blasi}, \&
  {Amato}}]{cristofari20}
{Cristofari}, P., {Blasi}, P., \& {Amato}, E. 2020, Astroparticle Physics, 123,
  102492, \dodoi{10.1016/j.astropartphys.2020.102492}

\bibitem[{{Cui} {et~al.}(2018){Cui}, {Yeung}, {Tam}, \&
  {P{\"u}hlhofer}}]{cui18}
{Cui}, Y., {Yeung}, P. K.~H., {Tam}, P.~H.~T., \& {P{\"u}hlhofer}, G. 2018,
  \apj, 860, 69, \dodoi{10.3847/1538-4357/aac37b}

\bibitem[{{de Palma} {et~al.}(2013){de Palma}, {Brandt}, {Johannesson}, \&
  {Tibaldo}}]{depalma13}
{de Palma}, F., {Brandt}, T.~J., {Johannesson}, G., \& {Tibaldo}, L. 2013,
  arXiv e-prints, arXiv:1304.1395.
\newblock \doarXiv{1304.1395}

\bibitem[{{Doroshenko} {et~al.}(2017){Doroshenko}, {P{\"u}hlhofer}, {Bamba},
  {Acero}, {Tian}, {Klochkov}, \& {Santangelo}}]{doroshenko17}
{Doroshenko}, V., {P{\"u}hlhofer}, G., {Bamba}, A., {et~al.} 2017, \aap, 608,
  A23, \dodoi{10.1051/0004-6361/201730983}

\bibitem[{{Ergin} {et~al.}(2014){Ergin}, {Sezer}, {Saha}, {Majumdar},
  {Chatterjee}, {Bayirli}, \& {Ercan}}]{ergin14}
{Ergin}, T., {Sezer}, A., {Saha}, L., {et~al.} 2014, \apj, 790, 65,
  \dodoi{10.1088/0004-637X/790/1/65}

\bibitem[{{Ergin} {et~al.}(2015){Ergin}, {Sezer}, {Saha}, {Majumdar},
  {G{\"o}k}, \& {Ercan}}]{ergin15}
---. 2015, \apj, 804, 124, \dodoi{10.1088/0004-637X/804/2/124}

\bibitem[{{Finley} \& {Oegelman}(1994)}]{finley94}
{Finley}, J.~P., \& {Oegelman}, H. 1994, \apjl, 434, L25,
  \dodoi{10.1086/187563}

\bibitem[{{Fraija} \& {Araya}(2016)}]{fraija16}
{Fraija}, N., \& {Araya}, M. 2016, \apj, 826, 31,
  \dodoi{10.3847/0004-637X/826/1/31}

\bibitem[{{Fukui} {et~al.}(2003){Fukui}, {Moriguchi}, {Tamura}, {Yamamoto},
  {Tawara}, {Mizuno}, {Onishi}, {Mizuno}, {Uchiyama}, {Hiraga}, {Takahashi},
  {Yamashita}, \& {Ikeuchi}}]{fukui03}
{Fukui}, Y., {Moriguchi}, Y., {Tamura}, K., {et~al.} 2003, \pasj, 55, L61,
  \dodoi{10.1093/pasj/55.5.L61}

\bibitem[{{Gelfand} {et~al.}(2013){Gelfand}, {Castro}, {Slane}, {Temim},
  {Hughes}, \& {Rakowski}}]{gelfand13}
{Gelfand}, J.~D., {Castro}, D., {Slane}, P.~O., {et~al.} 2013, \apj, 777, 148,
  \dodoi{10.1088/0004-637X/777/2/148}

\bibitem[{{Giacani} {et~al.}(2009){Giacani}, {Smith}, {Dubner}, {Loiseau},
  {Castelletti}, \& {Paron}}]{giacani09}
{Giacani}, E., {Smith}, M.~J.~S., {Dubner}, G., {et~al.} 2009, \aap, 507, 841,
  \dodoi{10.1051/0004-6361/200912253}

\bibitem[{{Giordano} {et~al.}(2012){Giordano}, {Naumann-Godo}, {Ballet},
  {Bechtol}, {Funk}, {Lande}, {Mazziotta}, {Rain{\`o}}, {Tanaka}, {Tibolla}, \&
  {Uchiyama}}]{giordano12}
{Giordano}, F., {Naumann-Godo}, M., {Ballet}, J., {et~al.} 2012, \apjl, 744,
  L2, \dodoi{10.1088/2041-8205/744/1/L2}

\bibitem[{{Gloeckler} \& {Jokipii}(1967)}]{gloeckler67}
{Gloeckler}, G., \& {Jokipii}, J.~R. 1967, \apjl, 148, L41,
  \dodoi{10.1086/180010}

\bibitem[{{Green} {et~al.}(1997){Green}, {Frail}, {Goss}, \&
  {Otrupcek}}]{green97}
{Green}, A.~J., {Frail}, D.~A., {Goss}, W.~M., \& {Otrupcek}, R. 1997, \aj,
  114, 2058, \dodoi{10.1086/118626}

\bibitem[{{Green}(2019{\natexlab{a}})}]{green19a}
{Green}, D.~A. 2019{\natexlab{a}}, Journal of Astrophysics and Astronomy, 40,
  36, \dodoi{10.1007/s12036-019-9601-6}

\bibitem[{{Green}(2019{\natexlab{b}})}]{green19b}
---. 2019{\natexlab{b}}, {\it A Catalogue of Galactic Supernova Remnants (2019
  June version)}, \url{http://www.mrao.cam.ac.uk/surveys/snrs/},  Cavendish
  Laboratory, Cambridge, United Kingdom

\bibitem[{{Green} \& {Stephenson}(2003)}]{green03}
{Green}, D.~A., \& {Stephenson}, F.~R. 2003, {Historical Supernovae}, ed.
  K.~{Weiler}, Vol. 598, 7--19, \dodoi{10.1007/3-540-45863-8_2}

\bibitem[{{H.~E.~S.~S. Collaboration} {et~al.}(2015{\natexlab{a}}){H.~E.~S.~S.
  Collaboration}, {Abramowski}, {Aharonian}, {Ait Benkhali}, {Akhperjanian},
  {Ang{\"u}ner}, {Backes}, {Balenderan}, {Balzer}, {Barnacka}, {Becherini},
  {Becker Tjus}, {Berge}, {Bernhard}, {Bernl{\"o}hr}, {Birsin}, {Biteau},
  {B{\"o}ttcher}, {Boisson}, {Bolmont}, {Bordas}, {Bregeon}, {Brun}, {Brun},
  {Bryan}, {Bulik}, {Carrigan}, {Casanova}, {Chadwick}, {Chakraborty},
  {Chalme-Calvet}, {Chaves}, {Chr{\'e}tien}, {Colafrancesco}, {Cologna},
  {Conrad}, {Couturier}, {Cui}, {Davids}, {Degrange}, {Deil}, {deWilt},
  {Djannati-Ata{\"\i}}, {Domainko}, {Donath}, {O'C. Drury}, {Dubus}, {Dutson},
  {Dyks}, {Dyrda}, {Edwards}, {Egberts}, {Eger}, {Espigat}, {Farnier}, {Fegan},
  {Feinstein}, {Fernandes}, {Fernand ez}, {Fiasson}, {Fontaine}, {F{\"o}rster},
  {F{\"u}{\ss}ling}, {Gabici}, {Gajdus}, {Gallant}, {Garrigoux}, {Giavitto},
  {Giebels}, {Glicenstein}, {Gottschall}, {Grondin}, {Grudzi{\'n}ska},
  {Hadasch}, {H{\"a}ffner}, {Hahn}, {Harris}, {Heinzelmann}, {Henri},
  {Hermann}, {Hervet}, {Hillert}, {Hinton}, {Hofmann}, {Hofverberg}, {Holler},
  {Horns}, {Ivascenko}, {Jacholkowska}, {Jahn}, {Jamrozy}, {Janiak},
  {Jankowsky}, {Jung-Richardt}, {Kastendieck}, {Katarzy{\'n}ski}, {Katz},
  {Kaufmann}, {Kh{\'e}lifi}, {Kieffer}, {Klepser}, {Klochkov}, {Klu{\'z}niak},
  {Kolitzus}, {Komin}, {Kosack}, {Krakau}, {Krayzel}, {Kr{\"u}ger}, {Laffon},
  {Lamanna}, {Lefaucheur}, {Lefranc}, {Lemi{\`e}re}, {Lemoine-Goumard},
  {Lenain}, {Lohse}, {Lopatin}, {Lu}, {Marandon}, {Marcowith}, {Marx},
  {Maurin}, {Maxted}, {Mayer}, {McComb}, {M{\'e}hault}, {Meintjes}, {Menzler},
  {Meyer}, {Mitchell}, {Moderski}, {Mohamed}, {Mor{\r{a}}}, {Moulin}, {Murach},
  {de Naurois}, {Niemiec}, {Nolan}, {Oakes}, {Odaka}, {Ohm}, {Opitz},
  {Ostrowski}, {Oya}, {Panter}, {Parsons}, {Arribas}, {Pekeur}, {Pelletier},
  {Petrucci}, {Peyaud}, {Pita}, {Poon}, {P{\"u}hlhofer}, {Punch},
  {Quirrenbach}, {Raab}, {Reichardt}, {Reimer}, {Reimer}, {Renaud}, {de los
  Reyes}, {Rieger}, {Romoli}, {Rosier-Lees}, {Rowell}, {Rudak}, {Rulten},
  {Sahakian}, {Salek}, {Sanchez}, {Santangelo}, {Schlickeiser},
  {Sch{\"u}ssler}, {Schulz}, {Schwanke}, {Schwarzburg}, {Schwemmer}, {Sol},
  {Spanier}, {Spengler}, {Spies}, {Stawarz}, {Steenkamp}, {Stegmann},
  {Stinzing}, {Stycz}, {Sushch}, {Tavernet}, {Tavernier}, {Taylor}, {Terrier},
  {Tluczykont}, {Trichard}, {Valerius}, {van Eldik}, {van Soelen},
  {Vasileiadis}, {Veh}, {Venter}, {Viana}, {Vincent}, {Vink}, {V{\"o}lk},
  {Volpe}, {Vorster}, {Vuillaume}, {Wagner}, {Wagner}, {Wagner}, {Ward},
  {Weidinger}, {Weitzel}, {White}, {Wierzcholska}, {Willmann}, {W{\"o}rnlein},
  {Wouters}, {Yang}, {Zabalza}, {Zaborov}, {Zacharias}, {Zdziarski}, {Zech}, \&
  {Zechlin}}]{hess15b}
{H.~E.~S.~S. Collaboration}, {Abramowski}, A., {Aharonian}, F., {et~al.}
  2015{\natexlab{a}}, \aap, 574, A100, \dodoi{10.1051/0004-6361/201425070}

\bibitem[{{H.~E.~S.~S. Collaboration} {et~al.}(2015{\natexlab{b}}){H.~E.~S.~S.
  Collaboration}, {Abramowski}, {Aharonian}, {Ait Benkhali}, {Akhperjanian},
  {Ang{\"u}ner}, {Backes}, {Balenderan}, {Balzer}, {Barnacka}, {Becherini},
  {Becker Tjus}, {Berge}, {Bernhard}, {Bernl{\"o}hr}, {Birsin}, {Biteau},
  {B{\"o}ttcher}, {Boisson}, {Bolmont}, {Bordas}, {Bregeon}, {Brun}, {Brun},
  {Bryan}, {Bulik}, {Carrigan}, {Casanova}, {Chadwick}, {Chakraborty},
  {Chalme-Calvet}, {Chaves}, {Chr{\'e}tien}, {Colafrancesco}, {Cologna},
  {Conrad}, {Couturier}, {Cui}, {Davids}, {Degrange}, {Deil}, {deWilt},
  {Djannati-Ata{\"\i}}, {Domainko}, {Donath}, {O'C. Drury}, {Dubus}, {Dutson},
  {Dyks}, {Dyrda}, {Edwards}, {Egberts}, {Eger}, {Espigat}, {Farnier}, {Fegan},
  {Feinstein}, {Fernandes}, {Fernand ez}, {Fiasson}, {Fontaine}, {F{\"o}rster},
  {F{\"u}{\ss}ling}, {Gabici}, {Gajdus}, {Gallant}, {Garrigoux}, {Giavitto},
  {Giebels}, {Glicenstein}, {Gottschall}, {Grondin}, {Grudzi{\'n}ska},
  {Hadasch}, {H{\"a}ffner}, {Hahn}, {Harris}, {Heinzelmann}, {Henri},
  {Hermann}, {Hervet}, {Hillert}, {Hinton}, {Hofmann}, {Hofverberg}, {Holler},
  {Horns}, {Ivascenko}, {Jacholkowska}, {Jahn}, {Jamrozy}, {Janiak},
  {Jankowsky}, {Jung-Richardt}, {Kastendieck}, {Katarzy{\'n}ski}, {Katz},
  {Kaufmann}, {Kh{\'e}lifi}, {Kieffer}, {Klepser}, {Klochkov}, {Klu{\'z}niak},
  {Kolitzus}, {Komin}, {Kosack}, {Krakau}, {Krayzel}, {Kr{\"u}ger}, {Laffon},
  {Lamanna}, {Lefaucheur}, {Lefranc}, {Lemi{\`e}re}, {Lemoine-Goumard},
  {Lenain}, {Lohse}, {Lopatin}, {Lu}, {Marandon}, {Marcowith}, {Marx},
  {Maurin}, {Maxted}, {Mayer}, {McComb}, {M{\'e}hault}, {Meintjes}, {Menzler},
  {Meyer}, {Mitchell}, {Moderski}, {Mohamed}, {Mor{\r{a}}}, {Moulin}, {Murach},
  {de Naurois}, {Niemiec}, {Nolan}, {Oakes}, {Odaka}, {Ohm}, {Opitz},
  {Ostrowski}, {Oya}, {Panter}, {Parsons}, {Arribas}, {Pekeur}, {Pelletier},
  {Petrucci}, {Peyaud}, {Pita}, {Poon}, {P{\"u}hlhofer}, {Punch},
  {Quirrenbach}, {Raab}, {Reichardt}, {Reimer}, {Reimer}, {Renaud}, {de los
  Reyes}, {Rieger}, {Romoli}, {Rosier-Lees}, {Rowell}, {Rudak}, {Rulten},
  {Sahakian}, {Salek}, {Sanchez}, {Santangelo}, {Schlickeiser},
  {Sch{\"u}ssler}, {Schulz}, {Schwanke}, {Schwarzburg}, {Schwemmer}, {Sol},
  {Spanier}, {Spengler}, {Spies}, {Stawarz}, {Steenkamp}, {Stegmann},
  {Stinzing}, {Stycz}, {Sushch}, {Tavernet}, {Tavernier}, {Taylor}, {Terrier},
  {Tluczykont}, {Trichard}, {Valerius}, {van Eldik}, {van Soelen},
  {Vasileiadis}, {Veh}, {Venter}, {Viana}, {Vincent}, {Vink}, {V{\"o}lk},
  {Volpe}, {Vorster}, {Vuillaume}, {Wagner}, {Wagner}, {Wagner}, {Ward},
  {Weidinger}, {Weitzel}, {White}, {Wierzcholska}, {Willmann}, {W{\"o}rnlein},
  {Wouters}, {Yang}, {Zabalza}, {Zaborov}, {Zacharias}, {Zdziarski}, {Zech}, \&
  {Zechlin}}]{hess15c}
---. 2015{\natexlab{b}}, \aap, 575, A81, \dodoi{10.1051/0004-6361/201424805}

\bibitem[{{H.~E.~S.~S. Collaboration} {et~al.}(2015{\natexlab{c}}){H.~E.~S.~S.
  Collaboration}, {Abramowski}, {Aharonian}, {Ait Benkhali}, {Akhperjanian},
  {Ang{\"u}ner}, {Anton}, {Backes}, {Balenderan}, {Balzer}, {Barnacka},
  {Becherini}, {Becker Tjus}, {Bernl{\"o}hr}, {Birsin}, {Bissaldi}, {Biteau},
  {B{\"o}ttcher}, {Boisson}, {Bolmont}, {Bordas}, {Brucker}, {Brun}, {Brun},
  {Bulik}, {Carrigan}, {Casanova}, {Chadwick}, {Chalme-Calvet}, {Chaves},
  {Cheesebrough}, {Chr{\'e}tien}, {Colafrancesco}, {Cologna}, {Conrad},
  {Couturier}, {Cui}, {Dalton}, {Daniel}, {Davids}, {Degrange}, {Deil},
  {deWilt}, {Dickinson}, {Djannati-Ata{\"\i}}, {Domainko}, {O'C. Drury},
  {Dubus}, {Dutson}, {Dyks}, {Dyrda}, {Edwards}, {Egberts}, {Eger}, {Espigat},
  {Farnier}, {Fegan}, {Feinstein}, {Fernand es}, {Fernandez}, {Fiasson},
  {Fontaine}, {F{\"o}rster}, {F{\"u}{\ss}ling}, {Gajdus}, {Gallant},
  {Garrigoux}, {Giavitto}, {Giebels}, {Glicenstein}, {Grondin},
  {Grudzi{\'n}ska}, {H{\"a}ffner}, {Hahn}, {Harris}, {Heinzelmann}, {Henri},
  {Hermann}, {Hervet}, {Hillert}, {Hinton}, {Hofmann}, {Hofverberg}, {Holler},
  {Horns}, {Jacholkowska}, {Jahn}, {Jamrozy}, {Janiak}, {Jankowsky}, {Jung},
  {Kastendieck}, {Katarzy{\'n}ski}, {Katz}, {Kaufmann}, {Kh{\'e}lifi},
  {Kieffer}, {Klepser}, {Klochkov}, {Klu{\'z}niak}, {Kneiske}, {Kolitzus},
  {Komin}, {Kosack}, {Krakau}, {Krayzel}, {Kr{\"u}ger}, {Laffon}, {Lamanna},
  {Lefaucheur}, {Lemi{\`e}re}, {Lemoine-Goumard}, {Lenain}, {Lohse}, {Lopatin},
  {Lu}, {Marandon}, {Marcowith}, {Marx}, {Maurin}, {Maxted}, {Mayer}, {McComb},
  {M{\'e}hault}, {Meintjes}, {Menzler}, {Meyer}, {Moderski}, {Mohamed},
  {Moulin}, {Murach}, {Naumann}, {de Naurois}, {Niemiec}, {Nolan}, {Oakes},
  {Odaka}, {Ohm}, {de O{\~n}a Wilhelmi}, {Opitz}, {Ostrowski}, {Oya}, {Panter},
  {Parsons}, {Paz Arribas}, {Pekeur}, {Pelletier}, {Perez}, {Petrucci},
  {Peyaud}, {Pita}, {Poon}, {P{\"u}hlhofer}, {Punch}, {Quirrenbach}, {Raab},
  {Raue}, {Reichardt}, {Reimer}, {Reimer}, {Renaud}, {de los Reyes}, {Rieger},
  {Rob}, {Romoli}, {Rosier-Lees}, {Rowell}, {Rudak}, {Rulten}, {Sahakian},
  {Sanchez}, {Santangelo}, {Schlickeiser}, {Sch{\"u}ssler}, {Schulz},
  {Schwanke}, {Schwarzburg}, {Schwemmer}, {Sol}, {Spengler}, {Spies},
  {Stawarz}, {Steenkamp}, {Stegmann}, {Stinzing}, {Stycz}, {Sushch},
  {Tavernet}, {Tavernier}, {Taylor}, {Terrier}, {Tluczykont}, {Trichard},
  {Valerius}, {van Eldik}, {van Soelen}, {Vasileiadis}, {Venter}, {Viana},
  {Vincent}, {V{\"o}lk}, {Volpe}, {Vorster}, {Vuillaume}, {Wagner}, {Wagner},
  {Wagner}, {Ward}, {Weidinger}, {Weitzel}, {White}, {Wierzcholska},
  {Willmann}, {W{\"o}rnlein}, {Wouters}, {Yang}, {Zabalza}, {Zacharias},
  {Zdziarski}, {Zech}, \& {Zechlin}}]{hess15a}
---. 2015{\natexlab{c}}, \aap, 574, A27, \dodoi{10.1051/0004-6361/201322694}

\bibitem[{{H.~E.~S.~S. Collaboration} {et~al.}(2018){H.~E.~S.~S.
  Collaboration}, {Abdalla}, {Abramowski}, {Aharonian}, {Ait Benkhali},
  {Akhperjanian}, {Andersson}, {Ang{\"u}ner}, {Arrieta}, {Aubert}, {Backes},
  {Balzer}, {Barnard}, {Becherini}, {Becker Tjus}, {Berge}, {Bernhard},
  {Bernl{\"o}hr}, {Blackwell}, {B{\"o}ttcher}, {Boisson}, {Bolmont}, {Bordas},
  {Bregeon}, {Brun}, {Brun}, {Bryan}, {Bulik}, {Capasso}, {Carr}, {Casanova},
  {Cerruti}, {Chakraborty}, {Chalme-Calvet}, {Chaves}, {Chen}, {Chevalier},
  {Chr{\'e}tien}, {Colafrancesco}, {Cologna}, {Condon}, {Conrad}, {Cui},
  {Davids}, {Decock}, {Degrange}, {Deil}, {Devin}, {deWilt}, {Dirson},
  {Djannati-Ata{\"\i}}, {Domainko}, {Donath}, {Drury}, {Dubus}, {Dutson},
  {Dyks}, {Edwards}, {Egberts}, {Eger}, {Ernenwein}, {Eschbach}, {Farnier},
  {Fegan}, {Fernand es}, {Fiasson}, {Fontaine}, {F{\"o}rster}, {Fukuyama},
  {Funk}, {F{\"u}{\ss}ling}, {Gabici}, {Gajdus}, {Gallant}, {Garrigoux},
  {Giavitto}, {Giebels}, {Glicenstein}, {Gottschall}, {Goyal}, {Grondin},
  {Hadasch}, {Hahn}, {Haupt}, {Hawkes}, {Heinzelmann}, {Henri}, {Hermann},
  {Hervet}, {Hinton}, {Hofmann}, {Hoischen}, {Holler}, {Horns}, {Ivascenko},
  {Jacholkowska}, {Jamrozy}, {Janiak}, {Jankowsky}, {Jankowsky}, {Jingo},
  {Jogler}, {Jouvin}, {Jung-Richardt}, {Kastendieck}, {Katarzy{\'n}ski},
  {Katz}, {Kerszberg}, {Kh{\'e}lifi}, {Kieffer}, {King}, {Klepser}, {Klochkov},
  {Klu{\'z}niak}, {Kolitzus}, {Komin}, {Kosack}, {Krakau}, {Kraus}, {Krayzel},
  {Kr{\"u}ger}, {Laffon}, {Lamanna}, {Lau}, {Lees}, {Lefaucheur}, {Lefranc},
  {Lemi{\`e}re}, {Lemoine-Goumard}, {Lenain}, {Leser}, {Lohse}, {Lorentz},
  {Liu}, {L{\'o}pez-Coto}, {Lypova}, {Marandon}, {Marcowith}, {Mariaud},
  {Marx}, {Maurin}, {Maxted}, {Mayer}, {Meintjes}, {Meyer}, {Mitchell},
  {Moderski}, {Mohamed}, {Mohrmann}, {Mor{\r{a}}}, {Moulin}, {Murach}, {de
  Naurois}, {Niederwanger}, {Niemiec}, {Oakes}, {O'Brien}, {Odaka}, {{\"O}ttl},
  {Ohm}, {Ostrowski}, {Oya}, {Padovani}, {Panter}, {Parsons}, {Pekeur},
  {Pelletier}, {Perennes}, {Petrucci}, {Peyaud}, {Piel}, {Pita}, {Poon},
  {Prokhorov}, {Prokoph}, {P{\"u}hlhofer}, {Punch}, {Quirrenbach}, {Raab},
  {Reimer}, {Reimer}, {Renaud}, {de los Reyes}, {Rieger}, {Romoli},
  {Rosier-Lees}, {Rowell}, {Rudak}, {Rulten}, {Sahakian}, {Salek}, {Sanchez},
  {Santangelo}, {Sasaki}, {Schlickeiser}, {Sch{\"u}ssler}, {Schulz},
  {Schwanke}, {Schwemmer}, {Settimo}, {Seyffert}, {Shafi}, {Shilon}, {Simoni},
  {Sol}, {Spanier}, {Spengler}, {Spies}, {Stawarz}, {Steenkamp}, {Stegmann},
  {Stinzing}, {Stycz}, {Sushch}, {Takahashi}, {Tavernet}, {Tavernier},
  {Taylor}, {Terrier}, {Tibaldo}, {Tiziani}, {Tluczykont}, {Trichard}, {Tuffs},
  {Uchiyama}, {van der Walt}, {van Eldik}, {van Rensburg}, {van Soelen},
  {Vasileiadis}, {Veh}, {Venter}, {Viana}, {Vincent}, {Vink}, {Voisin},
  {V{\"o}lk}, {Volpe}, {Vuillaume}, {Wadiasingh}, {Wagner}, {Wagner}, {Wagner},
  {White}, {Wierzcholska}, {Willmann}, {W{\"o}rnlein}, {Wouters}, {Yang},
  {Zabalza}, {Zaborov}, {Zacharias}, {Zdziarski}, {Zech}, {Zefi}, {Ziegler}, \&
  {{\.Z}ywucka}}]{hess18b}
{H.~E.~S.~S. Collaboration}, {Abdalla}, H., {Abramowski}, A., {et~al.} 2018,
  \aap, 612, A6, \dodoi{10.1051/0004-6361/201629790}

\bibitem[{{Hayato} {et~al.}(2010){Hayato}, {Yamaguchi}, {Tamagawa}, {Katsuda},
  {Hwang}, {Hughes}, {Ozawa}, {Bamba}, {Kinugasa}, {Terada}, {Furuzawa},
  {Kunieda}, \& {Makishima}}]{hayato10}
{Hayato}, A., {Yamaguchi}, H., {Tamagawa}, T., {et~al.} 2010, \apj, 725, 894,
  \dodoi{10.1088/0004-637X/725/1/894}

\bibitem[{{HEASARC}(2014)}]{heasarc14}
{HEASARC}. 2014, {HEAsoft: Unified Release of FTOOLS and XANADU}.
\newblock \doeprint{1408.004}

\bibitem[{{H.E.S.S. Collaboration} {et~al.}(2018){H.E.S.S. Collaboration},
  {Abdalla, H.}, {Abramowski, A.}, {Aharonian, F.}, {Ait Benkhali, F.},
  {Akhperjanian, A. G.}, {Andersson, T.}, {Ang\"uner, E. O.}, {Arrieta, M.},
  {Aubert, P.}, {Backes, M.}, {Balzer, A.}, {Barnard, M.}, {Becherini, Y.},
  {Becker Tjus, J.}, {Berge, D.}, {Bernhard, S.}, {Bernl\"ohr, K.}, {Blackwell,
  R.}, {B\"ottcher, M.}, {Boisson, C.}, {Bolmont, J.}, {Bordas, P.}, {Bregeon,
  J.}, {Brun, F.}, {Brun, P.}, {Bryan, M.}, {Bulik, T.}, {Capasso, M.}, {Carr,
  J.}, {Casanova, S.}, {Cerruti, M.}, {Chakraborty, N.}, {Chalme-Calvet, R.},
  {Chaves, R.C. G.}, {Chen, A.}, {Chevalier, J.}, {Chr\'etien, M.},
  {Colafrancesco, S.}, {Cologna, G.}, {Condon, B.}, {Conrad, J.}, {Cui, Y.},
  {Davids, I. D.}, {Decock, J.}, {Degrange, B.}, {Deil, C.}, {Devin, J.},
  {deWilt, P.}, {Dirson, L.}, {Djannati-Ata\"{\i}, A.}, {Domainko, W.},
  {Donath, A.}, {Drury, L. O\'{}C.}, {Dubus, G.}, {Dutson, K.}, {Dyks, J.},
  {Edwards, T.}, {Egberts, K.}, {Eger, P.}, {Ernenwein, J.-P.}, {Eschbach, S.},
  {Farnier, C.}, {Fegan, S.}, {Fernandes, M. V.}, {Fiasson, A.}, {Fontaine,
  G.}, {F\"orster, A.}, {Funk, S.}, {F\"u\ss{}ling, M.}, {Gabici, S.}, {Gajdus,
  M.}, {Gallant, Y. A.}, {Garrigoux, T.}, {Giavitto, G.}, {Giebels, B.},
  {Glicenstein, J. F.}, {Gottschall, D.}, {Goyal, A.}, {Grondin, M.-H.},
  {Hadasch, D.}, {Hahn, J.}, {Haupt, M.}, {Hawkes, J.}, {Heinzelmann, G.},
  {Henri, G.}, {Hermann, G.}, {Hervet, O.}, {Hinton, J. A.}, {Hofmann, W.},
  {Hoischen, C.}, {Holler, M.}, {Horns, D.}, {Ivascenko, A.}, {Jacholkowska,
  A.}, {Jamrozy, M.}, {Janiak, M.}, {Jankowsky, D.}, {Jankowsky, F.}, {Jingo,
  M.}, {Jogler, T.}, {Jouvin, L.}, {Jung-Richardt, I.}, {Kastendieck, M. A.},
  {Katarzy\'{}nski, K.}, {Katz, U.}, {Kerszberg, D.}, {Kh\'elifi, B.},
  {Kieffer, M.}, {King, J.}, {Klepser, S.}, {Klochkov, D.}, {Klu\'{}zniak, W.},
  {Kolitzus, D.}, {Komin, Nu.}, {Kosack, K.}, {Krakau, S.}, {Kraus, M.},
  {Krayzel, F.}, {Kr\"uger, P. P.}, {Laffon, H.}, {Lamanna, G.}, {Lau, J.},
  {Lees, J.-P.}, {Lefaucheur, J.}, {Lefranc, V.}, {Lemi\`ere, A.},
  {Lemoine-Goumard, M.}, {Lenain, J.-P.}, {Leser, E.}, {Lohse, T.}, {Lorentz,
  M.}, {Liu, R.}, {L\'opez-Coto, R.}, {Lypova, I.}, {Marandon, V.}, {Marcowith,
  A.}, {Mariaud, C.}, {Marx, R.}, {Maurin, G.}, {Maxted, N.}, {Mayer, M.},
  {Meintjes, P. J.}, {Meyer, M.}, {Mitchell, A. M. W.}, {Moderski, R.},
  {Mohamed, M.}, {Mohrmann, L.}, {Mor\aa{}, K.}, {Moulin, E.}, {Murach, T.},
  {de Naurois, M.}, {Niederwanger, F.}, {Niemiec, J.}, {Oakes, L.},
  {O\'{}Brien, P.}, {Odaka, H.}, {\"Ottl, S.}, {Ohm, S.}, {Ostrowski, M.},
  {Oya, I.}, {Padovani, M.}, {Panter, M.}, {Parsons, R. D.}, {Pekeur, N. W.},
  {Pelletier, G.}, {Perennes, C.}, {Petrucci, P.-O.}, {Peyaud, B.}, {Piel, Q.},
  {Pita, S.}, {Poon, H.}, {Prokhorov, D.}, {Prokoph, H.}, {P\"uhlhofer, G.},
  {Punch, M.}, {Quirrenbach, A.}, {Raab, S.}, {Reimer, A.}, {Reimer, O.},
  {Renaud, M.}, {de los Reyes, R.}, {Rieger, F.}, {Romoli, C.}, {Rosier-Lees,
  S.}, {Rowell, G.}, {Rudak, B.}, {Rulten, C. B.}, {Sahakian, V.}, {Salek, D.},
  {Sanchez, D. A.}, {Santangelo, A.}, {Sasaki, M.}, {Schlickeiser, R.},
  {Sch\"ussler, F.}, {Schulz, A.}, {Schwanke, U.}, {Schwemmer, S.}, {Settimo,
  M.}, {Seyffert, A. S.}, {Shafi, N.}, {Shilon, I.}, {Simoni, R.}, {Sol, H.},
  {Spanier, F.}, {Spengler, G.}, {Spies, F.}, {Stawarz, L.}, {Steenkamp, R.},
  {Stegmann, C.}, {Stinzing, F.}, {Stycz, K.}, {Sushch, I.}, {Tavernet, J.-P.},
  {Tavernier, T.}, {Taylor, A. M.}, {Terrier, R.}, {Tibaldo, L.}, {Tiziani,
  D.}, {Tluczykont, M.}, {Trichard, C.}, {Tuffs, R.}, {Uchiyama, Y.}, {van der
  Walt, D. J.}, {van Eldik, C.}, {van Rensburg, C.}, {van Soelen, B.},
  {Vasileiadis, G.}, {Veh, J.}, {Venter, C.}, {Viana, A.}, {Vincent, P.},
  {Vink, J.}, {Voisin, F.}, {V\"olk, H. J.}, {Vuillaume, T.}, {Wadiasingh, Z.},
  {Wagner, S. J.}, {Wagner, P.}, {Wagner, R.M.}, {White, R.}, {Wierzcholska,
  A.}, {Willmann, P.}, {W\"ornlein, A.}, {Wouters, D.}, {Yang, R.}, {Zabalza,
  V.}, {Zaborov, D.}, {Zacharias, M.}, {Zdziarski, A. A.}, {Zech, A.}, {Zefi,
  F.}, {Ziegler, A.}, {Zywucka, N.}, {From Fermi-LAT Collaboration}, \&
  {Katsuta, J.}}]{hess18a}
{H.E.S.S. Collaboration}, {Abdalla, H.}, {Abramowski, A.}, {et~al.} 2018, A\&A,
  612, A5, \dodoi{10.1051/0004-6361/201527843}

\bibitem[{{Higgs} {et~al.}(1977){Higgs}, {Landecker}, \& {Roger}}]{higgs77}
{Higgs}, L.~A., {Landecker}, T.~L., \& {Roger}, R.~S. 1977, \aj, 82, 718,
  \dodoi{10.1086/112114}

\bibitem[{{Hughes}(2000)}]{hughes00}
{Hughes}, J.~P. 2000, \apjl, 545, L53, \dodoi{10.1086/317337}

\bibitem[{{Hui} {et~al.}(2015){Hui}, {Seo}, {Lin}, {Huang}, {Hu}, {Wu},
  {Trepl}, {Takata}, {Wang}, {Chou}, {Cheng}, \& {Kong}}]{hui15}
{Hui}, C.~Y., {Seo}, K.~A., {Lin}, L.~C.~C., {et~al.} 2015, \apj, 799, 76,
  \dodoi{10.1088/0004-637X/799/1/76}

\bibitem[{{Hui} {et~al.}(2016){Hui}, {Yeung}, {Ng}, {Lin}, {Tam}, {Cheng},
  {Kong}, {Chernyshov}, \& {Dogiel}}]{hui16}
{Hui}, C.~Y., {Yeung}, P.~K.~H., {Ng}, C.~W., {et~al.} 2016, \mnras, 457, 4262,
  \dodoi{10.1093/mnras/stw209}

\bibitem[{{Hwang} {et~al.}(2002){Hwang}, {Decourchelle}, {Holt}, \&
  {Petre}}]{hwang02}
{Hwang}, U., {Decourchelle}, A., {Holt}, S.~S., \& {Petre}, R. 2002, \apj, 581,
  1101, \dodoi{10.1086/344366}

\bibitem[{{Inoue} {et~al.}(2021){Inoue}, {Marcowith}, {Giacinti}, {van Marle},
  \& {Nishino}}]{inoue21}
{Inoue}, T., {Marcowith}, A., {Giacinti}, G., {van Marle}, A.~J., \& {Nishino},
  S. 2021, arXiv e-prints, arXiv:2108.13433.
\newblock \doarXiv{2108.13433}

\bibitem[{{Jogler} \& {Funk}(2016)}]{jogler16}
{Jogler}, T., \& {Funk}, S. 2016, \apj, 816, 100,
  \dodoi{10.3847/0004-637X/816/2/100}

\bibitem[{{Katagiri} {et~al.}(2016{\natexlab{a}}){Katagiri}, {Yoshida},
  {Ballet}, {Grondin}, {Hanabata}, {Hewitt}, {Kubo}, \&
  {Lemoine-Goumard}}]{katagiri16b}
{Katagiri}, H., {Yoshida}, K., {Ballet}, J., {et~al.} 2016{\natexlab{a}}, \apj,
  818, 114, \dodoi{10.3847/0004-637X/818/2/114}

\bibitem[{{Katagiri} {et~al.}(2011){Katagiri}, {Tibaldo}, {Ballet}, {Giordano},
  {Grenier}, {Porter}, {Roth}, {Tibolla}, {Uchiyama}, \&
  {Yamazaki}}]{katagiri11}
{Katagiri}, H., {Tibaldo}, L., {Ballet}, J., {et~al.} 2011, \apj, 741, 44,
  \dodoi{10.1088/0004-637X/741/1/44}

\bibitem[{{Katagiri} {et~al.}(2016{\natexlab{b}}){Katagiri}, {Sugiyama},
  {Ackermann}, {Ballet}, {Casandjian}, {Hanabata}, {Hewitt}, {Kerr}, {Kubo},
  {Lemoine-Goumard}, \& {Ray}}]{katagiri16a}
{Katagiri}, H., {Sugiyama}, S., {Ackermann}, M., {et~al.} 2016{\natexlab{b}},
  \apj, 831, 106, \dodoi{10.3847/0004-637X/831/1/106}

\bibitem[{{Katsuda} {et~al.}(2008){Katsuda}, {Tsunemi}, \& {Mori}}]{katsuda08}
{Katsuda}, S., {Tsunemi}, H., \& {Mori}, K. 2008, \apjl, 678, L35,
  \dodoi{10.1086/588499}

\bibitem[{{Katsuda} {et~al.}(2015){Katsuda}, {Acero}, {Tominaga}, {Fukui},
  {Hiraga}, {Koyama}, {Lee}, {Mori}, {Nagataki}, {Ohira}, {Petre}, {Sano},
  {Takeuchi}, {Tamagawa}, {Tsuji}, {Tsunemi}, \& {Uchiyama}}]{katsuda15}
{Katsuda}, S., {Acero}, F., {Tominaga}, N., {et~al.} 2015, \apj, 814, 29,
  \dodoi{10.1088/0004-637X/814/1/29}

\bibitem[{{Katsuta} {et~al.}(2012){Katsuta}, {Uchiyama}, {Tanaka}, {Tajima},
  {Bechtol}, {Funk}, {Lande}, {Ballet}, {Hanabata}, {Lemoine-Goumard}, \&
  {Takahashi}}]{katsuta12}
{Katsuta}, J., {Uchiyama}, Y., {Tanaka}, T., {et~al.} 2012, \apj, 752, 135,
  \dodoi{10.1088/0004-637X/752/2/135}

\bibitem[{{Koo} {et~al.}(1995){Koo}, {Kim}, \& {Seward}}]{koo95}
{Koo}, B.-C., {Kim}, K.-T., \& {Seward}, F.~D. 1995, \apj, 447, 211,
  \dodoi{10.1086/175867}

\bibitem[{{Koo} \& {Moon}(1997{\natexlab{a}})}]{koo97a}
{Koo}, B.-C., \& {Moon}, D.-S. 1997{\natexlab{a}}, \apj, 475, 194,
  \dodoi{10.1086/303527}

\bibitem[{{Koo} \& {Moon}(1997{\natexlab{b}})}]{koo97b}
---. 1997{\natexlab{b}}, \apj, 485, 263, \dodoi{10.1086/304391}

\bibitem[{{Lagage} \& {Cesarsky}(1983)}]{lagage83}
{Lagage}, P.~O., \& {Cesarsky}, C.~J. 1983, \aap, 125, 249

\bibitem[{{Lazendic} \& {Slane}(2006)}]{lazendic06}
{Lazendic}, J.~S., \& {Slane}, P.~O. 2006, \apj, 647, 350,
  \dodoi{10.1086/505380}

\bibitem[{{Leahy}(1989)}]{leahy89}
{Leahy}, D.~A. 1989, \aap, 216, 193

\bibitem[{{Leahy} \& {Aschenbach}(1995)}]{leahy95}
{Leahy}, D.~A., \& {Aschenbach}, B. 1995, \aap, 293, 853

\bibitem[{{Leahy} {et~al.}(2013){Leahy}, {Green}, \& {Ranasinghe}}]{leahy13}
{Leahy}, D.~A., {Green}, K., \& {Ranasinghe}, S. 2013, \mnras, 436, 968,
  \dodoi{10.1093/mnras/stt1596}

\bibitem[{{Leahy} {et~al.}(1986){Leahy}, {Naranan}, \& {Singh}}]{leahy86}
{Leahy}, D.~A., {Naranan}, S., \& {Singh}, K.~P. 1986, \mnras, 220, 501,
  \dodoi{10.1093/mnras/220.3.501}

\bibitem[{{Leahy} \& {Tian}(2008)}]{leahy08}
{Leahy}, D.~A., \& {Tian}, W.~W. 2008, \aj, 135, 167,
  \dodoi{10.1088/0004-6256/135/1/167}

\bibitem[{{Lemoine-Goumard} {et~al.}(2012){Lemoine-Goumard}, {Renaud}, {Vink},
  {Allen}, {Bamba}, {Giordano}, \& {Uchiyama}}]{lemoine12}
{Lemoine-Goumard}, M., {Renaud}, M., {Vink}, J., {et~al.} 2012, \aap, 545, A28,
  \dodoi{10.1051/0004-6361/201219896}

\bibitem[{{Li} \& {Chen}(2010)}]{li10}
{Li}, H., \& {Chen}, Y. 2010, \mnras, 409, L35,
  \dodoi{10.1111/j.1745-3933.2010.00944.x}

\bibitem[{{Lozinskaya} {et~al.}(2000){Lozinskaya}, {Pravdikova}, \&
  {Finoguenov}}]{lozinskaya00}
{Lozinskaya}, T.~A., {Pravdikova}, V.~V., \& {Finoguenov}, A.~V. 2000,
  Astronomy Letters, 26, 77, \dodoi{10.1134/1.20371}

\bibitem[{{Lozinskaya} {et~al.}(1993){Lozinskaya}, {Sitnik}, \&
  {Pravdikova}}]{lozinskaya93}
{Lozinskaya}, T.~A., {Sitnik}, T.~G., \& {Pravdikova}, V.~V. 1993, Astronomy
  Reports, 37, 240

\bibitem[{{MAGIC Collaboration} {et~al.}(2019){MAGIC Collaboration}, {Acciari},
  {Ansoldi}, {Antonelli}, {Arbet Engels}, {Arcaro}, {Baack}, {Babi{\'c}}, {},
  {Banerjee}, {Bangale}, {de Almeida}, {Barrio}, {Becerra Gonz{\'a}lez},
  {Bednarek}, {Bernardini}, {Berti}, {Besenrieder}, {Bhattacharyya},
  {Bigongiari}, {Biland}, {Blanch}, {Bonnoli}, {Carosi}, {Ceribella},
  {Chatterjee}, {Colak}, {Colin}, {Colombo}, {Contreras}, {Cortina}, {Covino},
  {Cumani}, {D'Elia}, {da Vela}, {Dazzi}, {de Angelis}, {de Lotto}, {Delfino},
  {Delgado}, {di Pierro}, {Dom{\'\i}nguez}, {Dominis Prester}, {Dorner},
  {Doro}, {Einecke}, {Elsaesser}, {Fallah Ramazani}, {Fattorini},
  {Fern{\'a}ndez-Barral}, {Ferrara}, {Fidalgo}, {Foffano}, {Fonseca}, {Font},
  {Fruck}, {Galindo}, {Gallozzi}, {Garc{\'\i}a L{\'o}pez}, {Garczarczyk},
  {Gaug}, {Giammaria}, {Godinovi{\'c}}, {}, {Guberman}, {Hadasch}, {Hahn},
  {Hassan}, {Herrera}, {Hoang}, {Hrupec}, {Inoue}, {Ishio}, {Iwamura}, {Kubo},
  {Kushida}, {Kuve{\v{z}}di{\'c}}, {}, {Lamastra}, {Lelas}, {Leone},
  {Lindfors}, {Lombardi}, {Longo}, {L{\'o}pez}, {L{\'o}pez-Oramas}, {Maggio},
  {Majumdar}, {Makariev}, {Maneva}, {Manganaro}, {Mannheim}, {Maraschi},
  {Mariotti}, {Mart{\'\i}nez}, {Masuda}, {Mazin}, {Minev}, {Miranda},
  {Mirzoyan}, {Molina}, {Moralejo}, {Moreno}, {Moretti}, {Neustroev},
  {Niedzwiecki}, {Nievas Rosillo}, {Nigro}, {Nilsson}, {Ninci}, {Nishijima},
  {Noda}, {Nogu{\'e}s}, {Paiano}, {Palacio}, {Paneque}, {Paoletti}, {Paredes},
  {Pedaletti}, {Pe{\~n}il}, {Peresano}, {Persic}, {Prada Moroni}, {Prand ini},
  {Puljak}, {Garcia}, {Rhode}, {Rib{\'o}}, {Rico}, {Righi}, {Rugliancich},
  {Saha}, {Saito}, {Satalecka}, {Schweizer}, {Sitarek}, {{\v{S}}nidari{\'c}},
  {}, {Sobczynska}, {Somero}, {Stamerra}, {Strzys}, {Suri{\'c}}, {},
  {Tavecchio}, {Temnikov}, {Terzi{\'c}}, {}, {Teshima}, {Torres-Alb{\`a}},
  {Tsujimoto}, {Vanzo}, {Vazquez Acosta}, {Vovk}, {Ward}, {Will}, {Zari{\'c}},
  {}, {de O{\~n}a Wilhelmi}, {Torres}, \& {Zanin}}]{magic19}
{MAGIC Collaboration}, {Acciari}, V.~A., {Ansoldi}, S., {et~al.} 2019, \mnras,
  483, 4578, \dodoi{10.1093/mnras/sty3387}

\bibitem[{{Malkov} \& {Voelk}(1995)}]{malkov95}
{Malkov}, M.~A., \& {Voelk}, H.~J. 1995, \aap, 300, 605

\bibitem[{{Marcowith} {et~al.}(2018){Marcowith}, {Dwarkadas}, {Renaud},
  {Tatischeff}, \& {Giacinti}}]{marcowith18}
{Marcowith}, A., {Dwarkadas}, V.~V., {Renaud}, M., {Tatischeff}, V., \&
  {Giacinti}, G. 2018, \mnras, 479, 4470, \dodoi{10.1093/mnras/sty1743}

\bibitem[{{Matsumura}(2018)}]{matsumura18}
{Matsumura}, H. 2018, PhD thesis, Kyoto University, Graduate School of Science

\bibitem[{{Matsumura} {et~al.}(2017){Matsumura}, {Uchida}, {Tanaka}, {Tsuru},
  {Nobukawa}, {Nobukawa}, \& {Itou}}]{matsumura17}
{Matsumura}, H., {Uchida}, H., {Tanaka}, T., {et~al.} 2017, \pasj, 69, 30,
  \dodoi{10.1093/pasj/psx001}

\bibitem[{{Mattox} {et~al.}(1996){Mattox}, {Bertsch}, {Chiang}, {Dingus},
  {Digel}, {Esposito}, {Fierro}, {Hartman}, {Hunter}, {Kanbach}, {Kniffen},
  {Lin}, {Macomb}, {Mayer-Hasselwander}, {Michelson}, {von Montigny},
  {Mukherjee}, {Nolan}, {Ramanamurthy}, {Schneid}, {Sreekumar}, {Thompson}, \&
  {Willis}}]{mattox96}
{Mattox}, J.~R., {Bertsch}, D.~L., {Chiang}, J., {et~al.} 1996, \apj, 461, 396,
  \dodoi{10.1086/177068}

\bibitem[{{Misanovic} {et~al.}(2011){Misanovic}, {Kargaltsev}, \&
  {Pavlov}}]{misanovic11}
{Misanovic}, Z., {Kargaltsev}, O., \& {Pavlov}, G.~G. 2011, \apj, 735, 33,
  \dodoi{10.1088/0004-637X/735/1/33}

\bibitem[{{Miyata} {et~al.}(1994){Miyata}, {Tsunemi}, {Pisarski}, \&
  {Kissel}}]{miyata94}
{Miyata}, E., {Tsunemi}, H., {Pisarski}, R., \& {Kissel}, S.~E. 1994, \pasj,
  46, L101

\bibitem[{{Moffett} {et~al.}(2001){Moffett}, {Gaensler}, \&
  {Green}}]{moffett01}
{Moffett}, D., {Gaensler}, B., \& {Green}, A. 2001, in American Institute of
  Physics Conference Series, Vol. 565, Young Supernova Remnants, ed. S.~S.
  {Holt} \& U.~{Hwang}, 333--336, \dodoi{10.1063/1.1377115}

\bibitem[{{Moffett} {et~al.}(2002){Moffett}, {Gaensler}, {Green}, {Slane},
  {Harrus}, \& {Dodson}}]{moffett02}
{Moffett}, D., {Gaensler}, B., {Green}, A., {et~al.} 2002, in Astronomical
  Society of the Pacific Conference Series, Vol. 271, Neutron Stars in
  Supernova Remnants, ed. P.~O. {Slane} \& B.~M. {Gaensler}, 221

\bibitem[{{Moffett} \& {Reynolds}(1994)}]{moffett94}
{Moffett}, D.~A., \& {Reynolds}, S.~P. 1994, \apj, 437, 705,
  \dodoi{10.1086/175033}

\bibitem[{{Murray} {et~al.}(1979){Murray}, {Fabbiano}, {Fabian}, {Epstein}, \&
  {Giacconi}}]{murray79}
{Murray}, S.~S., {Fabbiano}, G., {Fabian}, A.~C., {Epstein}, A., \& {Giacconi},
  R. 1979, \apjl, 234, L69, \dodoi{10.1086/183111}

\bibitem[{{Nava} {et~al.}(2016){Nava}, {Gabici}, {Marcowith}, {Morlino}, \&
  {Ptuskin}}]{nava16}
{Nava}, L., {Gabici}, S., {Marcowith}, A., {Morlino}, G., \& {Ptuskin}, V.~S.
  2016, \mnras, 461, 3552, \dodoi{10.1093/mnras/stw1592}

\bibitem[{{Ohira} {et~al.}(2018){Ohira}, {Kisaka}, \& {Yamazaki}}]{ohira18}
{Ohira}, Y., {Kisaka}, S., \& {Yamazaki}, R. 2018, \mnras, 478, 926,
  \dodoi{10.1093/mnras/sty1159}

\bibitem[{{Ohira} {et~al.}(2010){Ohira}, {Murase}, \& {Yamazaki}}]{ohira10}
{Ohira}, Y., {Murase}, K., \& {Yamazaki}, R. 2010, \aap, 513, A17,
  \dodoi{10.1051/0004-6361/200913495}

\bibitem[{{Ohira} {et~al.}(2011){Ohira}, {Murase}, \& {Yamazaki}}]{ohira11a}
---. 2011, \mnras, 410, 1577, \dodoi{10.1111/j.1365-2966.2010.17539.x}

\bibitem[{{Ohira} {et~al.}(2012){Ohira}, {Yamazaki}, {Kawanaka}, \&
  {Ioka}}]{ohira12}
{Ohira}, Y., {Yamazaki}, R., {Kawanaka}, N., \& {Ioka}, K. 2012, \mnras, 427,
  91, \dodoi{10.1111/j.1365-2966.2012.21908.x}

\bibitem[{{Ohnishi} {et~al.}(2011){Ohnishi}, {Koyama}, {Tsuru}, {Masai},
  {Yamaguchi}, \& {Ozawa}}]{ohnishi11}
{Ohnishi}, T., {Koyama}, K., {Tsuru}, T.~G., {et~al.} 2011, \pasj, 63, 527,
  \dodoi{10.1093/pasj/63.3.527}

\bibitem[{{Okon} {et~al.}(2018){Okon}, {Uchida}, {Tanaka}, {Matsumura}, \&
  {Tsuru}}]{okon18}
{Okon}, H., {Uchida}, H., {Tanaka}, T., {Matsumura}, H., \& {Tsuru}, T.~G.
  2018, \pasj, 70, 35, \dodoi{10.1093/pasj/psy022}

\bibitem[{{Patnaude} \& {Fesen}(2009)}]{patnaude09}
{Patnaude}, D.~J., \& {Fesen}, R.~A. 2009, \apj, 697, 535,
  \dodoi{10.1088/0004-637X/697/1/535}

\bibitem[{{Pavlovi{\'c}} {et~al.}(2013){Pavlovi{\'c}}, {Uro{\v{s}}evi{\'c}},
  {Vukoti{\'c}}, {Arbutina}, \& {G{\"o}ker}}]{pavlovic13}
{Pavlovi{\'c}}, M.~Z., {Uro{\v{s}}evi{\'c}}, D., {Vukoti{\'c}}, B., {Arbutina},
  B., \& {G{\"o}ker}, {\"U}.~D. 2013, \apjs, 204, 4,
  \dodoi{10.1088/0067-0049/204/1/4}

\bibitem[{{Petre} {et~al.}(1982){Petre}, {Kriss}, {Winkler}, \&
  {Canizares}}]{petre82}
{Petre}, R., {Kriss}, G.~A., {Winkler}, P.~F., \& {Canizares}, C.~R. 1982,
  \apj, 258, 22, \dodoi{10.1086/160045}

\bibitem[{{Ptuskin} {et~al.}(2006){Ptuskin}, {Moskalenko}, {Jones}, {Strong},
  \& {Zirakashvili}}]{ptuskin06}
{Ptuskin}, V.~S., {Moskalenko}, I.~V., {Jones}, F.~C., {Strong}, A.~W., \&
  {Zirakashvili}, V.~N. 2006, \apj, 642, 902, \dodoi{10.1086/501117}

\bibitem[{{Ptuskin} \& {Zirakashvili}(2003)}]{ptuskin03}
{Ptuskin}, V.~S., \& {Zirakashvili}, V.~N. 2003, \aap, 403, 1,
  \dodoi{10.1051/0004-6361:20030323}

\bibitem[{{Ptuskin} \& {Zirakashvili}(2005)}]{ptuskin05}
---. 2005, \aap, 429, 755, \dodoi{10.1051/0004-6361:20041517}

\bibitem[{{Ptuskin} {et~al.}(2008){Ptuskin}, {Zirakashvili}, \&
  {Plesser}}]{ptuskin08}
{Ptuskin}, V.~S., {Zirakashvili}, V.~N., \& {Plesser}, A.~A. 2008, Advances in
  Space Research, 42, 486, \dodoi{10.1016/j.asr.2007.12.007}

\bibitem[{{Radhakrishnan} {et~al.}(1972){Radhakrishnan}, {Goss}, {Murray}, \&
  {Brooks}}]{radhak72}
{Radhakrishnan}, V., {Goss}, W.~M., {Murray}, J.~D., \& {Brooks}, J.~W. 1972,
  \apjs, 24, 49, \dodoi{10.1086/190249}

\bibitem[{{Rappaport} {et~al.}(1974){Rappaport}, {Doxsey}, {Solinger}, \&
  {Borken}}]{rappaport74}
{Rappaport}, S., {Doxsey}, R., {Solinger}, A., \& {Borken}, R. 1974, \apj, 194,
  329, \dodoi{10.1086/153249}

\bibitem[{{Recchia} {et~al.}(2021){Recchia}, {Galli}, {Nava}, {Padovani},
  {Gabici}, {Marcowith}, {Ptuskin}, \& {Morlino}}]{recchia21}
{Recchia}, S., {Galli}, D., {Nava}, L., {et~al.} 2021, arXiv e-prints,
  arXiv:2106.04948.
\newblock \doarXiv{2106.04948}

\bibitem[{{Reed} {et~al.}(1995){Reed}, {Hester}, {Fabian}, \&
  {Winkler}}]{reed95}
{Reed}, J.~E., {Hester}, J.~J., {Fabian}, A.~C., \& {Winkler}, P.~F. 1995,
  \apj, 440, 706, \dodoi{10.1086/175308}

\bibitem[{{Reynoso} {et~al.}(2003){Reynoso}, {Green}, {Johnston}, {Dubner},
  {Giacani}, \& {Goss}}]{reynoso03}
{Reynoso}, E.~M., {Green}, A.~J., {Johnston}, S., {et~al.} 2003, \mnras, 345,
  671, \dodoi{10.1046/j.1365-8711.2003.06978.x}

\bibitem[{{Rosado} {et~al.}(1996){Rosado}, {Ambrocio-Cruz}, {Le Coarer}, \&
  {Marcelin}}]{rosado96}
{Rosado}, M., {Ambrocio-Cruz}, P., {Le Coarer}, E., \& {Marcelin}, M. 1996,
  \aap, 315, 243

\bibitem[{{Routledge} {et~al.}(1991){Routledge}, {Dewdney}, {Landecker}, \&
  {Vaneldik}}]{routledge91}
{Routledge}, D., {Dewdney}, P.~E., {Landecker}, T.~L., \& {Vaneldik}, J.~F.
  1991, \aap, 247, 529

\bibitem[{{Sano} {et~al.}(2021{\natexlab{a}}){Sano}, {Suzuki}, {Nobukawa},
  {Filipovi{\'c}}, {Fukui}, \& {Moriya}}]{sano21b}
{Sano}, H., {Suzuki}, H., {Nobukawa}, K.~K., {et~al.} 2021{\natexlab{a}}, arXiv
  e-prints, arXiv:2108.03392.
\newblock \doarXiv{2108.03392}

\bibitem[{{Sano} {et~al.}(2021{\natexlab{b}}){Sano}, {Yoshiike}, {Yamane},
  {Hayashi}, {Enokiya}, {Tokuda}, {Tachihara}, {Rowell}, {Filipovic}, \&
  {Fukui}}]{sano21a}
{Sano}, H., {Yoshiike}, S., {Yamane}, Y., {et~al.} 2021{\natexlab{b}}, arXiv
  e-prints, arXiv:2106.12009.
\newblock \doarXiv{2106.12009}

\bibitem[{{Sarma} {et~al.}(1997){Sarma}, {Goss}, {Green}, \& {Frail}}]{sarma97}
{Sarma}, A.~P., {Goss}, W.~M., {Green}, A.~J., \& {Frail}, D.~A. 1997, \apj,
  483, 335, \dodoi{10.1086/304246}

\bibitem[{{Sasaki} {et~al.}(2014){Sasaki}, {Heinitz}, {Warth}, \&
  {P{\"u}hlhofer}}]{sasaki14}
{Sasaki}, M., {Heinitz}, C., {Warth}, G., \& {P{\"u}hlhofer}, G. 2014, \aap,
  563, A9, \dodoi{10.1051/0004-6361/201323145}

\bibitem[{{Sasaki} {et~al.}(2013){Sasaki}, {Plucinsky}, {Gaetz}, \&
  {Bocchino}}]{sasaki13}
{Sasaki}, M., {Plucinsky}, P.~P., {Gaetz}, T.~J., \& {Bocchino}, F. 2013, \aap,
  552, A45, \dodoi{10.1051/0004-6361/201220836}

\bibitem[{{Sato} {et~al.}(2014){Sato}, {Koyama}, {Takahashi}, {Odaka}, \&
  {Nakashima}}]{sato14}
{Sato}, T., {Koyama}, K., {Takahashi}, T., {Odaka}, H., \& {Nakashima}, S.
  2014, \pasj, 66, 124, \dodoi{10.1093/pasj/psu120}

\bibitem[{{Schure} \& {Bell}(2013)}]{schure13}
{Schure}, K.~M., \& {Bell}, A.~R. 2013, \mnras, 435, 1174,
  \dodoi{10.1093/mnras/stt1371}

\bibitem[{{Sezer} {et~al.}(2019){Sezer}, {Ergin}, {Yamazaki}, {Sano}, \&
  {Fukui}}]{sezer19}
{Sezer}, A., {Ergin}, T., {Yamazaki}, R., {Sano}, H., \& {Fukui}, Y. 2019,
  \mnras, 489, 4300, \dodoi{10.1093/mnras/stz2461}

\bibitem[{{Slane} {et~al.}(2002){Slane}, {Chen}, {Lazendic}, \&
  {Hughes}}]{slane02}
{Slane}, P., {Chen}, Y., {Lazendic}, J.~S., \& {Hughes}, J.~P. 2002, \apj, 580,
  904, \dodoi{10.1086/343891}

\bibitem[{{Slane} {et~al.}(2001){Slane}, {Hughes}, {Edgar}, {Plucinsky},
  {Miyata}, {Tsunemi}, \& {Aschenbach}}]{slane01}
{Slane}, P., {Hughes}, J.~P., {Edgar}, R.~J., {et~al.} 2001, \apj, 548, 814,
  \dodoi{10.1086/319033}

\bibitem[{{Slane} {et~al.}(2012){Slane}, {Hughes}, {Temim}, {Rousseau},
  {Castro}, {Foight}, {Gaensler}, {Funk}, {Lemoine-Goumard}, {Gelfand},
  {Moffett}, {Dodson}, \& {Bernstein}}]{slane12}
{Slane}, P., {Hughes}, J.~P., {Temim}, T., {et~al.} 2012, \apj, 749, 131,
  \dodoi{10.1088/0004-637X/749/2/131}

\bibitem[{{Slavin} {et~al.}(2017){Slavin}, {Smith}, {Foster}, {Winter},
  {Raymond}, {Slane}, \& {Yamaguchi}}]{slavin17}
{Slavin}, J.~D., {Smith}, R.~K., {Foster}, A., {et~al.} 2017, \apj, 846, 77,
  \dodoi{10.3847/1538-4357/aa8552}

\bibitem[{{Sollerman} {et~al.}(2003){Sollerman}, {Ghavamian}, {Lundqvist}, \&
  {Smith}}]{sollerman03}
{Sollerman}, J., {Ghavamian}, P., {Lundqvist}, P., \& {Smith}, R.~C. 2003,
  \aap, 407, 249, \dodoi{10.1051/0004-6361:20030839}

\bibitem[{{Strong} \& {Moskalenko}(1998)}]{strong98}
{Strong}, A.~W., \& {Moskalenko}, I.~V. 1998, \apj, 509, 212,
  \dodoi{10.1086/306470}

\bibitem[{{Strong} {et~al.}(2000){Strong}, {Moskalenko}, \&
  {Reimer}}]{strong00}
{Strong}, A.~W., {Moskalenko}, I.~V., \& {Reimer}, O. 2000, \apj, 537, 763,
  \dodoi{10.1086/309038}

\bibitem[{{Sun} {et~al.}(2004){Sun}, {Seward}, {Smith}, \& {Slane}}]{sun04}
{Sun}, M., {Seward}, F.~D., {Smith}, R.~K., \& {Slane}, P.~O. 2004, \apj, 605,
  742, \dodoi{10.1086/382666}

\bibitem[{{Suzuki} {et~al.}(2020{\natexlab{a}}){Suzuki}, {Bamba}, {Enokiya},
  {Yamaguchi}, {Plucinsky}, \& {Odaka}}]{suzuki20a}
{Suzuki}, H., {Bamba}, A., {Enokiya}, R., {et~al.} 2020{\natexlab{a}}, \apj,
  893, 147, \dodoi{10.3847/1538-4357/ab80ba}

\bibitem[{{Suzuki} {et~al.}(2018){Suzuki}, {Bamba}, {Nakazawa}, {Furuta},
  {Sawada}, {Yamazaki}, \& {Koyama}}]{suzuki18}
{Suzuki}, H., {Bamba}, A., {Nakazawa}, K., {et~al.} 2018, \pasj, 70, 75,
  \dodoi{10.1093/pasj/psy069}

\bibitem[{{Suzuki} {et~al.}(2021){Suzuki}, {Bamba}, \& {Shibata}}]{suzuki21a}
{Suzuki}, H., {Bamba}, A., \& {Shibata}, S. 2021, \apj, 914, 103,
  \dodoi{10.3847/1538-4357/abfb02}

\bibitem[{{Suzuki} {et~al.}(2020{\natexlab{b}}){Suzuki}, {Bamba}, {Yamazaki},
  \& {Ohira}}]{suzuki20b}
{Suzuki}, H., {Bamba}, A., {Yamazaki}, R., \& {Ohira}, Y. 2020{\natexlab{b}},
  \pasj, \dodoi{10.1093/pasj/psaa061}

\bibitem[{{Tanaka} {et~al.}(2011){Tanaka}, {Allafort}, {Ballet}, {Funk},
  {Giordano}, {Hewitt}, {Lemoine-Goumard}, {Tajima}, {Tibolla}, \&
  {Uchiyama}}]{tanaka11}
{Tanaka}, T., {Allafort}, A., {Ballet}, J., {et~al.} 2011, \apjl, 740, L51,
  \dodoi{10.1088/2041-8205/740/2/L51}

\bibitem[{{Temim} {et~al.}(2013){Temim}, {Slane}, {Castro}, {Plucinsky},
  {Gelfand}, \& {Dickel}}]{temim13}
{Temim}, T., {Slane}, P., {Castro}, D., {et~al.} 2013, \apj, 768, 61,
  \dodoi{10.1088/0004-637X/768/1/61}

\bibitem[{{Tian} \& {Leahy}(2011)}]{tian11}
{Tian}, W.~W., \& {Leahy}, D.~A. 2011, \apjl, 729, L15,
  \dodoi{10.1088/2041-8205/729/2/L15}

\bibitem[{{Tian} \& {Leahy}(2012)}]{tian12}
---. 2012, \mnras, 421, 2593, \dodoi{10.1111/j.1365-2966.2012.20491.x}

\bibitem[{{Tian} \& {Leahy}(2014)}]{tian14}
---. 2014, \apjl, 783, L2, \dodoi{10.1088/2041-8205/783/1/L2}

\bibitem[{{Tian} {et~al.}(2007){Tian}, {Li}, {Leahy}, \& {Wang}}]{tian07}
{Tian}, W.~W., {Li}, Z., {Leahy}, D.~A., \& {Wang}, Q.~D. 2007, \apjl, 657,
  L25, \dodoi{10.1086/512544}

\bibitem[{{Troja} {et~al.}(2008){Troja}, {Bocchino}, {Miceli}, \&
  {Reale}}]{troja08}
{Troja}, E., {Bocchino}, F., {Miceli}, M., \& {Reale}, F. 2008, \aap, 485, 777,
  \dodoi{10.1051/0004-6361:20079123}

\bibitem[{{Tsuji} \& {Uchiyama}(2016)}]{tsuji16}
{Tsuji}, N., \& {Uchiyama}, Y. 2016, \pasj, 68, 108,
  \dodoi{10.1093/pasj/psw102}

\bibitem[{{Tsuji} {et~al.}(2021){Tsuji}, {Uchiyama}, {Khangulyan}, \&
  {Aharonian}}]{tsuji21}
{Tsuji}, N., {Uchiyama}, Y., {Khangulyan}, D., \& {Aharonian}, F. 2021, \apj,
  907, 117, \dodoi{10.3847/1538-4357/abce65}

\bibitem[{{Uchida} {et~al.}(2012){Uchida}, {Koyama}, {Yamaguchi}, {Sawada},
  {Ohnishi}, {Tsuru}, {Tanaka}, {Yoshiike}, \& {Fukui}}]{uchida12}
{Uchida}, H., {Koyama}, K., {Yamaguchi}, H., {et~al.} 2012, \pasj, 64, 141,
  \dodoi{10.1093/pasj/64.6.141}

\bibitem[{{Wang} {et~al.}(2020){Wang}, {Zhang}, {Jiang}, {Zhao}, {Chen},
  {Chen}, {Gao}, \& {Liu}}]{wang20}
{Wang}, S., {Zhang}, C., {Jiang}, B., {et~al.} 2020, \aap, 639, A72,
  \dodoi{10.1051/0004-6361/201936868}

\bibitem[{{Washino} {et~al.}(2016){Washino}, {Uchida}, {Nobukawa}, {Tsuru},
  {Tanaka}, {Nobukawa}, \& {Koyama}}]{washino16}
{Washino}, R., {Uchida}, H., {Nobukawa}, M., {et~al.} 2016, \pasj, 68, S4,
  \dodoi{10.1093/pasj/psv095}

\bibitem[{{White} \& {Long}(1991)}]{white91}
{White}, R.~L., \& {Long}, K.~S. 1991, \apj, 373, 543, \dodoi{10.1086/170073}

\bibitem[{{Williams} {et~al.}(2011){Williams}, {Blair}, {Blondin}, {Borkowski},
  {Ghavamian}, {Long}, {Raymond}, {Reynolds}, {Rho}, \& {Winkler}}]{williams11}
{Williams}, B.~J., {Blair}, W.~P., {Blondin}, J.~M., {et~al.} 2011, \apj, 741,
  96, \dodoi{10.1088/0004-637X/741/2/96}

\bibitem[{{Winkler} {et~al.}(2003){Winkler}, {Gupta}, \& {Long}}]{winkler03}
{Winkler}, P.~F., {Gupta}, G., \& {Long}, K.~S. 2003, \apj, 585, 324,
  \dodoi{10.1086/345985}

\bibitem[{{Winkler} {et~al.}(1988){Winkler}, {Tuttle}, {Kirshner}, \&
  {Irwin}}]{winkler88}
{Winkler}, P.~F., {Tuttle}, J.~H., {Kirshner}, R.~P., \& {Irwin}, M.~J. 1988,
  in IAU Colloq. 101: Supernova Remnants and the Interstellar Medium, ed. R.~S.
  {Roger} \& T.~L. {Landecker}, 65

\bibitem[{{Winkler} {et~al.}(2014){Winkler}, {Williams}, {Reynolds}, {Petre},
  {Long}, {Katsuda}, \& {Hwang}}]{winkler14}
{Winkler}, P.~F., {Williams}, B.~J., {Reynolds}, S.~P., {et~al.} 2014, \apj,
  781, 65, \dodoi{10.1088/0004-637X/781/2/65}

\bibitem[{{Wolfire} {et~al.}(1995){Wolfire}, {Hollenbach}, {McKee}, {Tielens},
  \& {Bakes}}]{wolfire95}
{Wolfire}, M.~G., {Hollenbach}, D., {McKee}, C.~F., {Tielens}, A.~G.~G.~M., \&
  {Bakes}, E.~L.~O. 1995, \apj, 443, 152, \dodoi{10.1086/175510}

\bibitem[{{Xin} {et~al.}(2016){Xin}, {Liang}, {Li}, {Yuan}, {Liu}, \&
  {Wei}}]{xin16}
{Xin}, Y.-L., {Liang}, Y.-F., {Li}, X., {et~al.} 2016, \apj, 817, 64,
  \dodoi{10.3847/0004-637X/817/1/64}

\bibitem[{{Xing} {et~al.}(2014){Xing}, {Wang}, {Zhang}, \& {Chen}}]{xing14}
{Xing}, Y., {Wang}, Z., {Zhang}, X., \& {Chen}, Y. 2014, \apj, 781, 64,
  \dodoi{10.1088/0004-637X/781/2/64}

\bibitem[{{Yamaguchi} {et~al.}(2008){Yamaguchi}, {Koyama}, {Katsuda},
  {Nakajima}, {Hughes}, {Bamba}, {Hiraga}, {Mori}, {Ozaki}, \&
  {Tsuru}}]{yamaguchi08}
{Yamaguchi}, H., {Koyama}, K., {Katsuda}, S., {et~al.} 2008, \pasj, 60, S141,
  \dodoi{10.1093/pasj/60.sp1.S141}

\bibitem[{{Yamauchi} {et~al.}(2014){Yamauchi}, {Minami}, {Ota}, \&
  {Koyama}}]{yamauchi14}
{Yamauchi}, S., {Minami}, S., {Ota}, N., \& {Koyama}, K. 2014, \pasj, 66, 2,
  \dodoi{10.1093/pasj/pst004}

\bibitem[{{Yasuda} \& {Lee}(2019)}]{yasuda19}
{Yasuda}, H., \& {Lee}, S.-H. 2019, \apj, 876, 27,
  \dodoi{10.3847/1538-4357/ab13ab}

\bibitem[{{Yuan} {et~al.}(2014){Yuan}, {Huang}, {Liu}, \& {Zhang}}]{yuan14}
{Yuan}, Q., {Huang}, X., {Liu}, S., \& {Zhang}, B. 2014, \apjl, 785, L22,
  \dodoi{10.1088/2041-8205/785/2/L22}

\bibitem[{{Zdziarski} {et~al.}(2016){Zdziarski}, {Malyshev}, {de O{\~n}a
  Wilhelmi}, {Pedaletti}, {Yang}, {Chernyakova}, {L{\'o}pez-Caniego},
  {Miko{\l}ajewska}, \& {Basak}}]{zdziarski16}
{Zdziarski}, A.~A., {Malyshev}, D., {de O{\~n}a Wilhelmi}, E., {et~al.} 2016,
  \mnras, 455, 1451, \dodoi{10.1093/mnras/stv2167}

\bibitem[{{Zeng} {et~al.}(2019){Zeng}, {Xin}, \& {Liu}}]{zeng19}
{Zeng}, H., {Xin}, Y., \& {Liu}, S. 2019, \apj, 874, 50,
  \dodoi{10.3847/1538-4357/aaf392}

\bibitem[{{Zhang} {et~al.}(2019){Zhang}, {Slavin}, {Foster}, {Smith}, {ZuHone},
  {Zhou}, \& {Chen}}]{zhang19}
{Zhang}, G.-Y., {Slavin}, J.~D., {Foster}, A., {et~al.} 2019, \apj, 875, 81,
  \dodoi{10.3847/1538-4357/ab0f9a}

\bibitem[{{Zhao} {et~al.}(2020){Zhao}, {Jiang}, {Li}, {Chen}, {Yu}, \&
  {Wang}}]{zhao20}
{Zhao}, H., {Jiang}, B., {Li}, J., {et~al.} 2020, \apj, 891, 137,
  \dodoi{10.3847/1538-4357/ab75ef}

\bibitem[{{Zhou} \& {Vink}(2018)}]{zhou18}
{Zhou}, P., \& {Vink}, J. 2018, \aap, 615, A150,
  \dodoi{10.1051/0004-6361/201731583}

\end{thebibliography}
\bibliographystyle{aasjournal}



\end{document}